\documentclass[a4paper,11pt]{article}
\usepackage{jheppub}
\usepackage{lineno}
\usepackage{txfonts}

\newlength{\fighskip} \fighskip=2pt
\newlength{\figvskip} \figvskip=3pt
\newcommand*{\figbox}[2]{{
\def\figscale{#1}
\def\arraystretch{0.8}
\arraycolsep=0pt
\begin{array}{c}
\vbox{\vskip\figscale\figvskip
\hbox{\hskip\figscale\fighskip
\includegraphics[scale=\figscale]{#2}}}
\end{array}}}

\title{\boldmath Quantum information recovery from black hole with projective measurement}

\author[a]{Ran Li,}
\author[b]{Jin Wang}
\affiliation[a]{Department of Physics, Qufu Normal University, Qufu, Shandong 273165, China}
\affiliation[b]{Department of Chemistry, and Department of Physics and Astronomy, The State University of New York at Stony Brook, Stony Brook, NY 11794, USA}
\emailAdd{liran@qfnu.edu.cn}
\emailAdd{jin.wang.1@stonybrook.edu}

\abstract{We studied the Hayden-Preskill thought experiment with the local projective measurement. Compared to the original model, the measurement is applied on the Hawking radiation that was emitted after throwing the quantum diary into the black hole. Within this setup, we explored various aspects of this model, including the information recovery from the black hole, the relation to the black hole final state proposal, the relation between the Yoshida-Kitaev protocol and Petz recovery map, the effects of the decoherence, and the quantum simulations of the decoding protocols. These aspects may provide us new insights into the non-perturbative nature of quantum black holes.}

\begin{document}
\maketitle
\flushbottom

\section{Introduction}
\label{sec:intro}

Whether the quantum information consumed by the black hole can be retrieved from the Hawking radiation is still a controversial problem, commonly referred to as the black hole information puzzle \cite{Hawking:1976ra}. Various proposals, including the holographic principle \cite{Maldacena:1997re}, the firewall hypothesis \cite{Almheiri:2012rt}, and the entanglement island \cite{Penington:2019npb,Almheiri:2019psf}, have been put forward to address this issue. However, a complete resolution is still challenging and realizing that may necessitate a deeper understanding of the UV-complete theory of quantum gravity.

It is generally believed that quantum information theory may provide insights into the black hole information puzzle. Page initiated the discussion on the entanglement entropy of the Hawking radiation \cite{Page:1993df,Page:1993wv}. By treating the black hole as an ordinary quantum system with $e^{S_{BH}}$ degrees of freedom, where the coarse-grained entropy $S_{BH}$ is proportional to the horizon area, he showed that the information is retrievable from the black hole mainly after the so-called Page time. Later, the issue of recovering information from the black hole was sharpened by Hayden and Preskill in a well known thought experiment \cite{Hayden:2007cs}, where a diary encoding the quantum information is thrown into the black hole that is maximally entangled with the Hawking radiation. By postulating that the interior dynamics of the black hole is both unitary and fast scrambling \cite{Sekino:2008he}, the study reveals that the black holes can release information remarkably quickly and decoding the quantum information from the Hawking radiation is information-theoretically possible. But the computational complexity of the decoding operations is not clearly resolved in their study. This issue was addressed later by Yoshida and Kitaev in \cite{Yoshida:2017non}, where two particular decoding strategies for reconstructing a quantum state from the Hawking radiation in the Hayden-Preskill thought experiment were proposed. It is shown that the probabilistic strategy can be promoted to be deterministic by increasing the circuit complexity. Yoshida-Kitaev decoding strategy has been attracted significant attentions, including the decoherence effects on the information recovery \cite{Yoshida:2018vly,Bao:2020zdo}, the deterministic decoding for the Clliford scrambling dynamics \cite{Yoshida:2021xyb,Oliviero:2022url}, the finite temperature effects \cite{Cheng:2019yib,Li:2021mnl}, the effects from local projective measurement or post-selection \cite{Yoshida:2021haf,Yoshida:2022srg,Li:2023rue} and realizing the circuit on the quantum processors \cite{Landsman:2018jpm}.

In the present work, we will investigate a revised version of Hayden-Preskill experiment firstly suggested by Yoshida in \cite{Yoshida:2022srg}. The difference from the original model lies in the introduction of the local projective measurement that is applied on the Hawking radiations at the late times, i.e. the radiations that were emitted after throwing the quantum diary into the black hole. This model was initially inspired by the monitored quantum circuits \cite{Li:2018mcv,Skinner:2018tjl,Chan:2018upn}, which consists of both the unitary dynamics and the local projective measurement. In \cite{Yoshida:2022srg}, it was pointed out that the Yoshida-Kitaev decoding strategies can be used to recover information from the Hayden-Preskill experiment with the projective measurement. However, the decoding probability and the fidelity were not calculated. We will utilize the graphical representation technique to calculate the decoding probability and the fidelity for this model. Because the projective measurement is closely related to the post-selection in quantum mechanics, we will also discuss the relation between the present model and the black hole final state model proposed by Horowitz and Maldacena \cite{Horowitz:2003he}, more precisely the generic final state model proposed by Lloyd and Preskill \cite{Lloyd:2013bza}. Recently, from the view point of quantum channel, the Petz recovery map is demonstrated to be equivalent to the Yoshida-Kitaev protocol for the Hayden-Preskill experiment \cite{Nakayama:2023kgr}. Inspired by this observation, we will establish that the Yoshida-Kitaev decoding protocol, specifically applied to the Hayden-Preskill experiment with the local projective measurement, is also equivalent to the Petz recovery channel. The equivalence can be nicely illustrated by using the graphical representations.

The aforementioned discussions are conducted under the assumption of an ideal scenario where noise or decoherence is not considered. In the real situations, the noise or the decoherence can not be avoided. Therefore, it is interesting to investigate the decoherence effects on the model. In quantum information theory, there exists various quantum channels that can used to model the decoherence or noise from the environments \cite{Wilde:2011npi}. We will focus on two types: the depolarizing channel and the dephasing channel. Towards explicit calculations, we show that the depolarizing channel can reduce the decoding probability as well as the fidelity. Conversely, the dephasing channel has no impact on the decoding procedure. Finally, we will conduct an experimental simulation of the decoding protocol by using the quantum processors. We execute two circuits on the quantum processor, one for teleporting the quantum state and another for teleporting the quantum entanglement. It is shown that the Yoshida-Kitaev protocol for teleporting quantum state is superior to the protocol for teleporting quantum entanglement.

This paper is arranged as follows. In section \ref{sec:IR_PM}, we study the quantum information recovery in the Hayden-Preskill experiment with the local projective measurement by using the Yoshida-Kitaev protocol. In section \ref{sec:relation_fs}, the relation between the model with the projective measurement and the black hole final state proposal is discussed. In section \ref{sec:petz_map}, by using the graphical representations, we show the Yoshida-Kitaev protocol as a quantum channel is equivalent to the Petz recovery map for the Hayden-Preskill experiment. In section \ref{sec:IR_noise}, the effects of the depolarizing noise on the information recovery are discussed. In section \ref{sec:q_sim}, we show the results of executing the quantum circuits for teleporting quantum state and quantum entanglement on the quantum processors. The conclusion and discussion are presented in the last section. In Appendix, we provide the detailed calculations of the Haar averages of some quantities used in the main text by invoking the graphical representations.

\section{Information recovery with projective measurement}
\label{sec:IR_PM}

\begin{figure}
    \centering
    \includegraphics[width=0.3\textwidth]{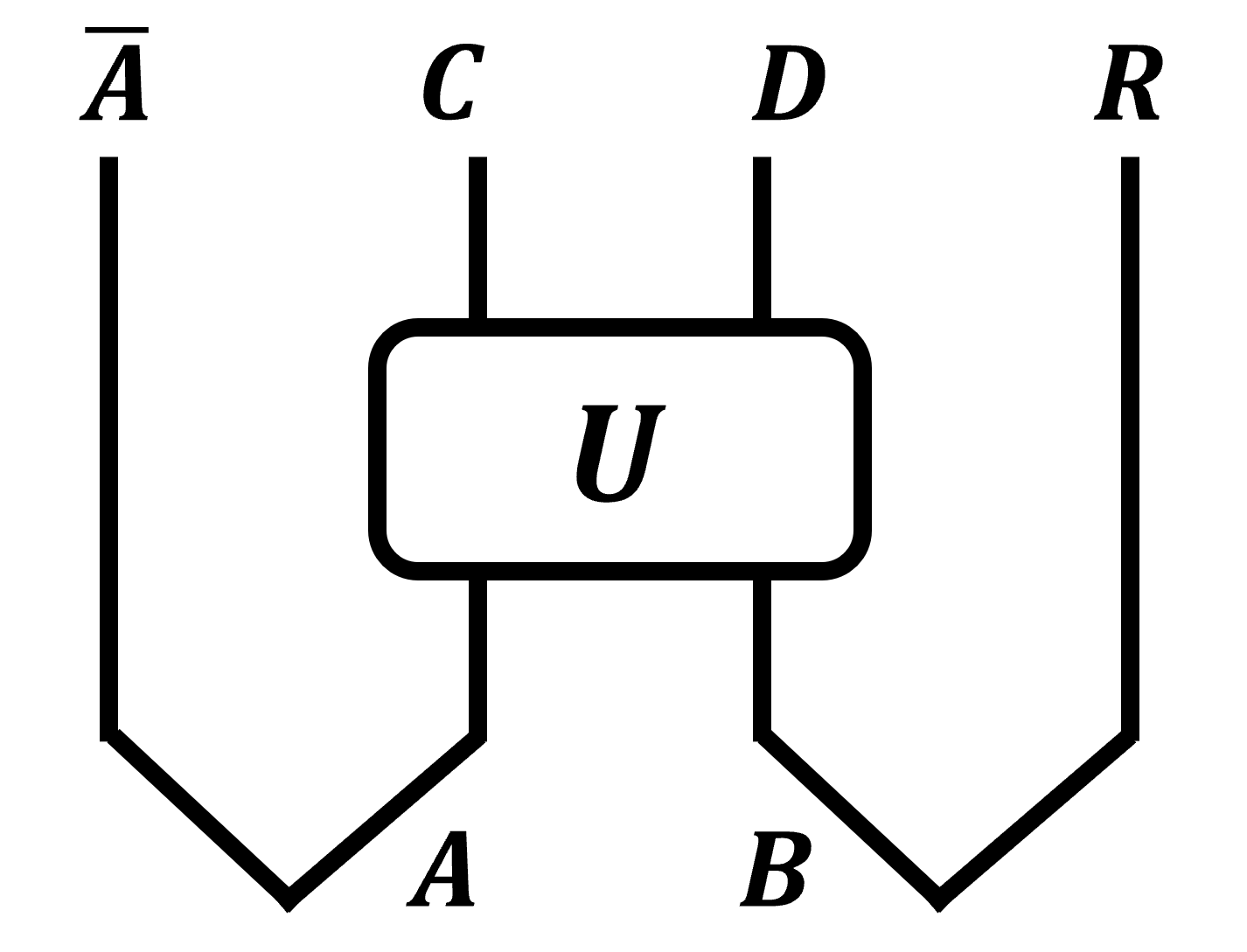}\quad
    \quad\quad
    \includegraphics[width=0.22\textwidth]{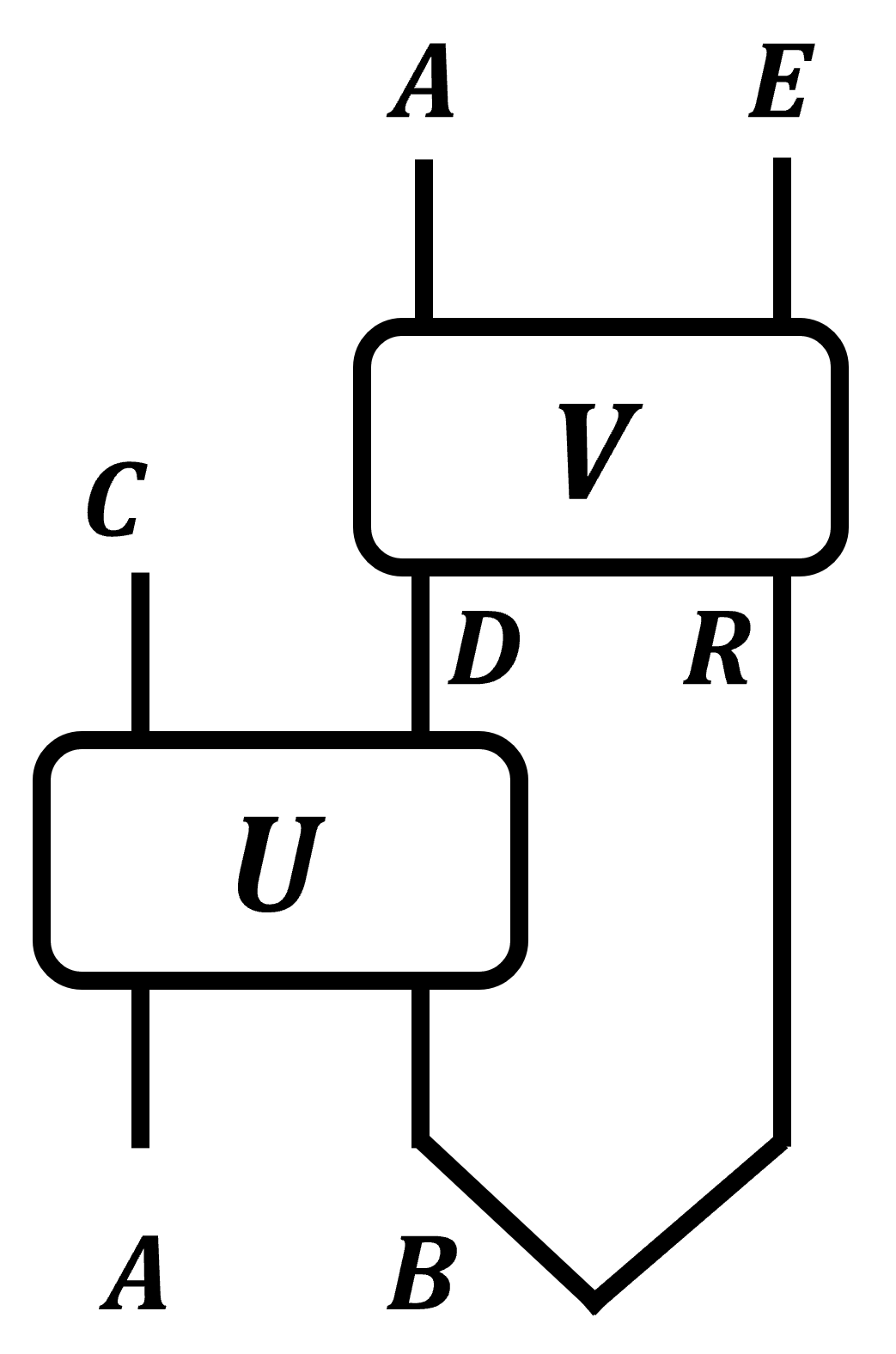}
    \caption{Hayden-Preskill thought experiment and its decoder. }
    \label{HP_experiment}
\end{figure}

The original Hayden-Preskill thought experiment considered the question of under which condition quantum information thrown into black hole can be retrieved. As an ideal model, which is illustrated in Figure \ref{HP_experiment}, quantum information encoded in system $A$ is thrown into an old black hole denoted as $B$. For convenience, a reference system $\overline{A}$ is also introduced, which is maximally entangled with system $A$. For an old black hole $B$, it is maximally entangled with the early Hawking radiation $R$. Assuming the internal dynamics of black hole are both unitary and fast scrambling, it becomes feasible to swiftly retrieve the quantum information from the black hole. In the left panel of Figure \ref{HP_experiment}, the internal unitary dynamics is represented by $U$. After some time, typically the scrambling time, the outcome of the black hole's internal dynamics consists of the late Hawking radiation $D$ and the remainder black hole $C$. Detailed calculations demonstrate that when $|D|\gg|A|$, the entanglement between $\overline{A}$ and black hole transfers to the entanglement between $\overline{A}$ and $DR$, i.e. there exists no entanglement between $\overline{A}$ and $C$. An external decoder, who has complete access to radiation $D$ and $R$ as shown in the left panel of Figure \ref{HP_experiment}, can theoretically retrieve the quantum information.

The Yoshida-Kitaev protocol provides a solution on how to decode the quantum information through local operations on $R$ and $D$. The essential of the protocol is to implementing an auxiliary entangled system and applying the operator $U^*$ on the early radiation $R$ with the auxiliary system to reverse the time evolution effectively. Then, post-selecting or projecting the outcome with the late radiation onto the EPR state enables the recovery of the initial state of $A$ from the auxiliary system. However, the post-selecting or the projecting is probabilistic and the protocol can be promoted into the deterministic one with the cost of increasing the computational complexity.

\subsection{Hayden-Preskill protocol with projective measurement}

In the present work, we consider a modified version of Hayden-Preskill protocol \cite{Yoshida:2022srg}. Consider a projective measurement is applied to $D$ after the scrambling time. Without loss of generality, we assume that the outcome of the measurement is the state $|0\rangle_{D}$. The projective measurement means the violation of the entanglement between $\overline{A}$ and $DR$. Specifically, the projective measurement destroys all quantum correlations between the subsystem $D$ and the rest of the system. Therefore, it is not apparent that the information can still be retrieved due to this kind of violation of entanglement. Now, if the decoder still wants to retrieve the quantum information swallowed by the black hole, he can only perform the local operations on the early radiation $R$. In this section, we study the question of how to retrieve the quantum information in this modified version of Hayden-Preskill thought experiment.

With the description of the model, the Hayden-Preskill protocol with the projective measurement can be graphically represented by 
\begin{eqnarray}\label{HP_state_eq}
    |\Psi_{HP}\rangle&=&\sqrt{|D|} ~_{D}\langle 0| I_{\overline{A}}\otimes U_{AB}\otimes I_{R}|EPR\rangle_{\overline{A}A}\otimes|EPR\rangle_{BR}    \nonumber\\
    &=&\sqrt{|D|}\quad \figbox{0.3}{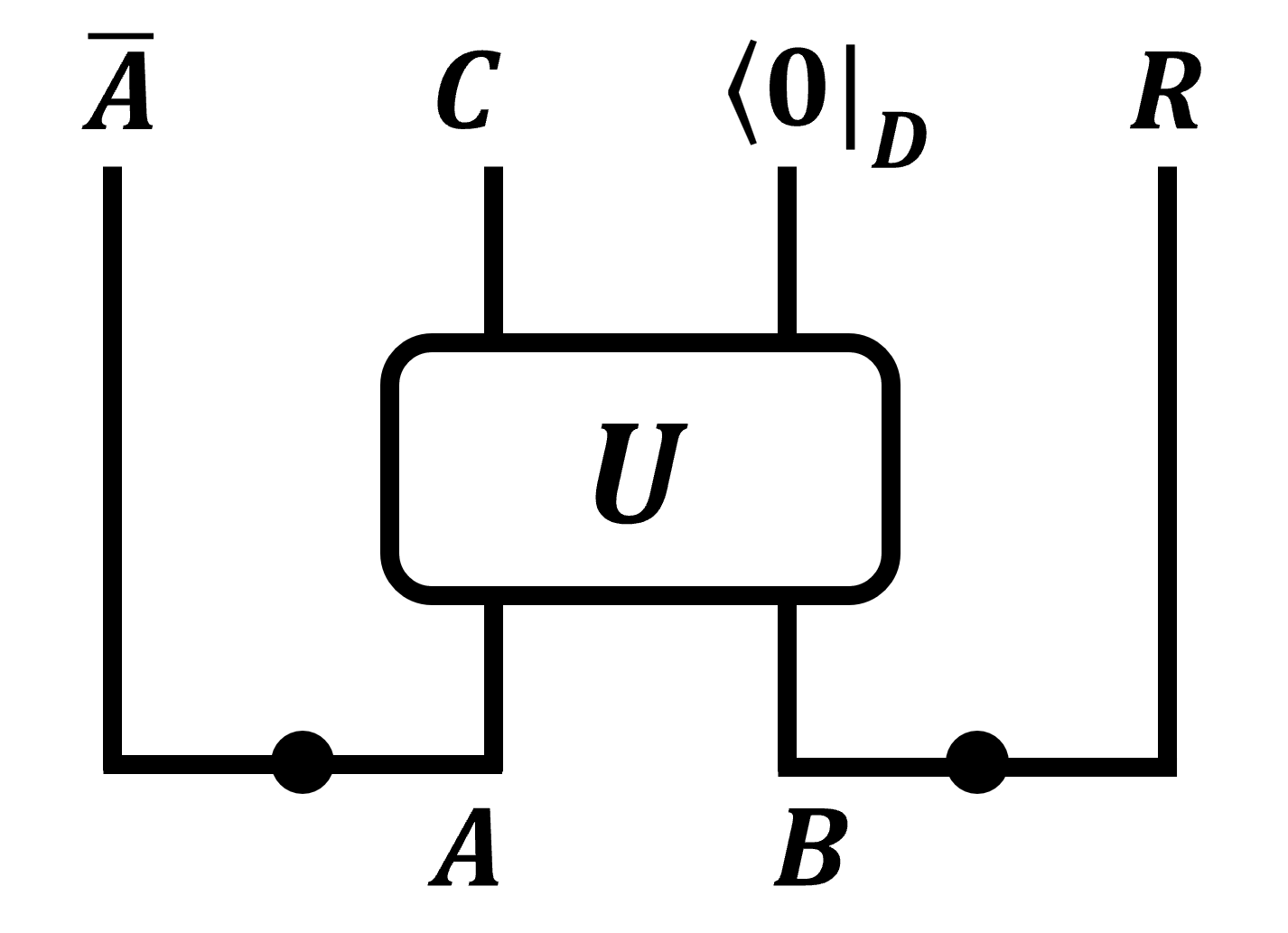}\;,
\end{eqnarray}
where $|D|$ denotes the dimension of Hilbert space for the subsystem $D$. In the following, we use the $|\cdot|$ to denote the dimension of Hilbert space for the corresponding system. In this graphical representation, $\figbox{0.2}{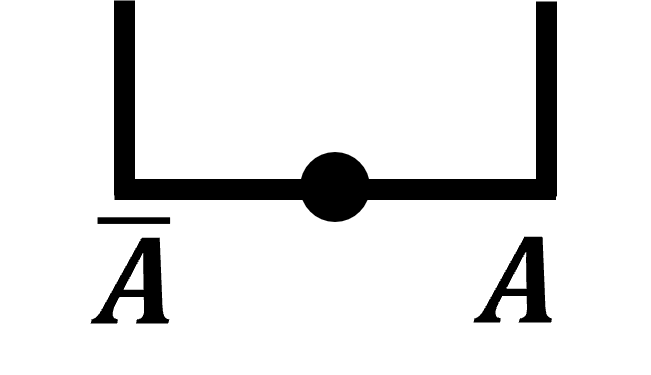}$ represents the $\textrm{EPR}$ state of the subsystems $A$ and $\overline{A}$ 
\begin{eqnarray}
    |EPR\rangle_{\overline{A}A}=\frac{1}{\sqrt{|A|}}\sum_{i=0}^{|A|-1}|i_{\overline{A}}\rangle\otimes|i_{A}\rangle\;.
\end{eqnarray} 
The black dot stands for the normalization factor $\frac{1}{\sqrt{|A|}}$. Similar rules applies to the system $B$ and $R$. It is obvoius that $|A|=|\overline{A}|$ and $|B|=|R|$. The prefactor $\sqrt{|D|}$ in Eq.\eqref{HP_state_eq} is introduced to guarantee the normalization condition
\begin{eqnarray}
    \langle \Psi_{HP}|\psi_{Hp}\rangle=1.
\end{eqnarray}
The calculation of the norm of the state $|\Psi_{HP}\rangle$ is given in the Appendix \ref{sec:HP_norm}.

We now consider under which condition the information carried by $A$ can be retrieved by decoding the radiation. This can be obtained by judging the disentanglement between the reference system $\overline{A}$ and the remainder black hole $C$. To this aim, we consider the following quantity
\begin{eqnarray}\label{quantity}
     \overline{\left\|\rho_{\overline{A}C}-\rho_{\overline{A}} \otimes \rho_C \right\|_1}=\int dU \left\|\rho_{\overline{A}C}-\rho_{\overline{A}} \otimes \rho_C \right\|_1 \;.
\end{eqnarray}
Here, $\left\|\mathcal{O}\right\|_1=\textrm{Tr}\sqrt{\mathcal{O}^\dagger\mathcal{O}}$ is defined as the trace distance of the operator $\mathcal{O}$. The reduced density matrices of the corresponding subsystems are denoted as $\rho_{\overline{A}C}$, $\rho_{\overline{A}}$ and $\rho_C$, respectively. The integral $\int dU$ stands for the integration of the unitary operator $U$ over the Haar measure. The quantity in Eq.\eqref{quantity} describes the Haar average distance between the reduced state $\rho_{\overline{A}C}$ and the direct product of $\rho_{\overline{A}}$ and $\rho_C$. If there is no entanglement between the subsystems $\overline{A}$ and $C$, this quantity should be small enough.

For the original Hayden-Preskill protocol, the reduced density matrices for the subsystem $\overline{A}$ and $C$ are just the maximally mixed density matrix. Here, with the projective measurement on $D$, $\rho_C$ is still maximally mixed, but $\rho_{\overline{A}}$ is not. Therefore, it is reasonable to consider the quantity in Eq.\eqref{quantity} instead of the quantity with $\rho_{\overline{A}}$ and $\rho_C$ are maximally mixed.

It can be shown that \cite{Harlow:2014yka}
\begin{eqnarray}
\overline{\left\|\rho_{\overline{A}C}-\rho_{\overline{A}} \otimes \rho_C \right\|_1}\leq  \sqrt{|\overline{A}||C|} \left[\overline{\textrm{Tr}\left(\rho_{\overline{A}C}^2\right)}-\overline{\textrm{Tr}\left(\rho_{\overline{A}C}\cdot\left(\rho_{\overline{A}}\otimes\rho_{C}\right)\right)} \right]^{\frac{1}{2}}\;,
\end{eqnarray}
where we have used the fact that 
\begin{eqnarray}
    \textrm{Tr}\left[\rho_{\overline{A}C}\cdot\left(\rho_{\overline{A}}\otimes\rho_{C}\right)\right]=\textrm{Tr}\left[\left(\rho_{\overline{A}}\otimes\rho_{C}\right)^2\right]\;.
\end{eqnarray}
Explicit calculations (see Appendix \ref{sec:Trac_rho2} and \ref{sec:Haar_3rho}) show that 
\begin{eqnarray}
    \overline{\textrm{Tr}\left(\rho_{\overline{A}C}^2\right)}&=&
    \frac{1}{(d^2-1)}\left(|A|d+ \frac{|D|}{|A|}d -\frac{|A||D|}{d} -\frac{d}{|A|} \right)\;,\label{trace_result1}\\
   \overline{\textrm{Tr}\left[\rho_{\overline{A}C}\cdot\left(\rho_{\overline{A}}\otimes\rho_{C}\right)\right]}&=&\frac{1}{(d^2-1)}\left(\frac{|A|}{|C|^2}d+ \frac{|B|}{|C|}d -\frac{|A|}{|C|} -\frac{|B|}{|C|^2} \right)\;.\label{trace_result2}
\end{eqnarray}
where $d$ is the dimension of the Hilbert space that the scrambling operator $U$ acts on. Note that $d=|A||B|=|C||D|$ for an unitary dynamics.

Combining the previous results, we have the following inequality  
\begin{eqnarray}\label{estimat_1}
   \overline{\left\|\rho_{\overline{A}C}-\rho_{\overline{A}} \otimes \rho_C \right\|_1} &\leq& \sqrt{\frac{\left(|A|^2-1\right)\left(|C|^2-1\right)|D|}{\left(d^2-1\right)}} \nonumber\\
   &\approx& \frac{|A|}{\sqrt{|D|}}\;,
\end{eqnarray}
where in the approximation we have used the assumption that all the subsystems are large enough.

One can perform the estimation of the quantity given in Eq.\eqref{quantity} by replacing the reduced density matrices $\rho_{\overline{A}}$ with the maximally mixed, one can obtain the result 
\begin{eqnarray}\label{estimat_2}
    \overline{\left\|\rho_{\overline{A}C}-\frac{1}{|\overline{A}||C|}I_{\overline{A}} \otimes I_C \right\|_1} &\leq& \sqrt{\frac{\left(|A|^2-1\right)\left(|C|^2|D|-1\right)}{\left(d^2-1\right)}}\;.
\end{eqnarray}
For sufficiently large subsystems, the results presented in Eq.\eqref{estimat_1} and \eqref{estimat_2} have no difference. The reason is that on the average, the reduced density matrices $\rho_{\overline{A}}$ has no difference compared with the maximally mixed one.

The inequality means that if the decoupling condition
\begin{eqnarray}\label{Decoupling_cond}
    |D|\gg |A|^2
\end{eqnarray}
is attained, the reference subsystem is not entangled with the black hole any longer and the information contained in the subsystem $A$ can be retrieved from decoding the early radiation $R$.

Let us give more discussions on the model and the results. Originally, this kind of model was suggested by Yoshida as a simply toy model that reproduces the features of monitored quantum circuits. The monitored quantum circuits consist of both the unitary scrambling dynamics and the local projective measurements. It is generally believed that the local projective measurement break the long range quantum entanglement. From the viewpoint of quantum error correction, the local quantum information is encoded into a larger physical space by the unitary scrambling dynamics. Therefore, the local projective measurement cannot destroy the initial quantum state easily, as we have shown by the explicit calculations.

In fact, by turning the graph in Eq.\eqref{HP_state_eq} upside down and adding a series of scrambling operators and local projective measurements between the initial state and the final state, it can be transformed into a monitored quantum circuits. The insight from the monitored quantum circuits indicates that as long as the qubits number $n_A$ of subsystem $A$ is small than the half of the qubits $n_D$ of subsystem $D$, the output subsystems $A$ and $B$ of the monitored quantum circuit are maximally entangled despite the detailed properties of the scrambling dynamics and the initial state. This condition is consistent with the decoupling condition. Therefore, the retrievable of quantum information from the black hole with projective measurement is equivalent to the entanglement preserving in monitored quantum circuits \cite{Yoshida:2022srg}.

\subsection{Yoshida-Kitaev decoding protocol}

Now we consider whether the Yoshida-Kitaev decoding protocol still works for the Hayden-Preskill protocol with the projective measurement. In this case, the late radiation subsystem $D$ has been projected onto a specific state. The decoder can only apply the local operations on the early radiation subsystem $R$.

Given the Hayden-Preskill state $|\Psi_{HP}\rangle$ as graphically represented in Eq.\eqref{HP_state_eq}, the Yoshida-Kitaev decoding protocol proceeds as follows:

\begin{itemize}
    \item[1.] Prepare a copy of $|EPR\rangle_{\overline{A}A}$, denoted as $|EPR\rangle_{A'\overline{A}'}$. 
    
    \item[2.] Apply $U^*$ on $RA'$. We denote the resultant state as the state $|\Psi_{in}\rangle$. It can be graphically represented as 
    \begin{eqnarray}
        |\Psi_{in}\rangle=\sqrt{|D|} \quad \figbox{0.3}{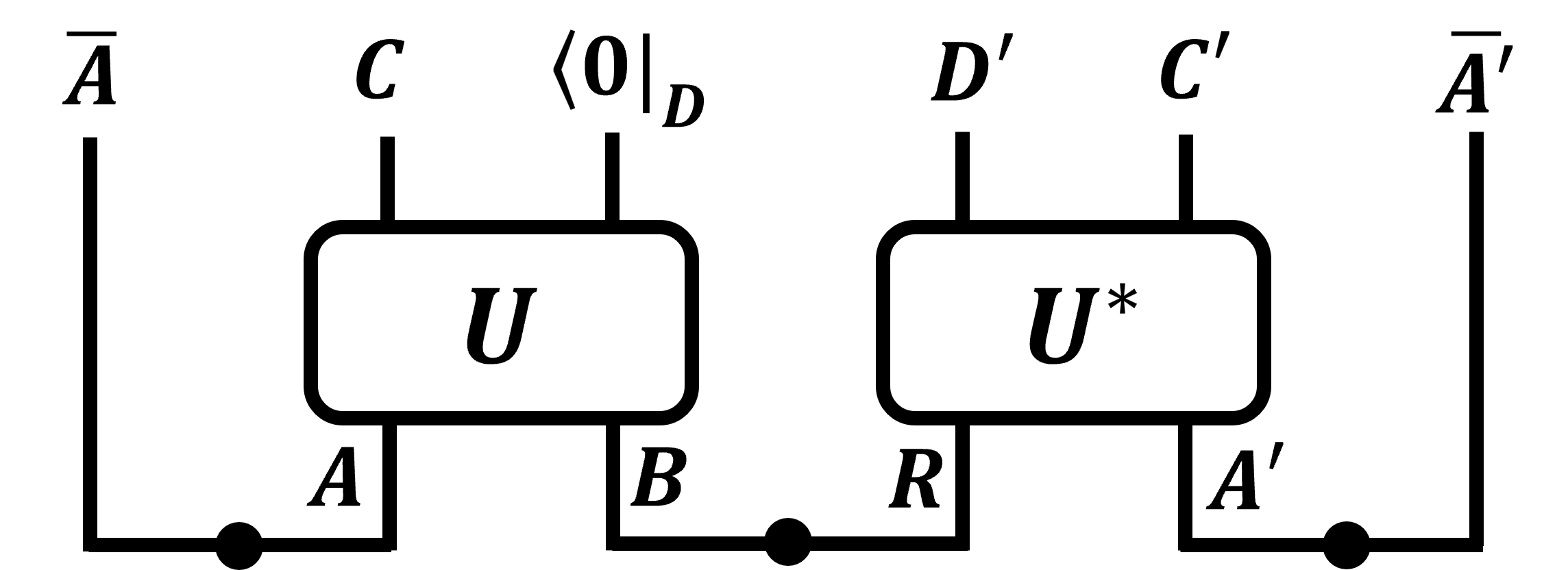}\;.
    \end{eqnarray}
    It can be checked that $|\Psi_{in}\rangle$ is normalized, i.e. $\langle\Psi_{in}|\Psi_{in}\rangle=1$. The calculation is presented in \ref{sec:psi_in_norm}. 

    \item[3.] Do the projective measurement on $D'$. If the outcome is $|0\rangle$, it means the successful decoding of of the initial information contained in $A$. We denote the resultant state as $|\Psi_{out}\rangle$, which can be graphically represented as
    \begin{eqnarray}\label{psi_out}
        |\Psi_{out}\rangle=\frac{1}{\sqrt{P}} ~_{D'}\langle 0|\Psi_{in}\rangle=\sqrt{\frac{|D|}{P}} \quad \figbox{0.3}{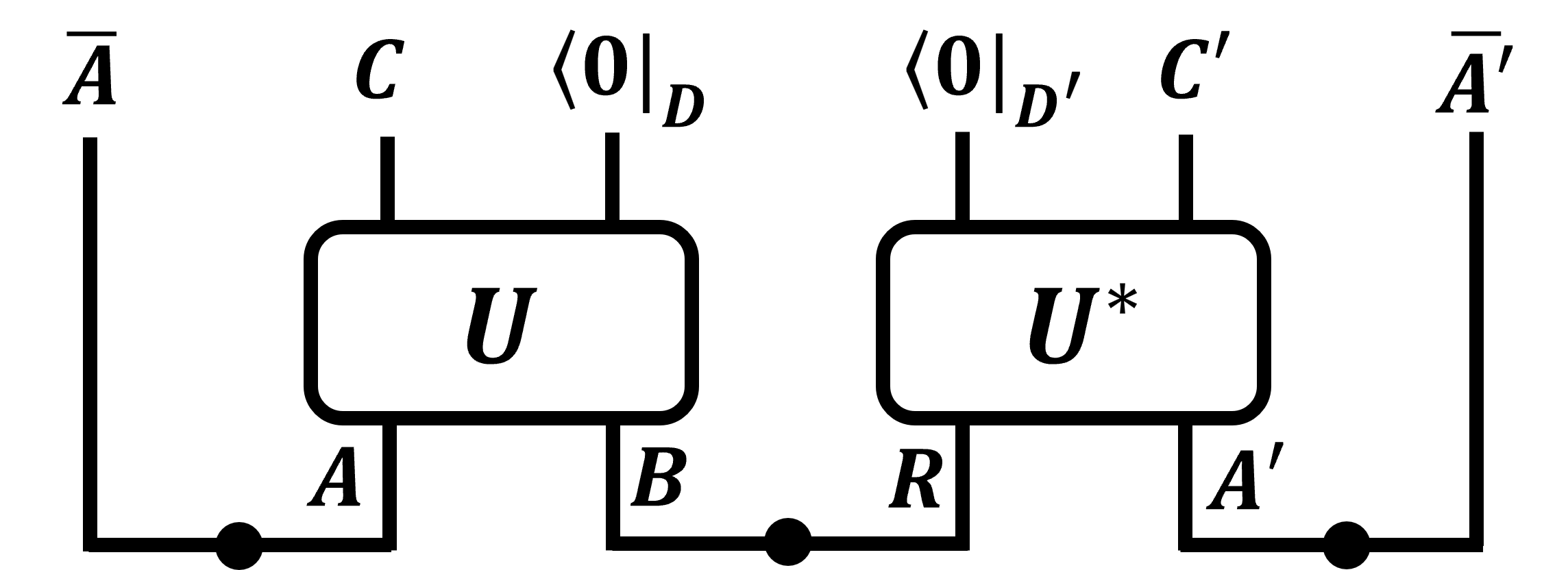}\;.
    \end{eqnarray}
An additional prefactor $\frac{1}{\sqrt{P}}$ is introduced to preserve the normalization of $|\Psi_{out}\rangle$. It is actually the projecting probability of the subsystem $D'$ onto $|0\rangle$.  
    
\end{itemize}

At this stage, let us discuss more on the local projective measurement. In the present model, the projective measurement is performed on the late radiation subsystem $D$, i.e. the measurement is done by the decoder outside of the black hole. After the measurement, the decoder can only operate the early radiation to recover the information. This is different from the model studied in \cite{Li:2023rue}. In that model, inspired by the non-isometric map \cite{Akers:2022qdl}, a portion of black hole degrees of freedom is post-selected onto fixed state. The post-selection happens in the black hole interior and the decoder can still operate the early and the late radiation to recover the information.

One can also imagine that the local projective measurement is performed by an intruder who has the access only to the late radiation $D$. Then the current model is similar to the model suggested by Yan et.al in \cite{Yan:2020fxu}. However, their model is not based on the framework of black hole scrambling. It is shown that the decoder who has the access to the partially destroyed system can still recover the initial information by applying the time reversal unitary dynamics and using the technique of quantum state tomography. Here, the situation is a little different. The subsystem $C$, which is the remainder black hole, is not available to the external decoder. The decoder can only apply the local operations on the early radiation and the local measurement on the outcome subsystem $D'$. In principle, the quantum state of the subsystem $D$ after the measurement of the intruder is not known to the decoder, although the intruder cannot get anything useful about the initial state. In this case, the decoder can also invoke the technique of quantum state tomography to perfectly reconstruct the state of the subsystem $D$. This will allow the decoder to compare his measurement outcome with the quantum state of the subsystem $D$. Therefore, the Yoshida-Kitaev decoding protocol can in principally be applied to recover the information in the case that the projective measurement is performed by an intruder.

The Yoshida-Kitaev decoding protocol is a probabilistic one. We confront with the following two questions: (1) the probability of successful decoding, i.e. the probability of the projecting the state of $D'$ onto $|0\rangle$; (2) the decoding fidelity measured by the matching of the state $|\Psi_{out}\rangle$ with the state $|EPR\rangle_{\overline{A}\overline{A}'}$, which quantifies the quality of the decoding.  

Using the graphical representation, the decoding probability is given by
\begin{eqnarray}
    P&=&\langle \Psi_{in}|0\rangle_{D'}\langle 0|\Psi_{in}\rangle=\frac{|D|}{|A|^2|B|} \quad\figbox{0.3}{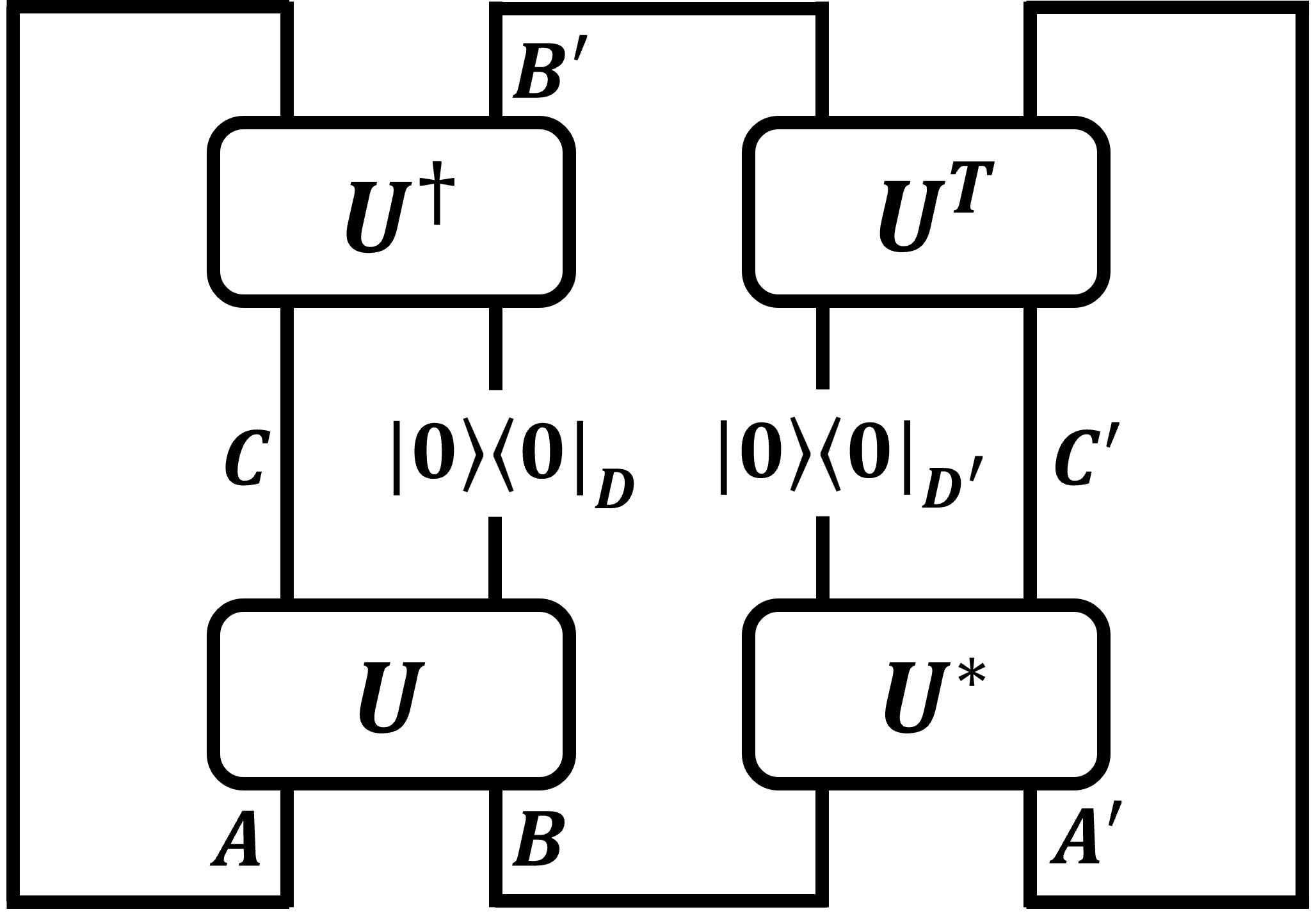}\;,
\end{eqnarray}
where the graph is just the representation of $\textrm{Tr}\left(\rho_{\overline{A}C}^2\right)$. Therefore, one can calculate the Haar average of the decoding probability as
\begin{eqnarray}\label{Decoding_Prob}
    \overline{P}=\int dU P&=& \frac{|B|}{|D|} \int dU \textrm{Tr}\left(\rho_{\overline{A}C}^2\right)\nonumber\\
    &=&\frac{1}{(d^2-1)}\left(\frac{d^2}{|A|^2}+\frac{d^2}{|D|} -\frac{d^2}{|A|^2|D|} -1 \right)\nonumber\\
    &\approx&\frac{1}{|A|^2}\;,
\end{eqnarray}
where in the last step we have used the decoupling condition Eq.\eqref{Decoupling_cond}. It shows that, in the current model, although we should collect more radiation from the black hole to complete the decoding task, the decoding probability in the ideal case is the same as the original Hayden-Preskill protocol, which only depends on the Hilbert space dimension of the initial infalling subsystem $A$.

The decoding fidelity can be calculated as 
\begin{eqnarray}\label{Decoding_Fid}
    F&=&\textrm{Tr}\left[|EPR\rangle_{\overline{A}\overline{A}'}\langle EPR|\Psi_{out}\rangle\langle \Psi_{out}|\right]\nonumber\\
    &=&\frac{|D|}{|A|^3|B|P} \quad \figbox{0.3}{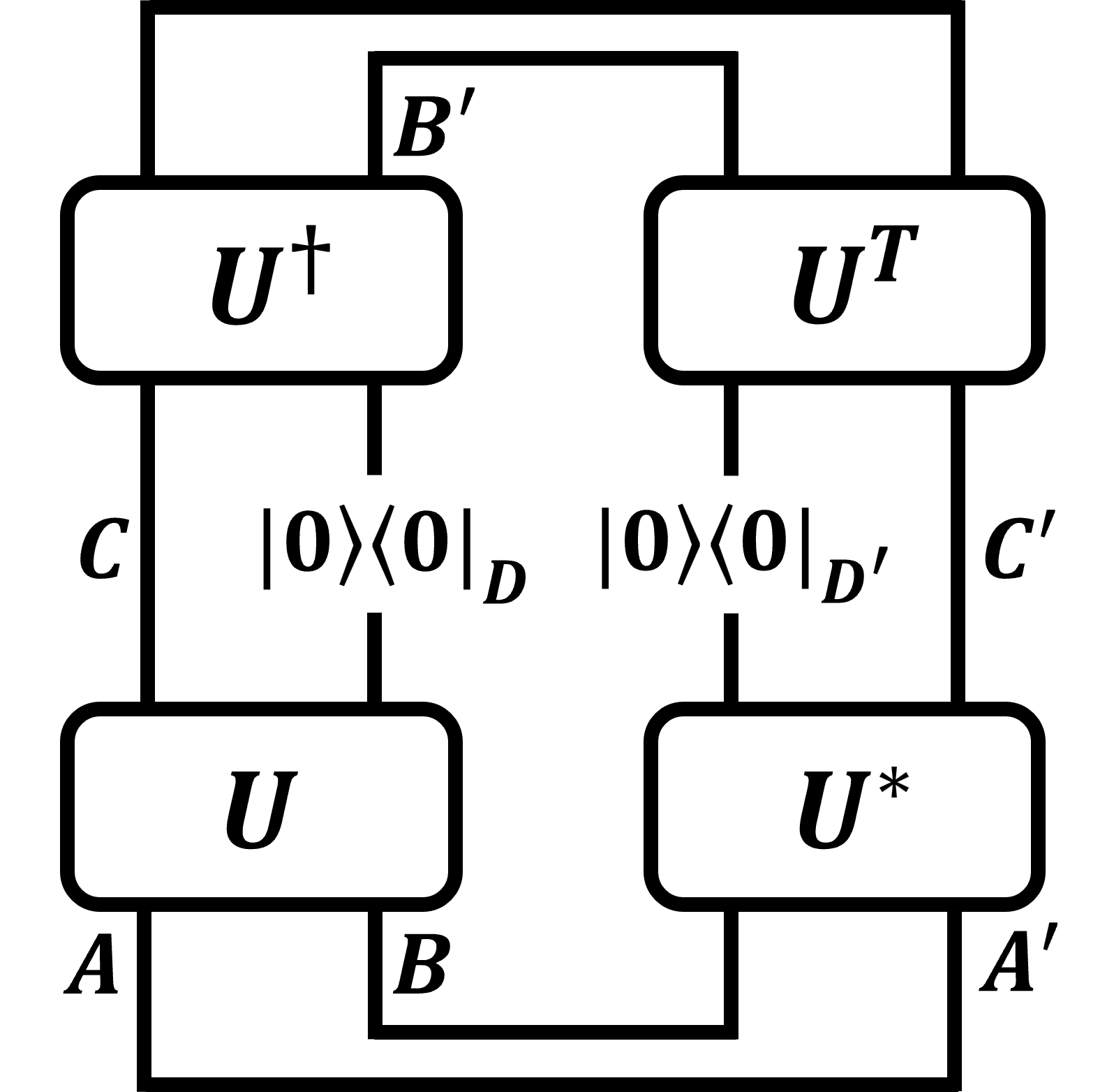}
    =\frac{|D|}{|A|^3|B|P} \quad \figbox{0.3}{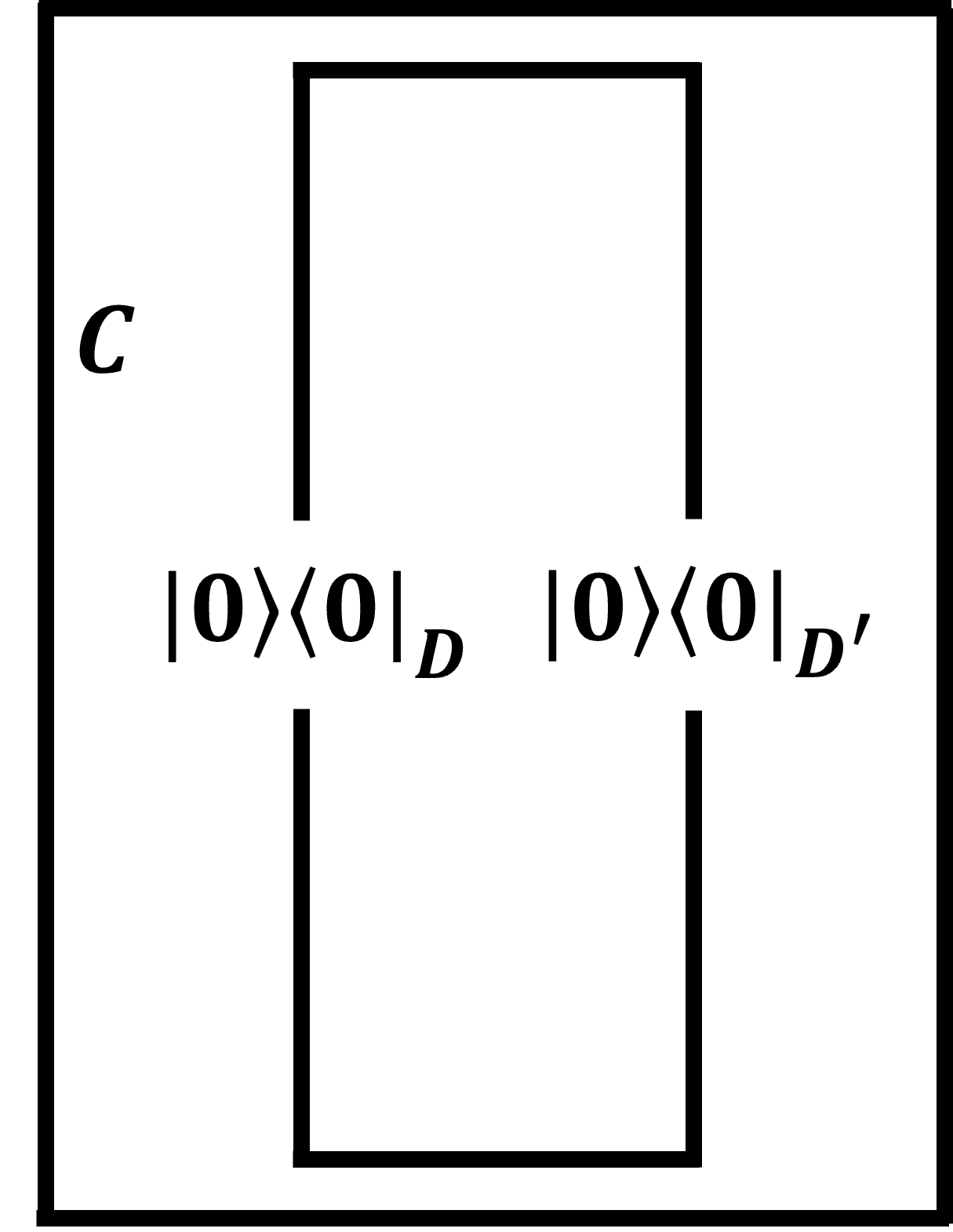}\nonumber\\
    &=&\frac{|C||D|}{|A|^3|B|P}\approx 1\;,
\end{eqnarray}
where in the last step we have used the average decoding probability $P\approx \frac{1}{|A|^2}$ when the decoupling condition is perfectly satisfied.

In the above equation, it also shows that the decoding fidelity is inversely proportional to the decoding probability as $F=\frac{1}{|A|^2 P}$, which means a small decoding probability resulting in the high decoding quality.

\section{Relation to black hole final state proposal}
\label{sec:relation_fs}

The final state projection model of black hole evaporation proposed by Horowitz and Maldacena (HM) aimed to resolve the paradox of the Hawking's semiclassical prediction with the unitary of black hole by post-selecting the quantum state of black hole interior onto the maximally entangled state at the singularity \cite{Horowitz:2003he}. This model also inspired the recent proposal of the non-isometric holographic map of the black hole interior \cite{Akers:2022qdl}, which states that at the late times of Hawking evaporation, a large portion of the degrees of freedom in the black hole interior from the effective field theory description is annihilated by the holographic map to the fundamental degrees of freedom.

\begin{figure}
    \centering
    \includegraphics[width=0.2\textwidth]{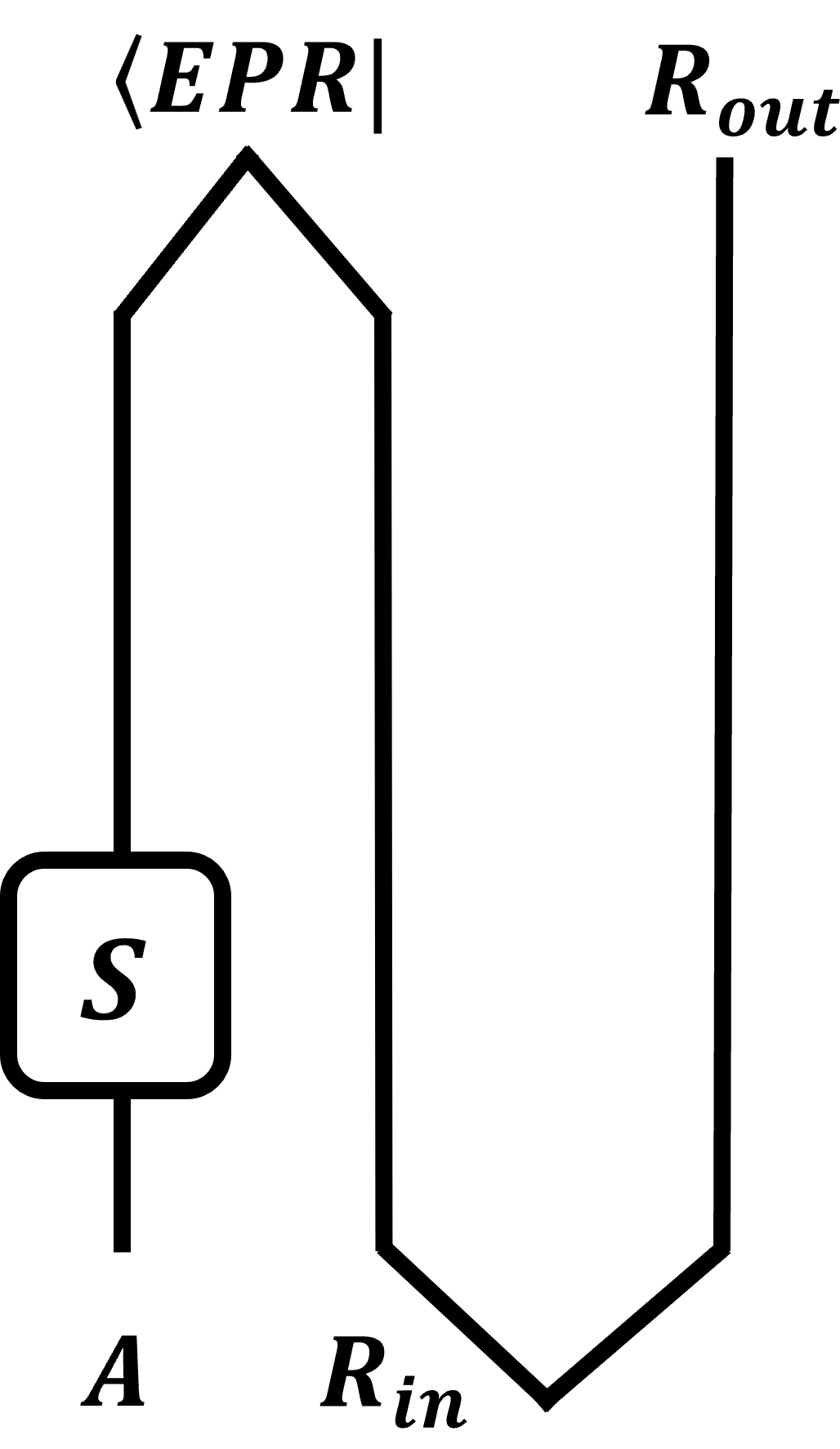}\quad\quad\quad  
    \includegraphics[width=0.2\textwidth]{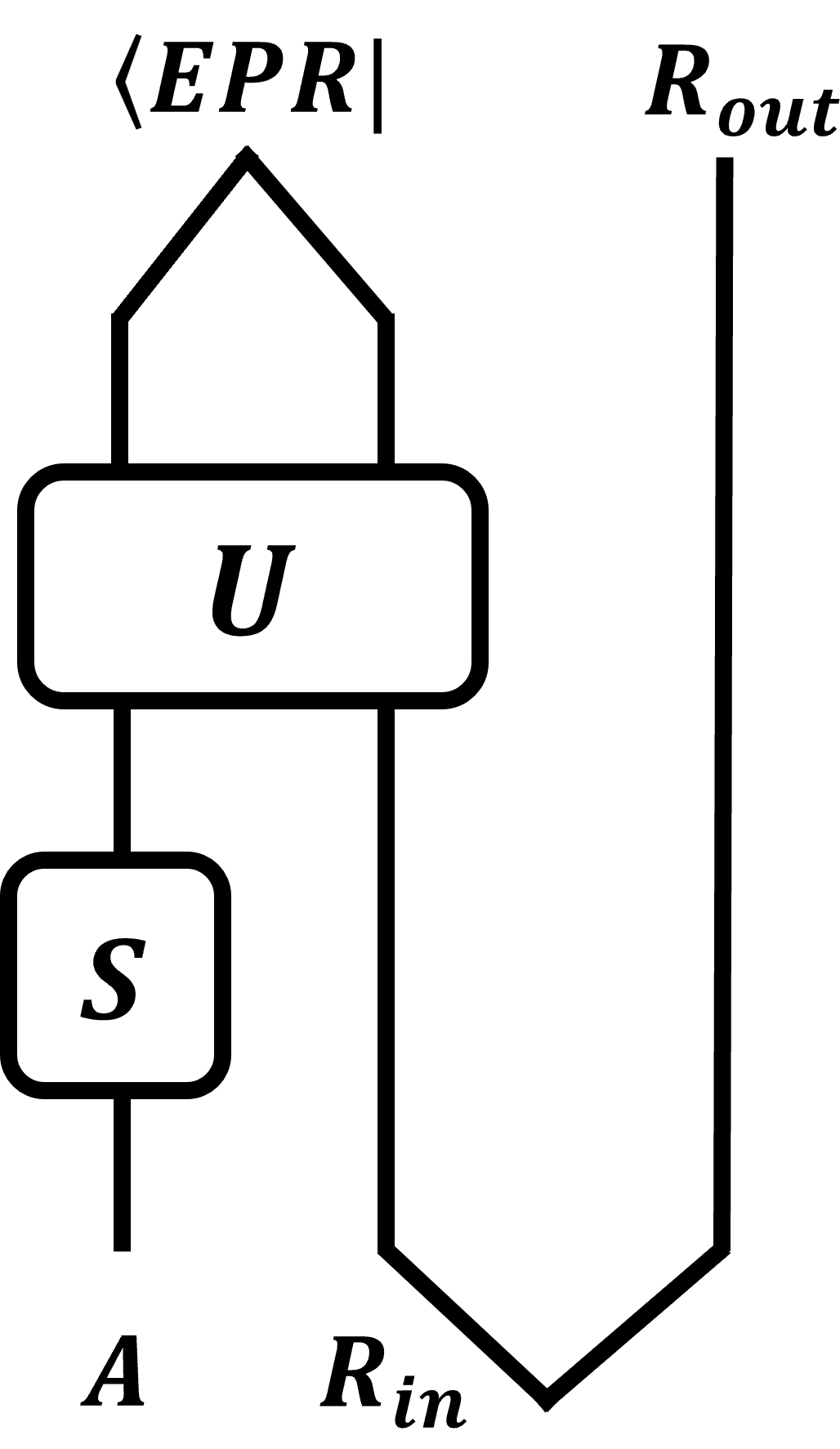}
    \caption{The original final state models proposed by Horowitz and Maldacena (left) and Gottesman and Preskill (right). The infalling matter system $A$ and the interior Hawking partner modes $R_{in}$ are post-selected onto a specific entangled state. In the left panel, $S$ is a unitary operator. In the right panel, $U$ is the scrambling operator that represents the interaction between in infalling matter and the interior modes. }
    \label{Final_state_HM_GK}
\end{figure}

The HM proposal (shown in the left panel of Figure \ref{Final_state_HM_GK}),  states that the infalling matter system $A$ and the interior Hawking partner modes $R_{in}$ are post-selected onto a specific entangled state. This is similar to quantum teleportation circuit, where the quantum information contained in the infalling matter system effectively flows backward in time and can be recovered from the radiation outside of the black hole. However, the original HM proposal does not take the scrambling dynamics in the black hole interior into account. In fact the interaction between the infalling matter system and the infalling partner system of the radiation inside of the black hole can effectively reduce the fidelity of the teleportation as discussed by Gottesman and Preskill in \cite{Gottesman:2003up}.

\begin{figure}
    \centering
    \includegraphics[width=0.3\textwidth]{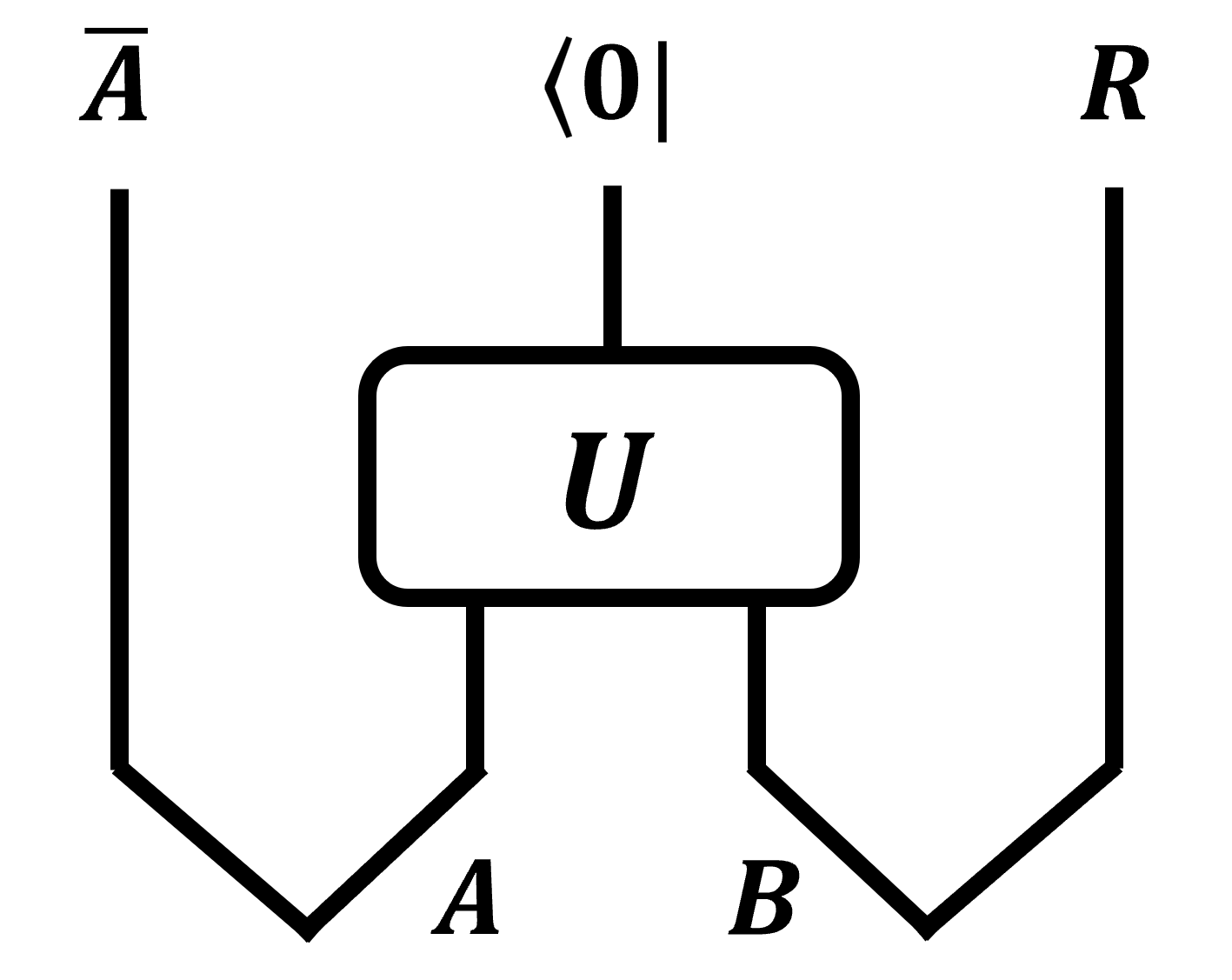}
    \caption{The generic final state model proposed by Lloyd and Preskill. In this circuit, the infalling matter $A$ is entangled with a reference system $\overline{A}$ and the black hole $B$ is entangled with the radiation $R$. }
    \label{Final_state}
\end{figure}

Lloyd and Preskill proposed a generic final state model \cite{Lloyd:2013bza}, which is illustrated in Figure \ref{Final_state}. Here, we consider a slightly different circuit but essentially the same version as the model proposed by Lloyd and Preskill. The meanings for the subsystem $A$, $\overline{A}$, $R$ and $B$ are interpreted in the caption of Figure \ref{Final_state}. After the scrambling dynamics between $A$ and $B$ inside the event horizon, at the singularity the interior degrees of freedom are all projected onto the specific state $|0\rangle$. The interaction can effectively result in that the state of $B$ and $R$ used in the teleportation protocol will not be maximally entangled.

\begin{figure}
    \centering
    \includegraphics[width=0.5\textwidth]{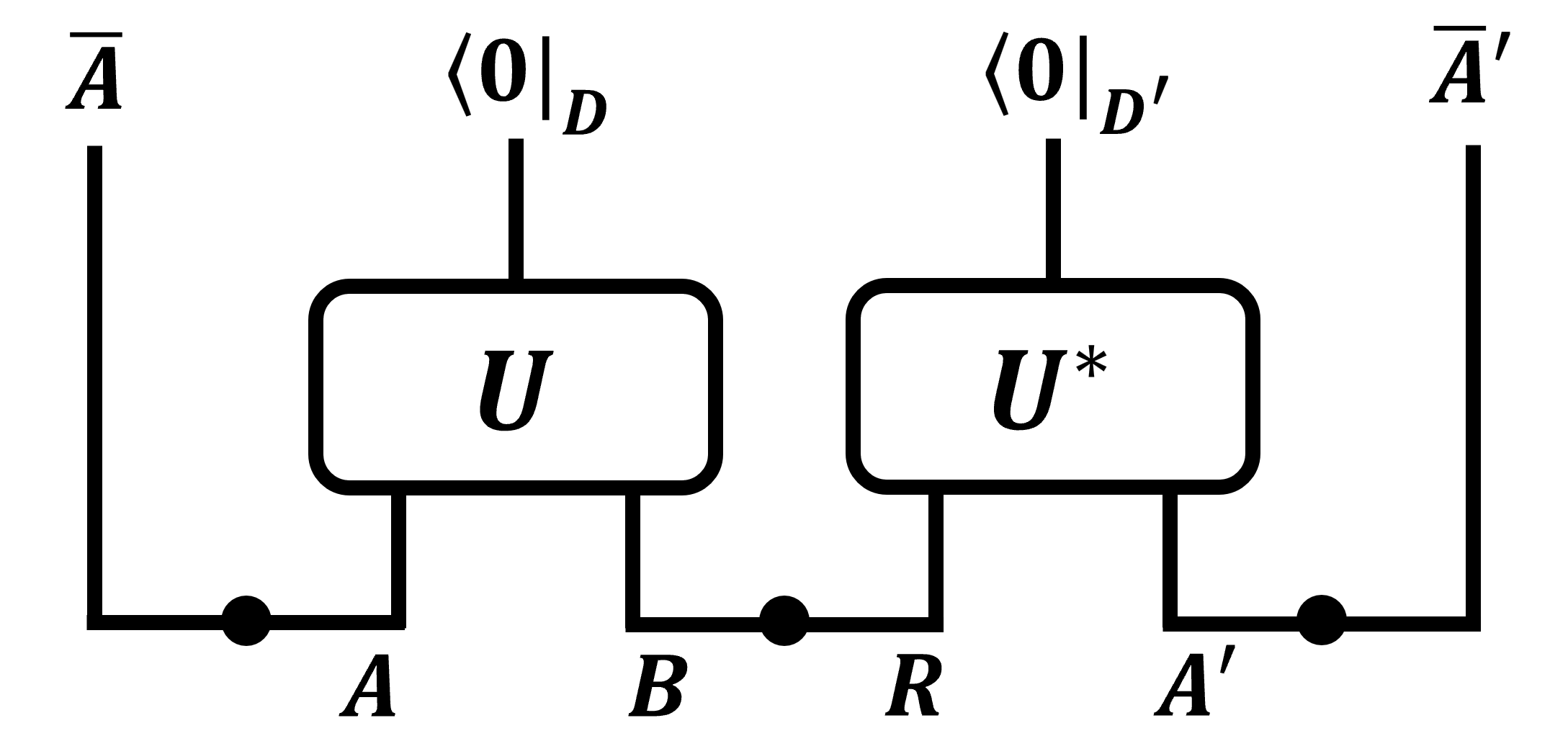}
    \caption{The decoding protocol for the generic final state model. }
    \label{Final_state_decoding}
\end{figure}

It is apparent that this version of the final state model can be properly interpreted as the Hayden-Preskill protocol where the projective measurement is performed on the whole outcome of the scrambling dynamics. Our previous discussions can be properly applied in this model by setting $|C|=1$. For the decoupling condition given in Eq.\eqref{Decoupling_cond} is converted into by noting that $|D|=|A||B|$ here
\begin{eqnarray}\label{decoupling_con_final}
    |B|\gg |A|\;.
\end{eqnarray}
Note that $|B|=|R|$. The above equation also implies that the quantum information can be retrieved from the radiation $R$ as long as the Hilbert space of the radiation $R$ is greater than that of the infalling matter system $A$. Therefore, the Yoshida-Kitaev decoding protocol can also be introduced to decode the initial quantum information contained in $A$. The decoding protocol is explicitly illustrated in Figure \ref{Final_state_decoding}. In this protocol, one may suspect that there is no way to compare the measurement outcomes of $D'$ with the final state $\langle 0|_{D}$ of the black hole because there is no classical channel to transport the measurement results inside of the event horizon out. The essential point is that the final state measurement or post-selection should be understand as the boundary condition at the singularity which is prior to the outside observers. Thus, there is no need of classical communication to convey the measurement results. The decoding probability and the fidelity are given by
\begin{eqnarray}
    \overline{P}&=&\frac{1}{(d^2-1)}\left(|B|^2+|A||B|-1-\frac{|B|}{|A|}\right)\;,\nonumber\\
    F&=&\frac{1}{P|A|^2}\;.
\end{eqnarray}

In the large $d$ and large $|B|$ limit, the decoding probability is still approximated by $P\approx \frac{1}{|A|^2}$ and the fidelity attains the maximal value $1$ in the ideal case.

Actually, the decoding problem is also related to entanglement distillation, which refers to a protocol in quantum information theory where high-quality entangled state is generated or extracted from a collection of lower-quality or mixed entangled state. The decoupling condition in Eq.\eqref{decoupling_con_final} indicates that the reference subsystem $\overline{A}$ and the radiation $R$ are in a lower-quality entangled state. The goal of entanglement distillation is to produce more pure and stronger entangled states. The distillation protocol involves local operations and classical communication (LOCC) among the entangled parties to purify the entangled states by filtering out noise and unwanted components, ultimately enhancing the overall entanglement content. In the current model, the problem is how to use the LOCC to decompose $R$ into a bipartite system $R_1$ and $R_2$ and make sure that $\overline{A}$ and $R_1$ are in the EPR state.

\begin{figure}
    \centering
    \includegraphics[width=0.5\textwidth]{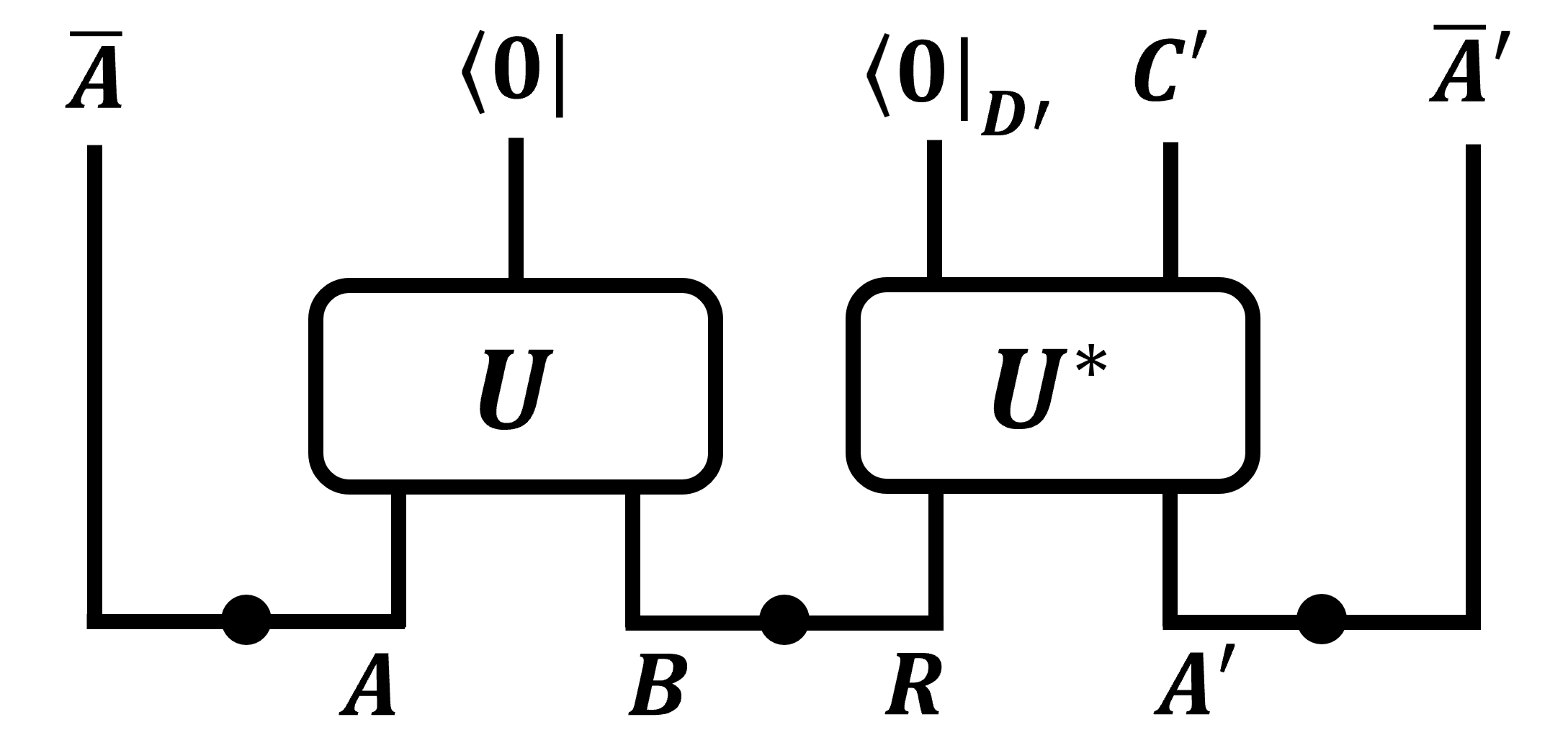}
    \caption{The decoding protocol for the generic final state model where only a portion of the output of scrambling operator $U^*$ is measured. }
    \label{Final_state_decoding_alter}
\end{figure}

Inspired by the spirit of entanglement distillation, we can revise the decoding protocol by only measuring a portion degrees of freedom for the output of the scrambling $U^*$, as shown in Figure \ref{Final_state_decoding_alter}. In this protocol, the scrambling $U^*$ and the measurement on the subsystem $D'$ can be viewed as a distillation protocol. Although the subsystem $D'$ is randomly selected, the Hilbert space dimension $|D'|$ is imposed to satisfy the relation 
\begin{eqnarray}
    |D'|=|A|^2\;, 
\end{eqnarray}
in order to guarantee $|C'||A'|=|R|$.

An explicit calculations show that the decoding probability and the fidelity are given by
\begin{eqnarray}
    \overline{P}&=&\frac{1}{(d^2-1)} \left(2|B|^2-1-\frac{|B|^2}{|A|^2}\right)\;,\nonumber\\
    F&=&\frac{1}{P|A|^2}\;.
\end{eqnarray}
Thus in the large $d$ and large $|B|$ limit, the probability is doubled compared with the protocol in Figure \ref{Final_state_decoding} with the loss of the decoding quality.

\section{Recovery channel with projective measurement as Petz map}\label{sec:petz_map}

Hayden-Preskill protocol can be viewed as a quantum channel from the subsystem $A$ to the subsystem $DR$. When the decoupling condition is satisfied, the channel is recoverable. The recovery channel is known as Petz map. Recent study revealed the relationship between the Yoshida-Kitaev protocol and Petz map \cite{Nakayama:2023kgr}. It shows that the Yoshida-Kitaev protocol as a quantum channel can be written in the form of Petz map. In this section, we will discuss whether the Yoshida-Kitaev protocol for the quantum information recovery in the Hayden-Preskill experiment with the projective measurement can be rewritten as the form of Petz map. We will mainly use the graphical representations to derive the result.

\begin{figure}
    \centering
    \includegraphics[width=0.5\textwidth]{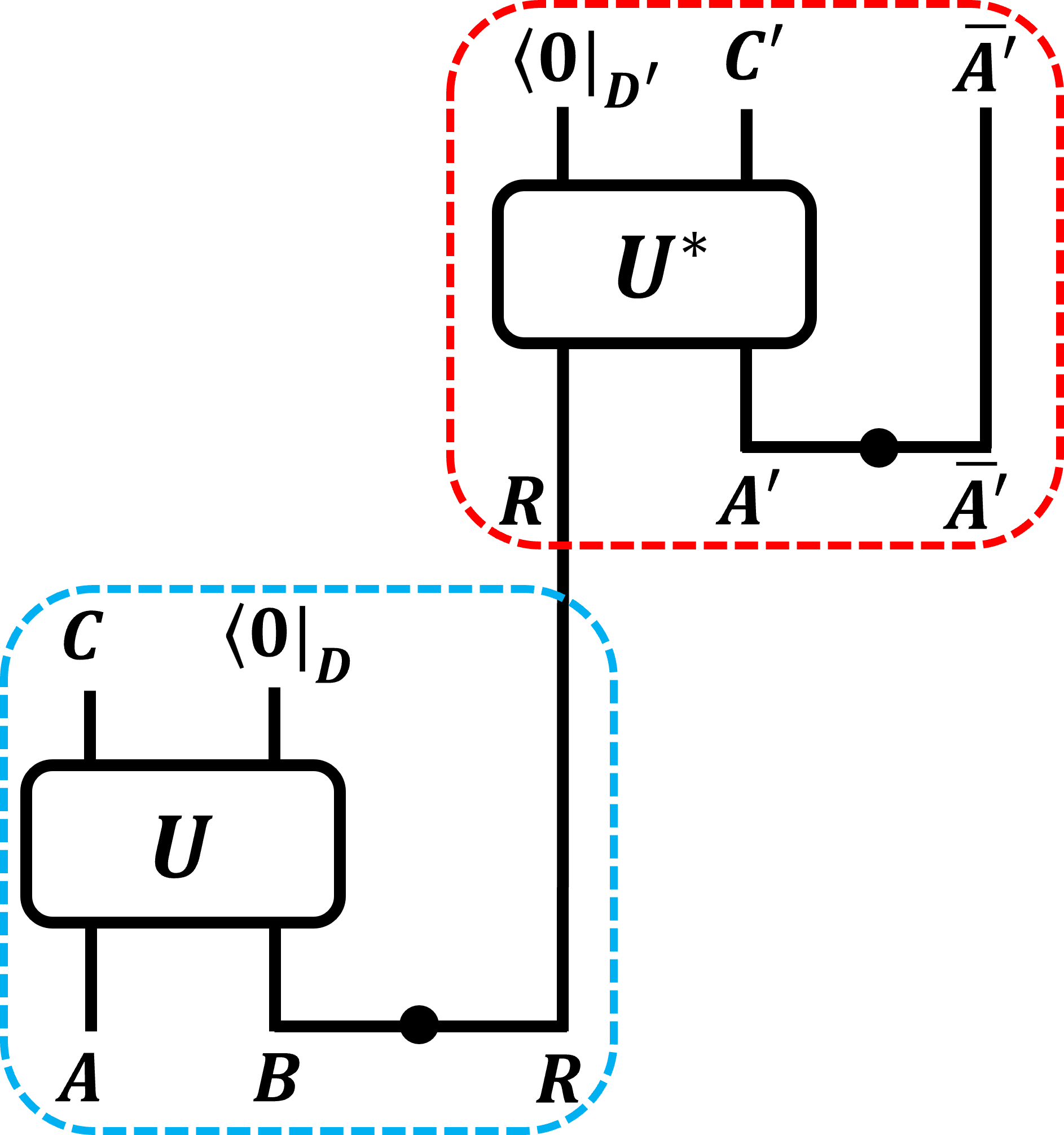}
    \caption{Hayden-Preskill protocol (contained in blue box) and Yoshida-Kitaev decoding protocol (contained in red box). }
    \label{YK_decoding_protocol}
\end{figure}

In this section, we consider the case that only the subsystem $A$ is thrown into the black hole $B$, i.e. the reference system $\overline{A}$ will not be presented for instance. In figure \ref{YK_decoding_protocol}, we have illustrated the Hayden-Preskill protocol and the Yoshida-Kitaev protocol explicitly. From the viewpoint of quantum error correction code, the state $\rho_{A}$ of the subsystem $A$ is the logical state, which is encoded in subsystem $R$ by the scrambling with the black hole $B$ and the local projective measurement on the subsystem $D$. In the present case, the Hayden-Preskill protocol can be treated as a quantum channel from $\mathcal{H}_A$ to $\mathcal{H}_R$. It can be written as the form of quantum channel
\begin{eqnarray}
    \mathcal{N}_{HP}[\rho_{A}]&=& |D| \textrm{Tr}_{C}\left[\langle 0|_{D} U_{AB}\otimes I_{R} \left(
    \rho_{A} |EPR\rangle_{BR}\langle EPR|\right) U_{AB}^\dagger \otimes I_{R}|0\rangle_{D}   \right]\nonumber\\
    &=& |D|\quad \figbox{0.3}{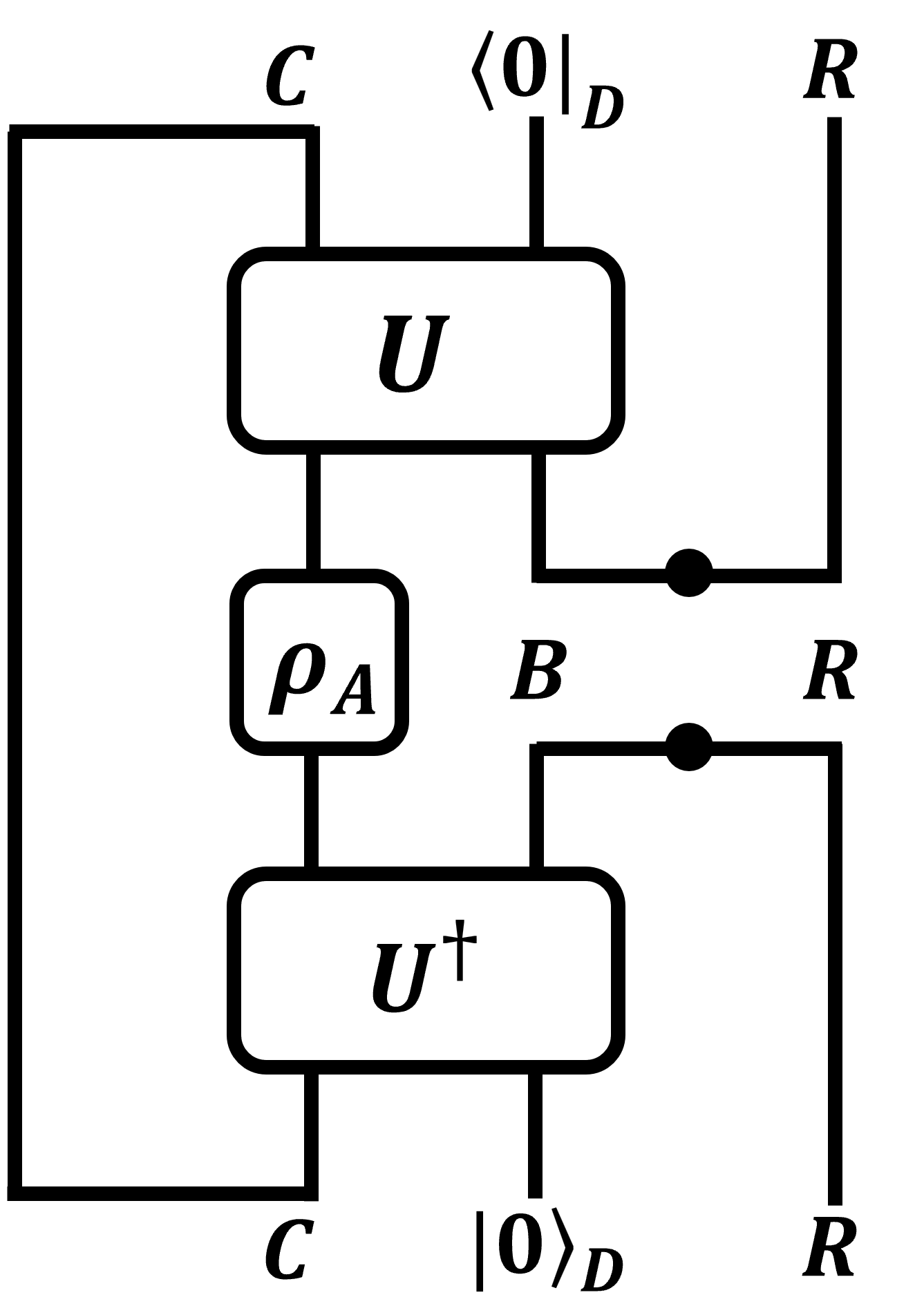}\;,
\end{eqnarray}
where the factor $|D|$ is introduced to guarantee that the channel is trace-preserving.

It was shown that the Hayden-Preskill channel is reversible if the decoupling condition is satisfied. The recovery channel is the well known Petz map \cite{Petz:1986tvy,Petz:2002eql,barnum2000reversing}. For a state $\omega_{R}$ in $\mathcal{H}_{R}$, it is explicitly given by 
\begin{eqnarray}
    \mathcal{R}_{Petz}[\omega_{R}]=\sigma^{\frac{1}{2}} \mathcal{N}_{HP}^{\dagger} \left[\mathcal{N}_{HP}[\sigma]^{-\frac{1}{2}}\omega_{R}\mathcal{N}_{HP}[\sigma]^{-\frac{1}{2}}\right]\sigma^{\frac{1}{2}}\;,
\end{eqnarray}
where $\sigma$ is an arbitrary full rank density matrix on $\mathcal{H}_{A}$ and $\mathcal{N}_{HP}^\dagger$ is the adjoint channel of $\mathcal{N}_{HP}$. For an arbitrary state $\rho_A$ in $\mathcal{H}_A$, the Petz map can recover the state $\mathcal{N}_{HP}[\rho_A]$ as 
\begin{eqnarray}
    \mathcal{R}_{Petz} \circ \mathcal{N}[\rho_A]=\rho_A\;.
\end{eqnarray}

For a scrambling channel $\mathcal{N}_{HP}$, $\mathcal{N}_{HP}[\sigma]$ has a nearly flat spectrum and can be approximated by a maximally mixed state. It turns out that the Petz recovery map can be simplified as 
\begin{eqnarray}
    \mathcal{R}_{Petz}\sim \mathcal{N}_{HP}^\dagger\;.
\end{eqnarray} 
In the following, we will show that the Yoshida-Kitaev protocol, which can also be treated as a quantum channel, is identical to the $\mathcal{N}_{HP}^\dagger$ up to a normalization factor.

The adjoint channel of $\mathcal{N}_{HP}$ is defined as  
\begin{eqnarray}\label{joint_def}
    \textrm{Tr}_{R}\left[\mathcal{N}_{HP}[\rho_A] \mathcal{O}_R \right]
    =\textrm{Tr}_{A}\left[\rho_A \mathcal{N}_{HP}^\dagger[\mathcal{O}_R]\right]\;,
\end{eqnarray}
where $\mathcal{O}_{R}$ is an arbitrary operator acting on $\mathcal{H}_{R}$. By using the graphical representation, we have
\begin{eqnarray}\label{joint_tra}
   \textrm{Tr}_{R}\left[\mathcal{N}_{HP}[\rho_A] \mathcal{O}_R \right]= |D|\quad \figbox{0.3}{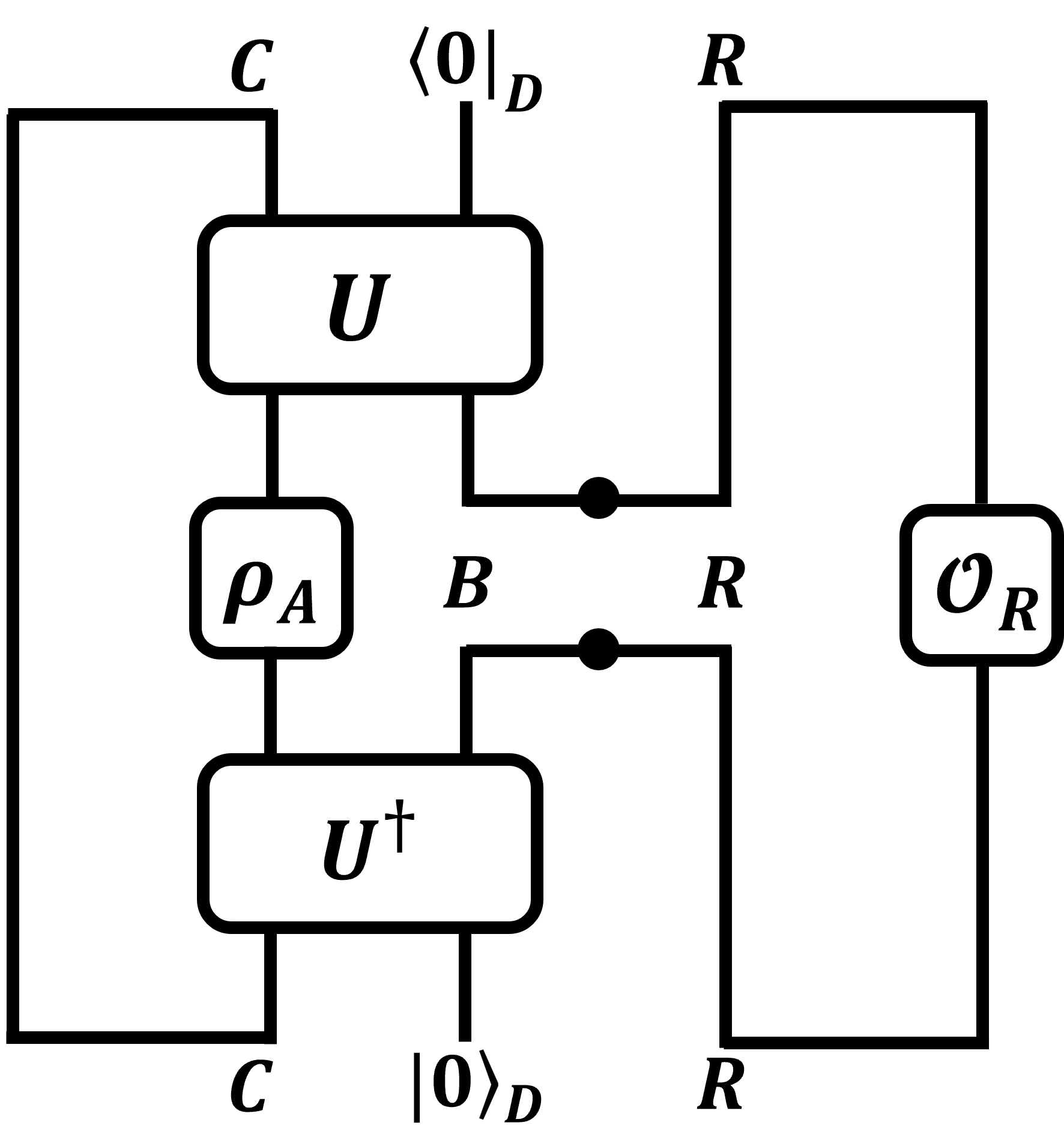}=|D|\quad \figbox{0.3}{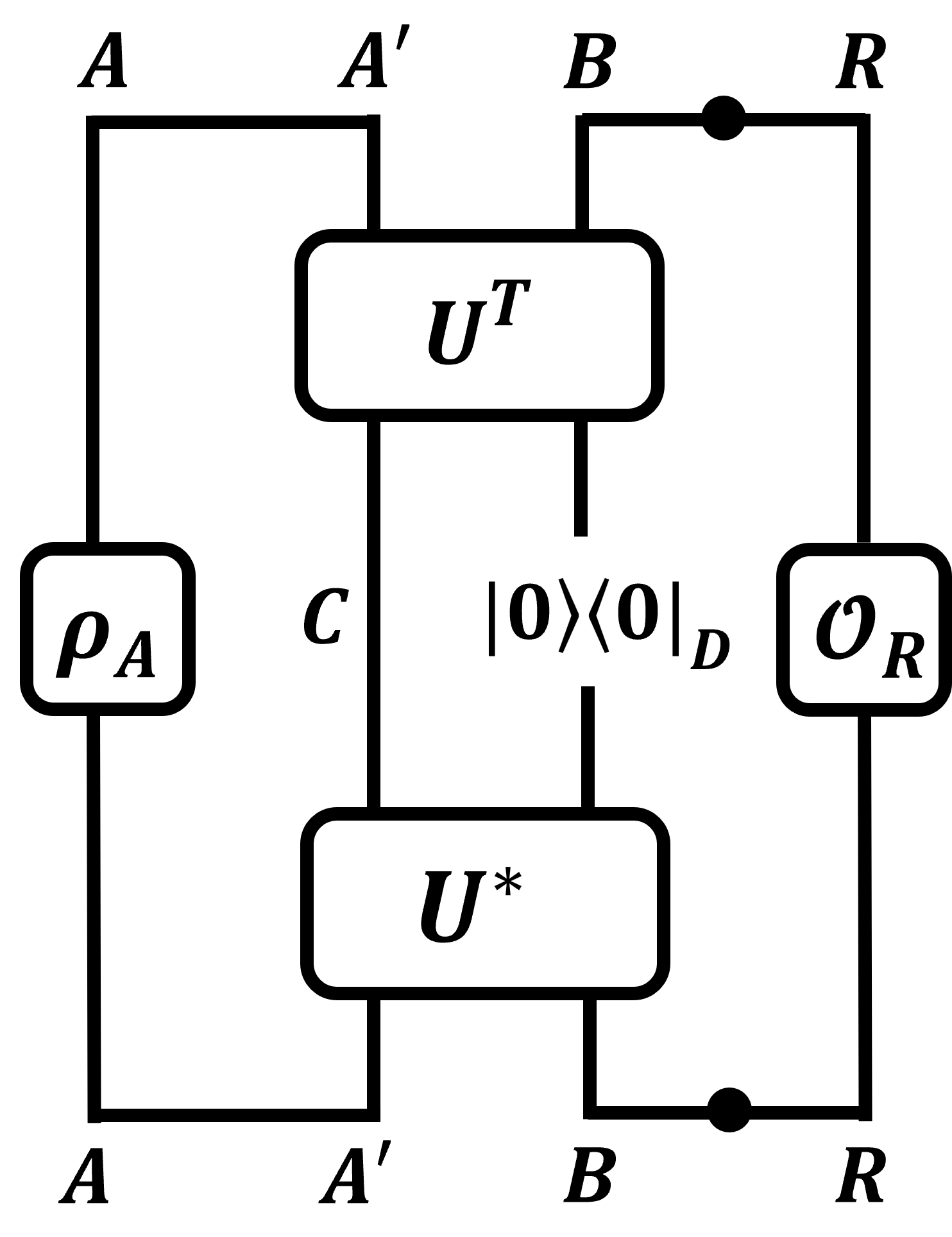}\;,
\end{eqnarray}
where an identical transformation between two graphs are performed. Combining Eq.\eqref{joint_def} and Eq.\eqref{joint_tra}, we can obtain the adjoint channel of $\mathcal{N}_{HP}$ as
\begin{eqnarray}\label{HP_ajg}
    \mathcal{N}_{HP}^\dagger[\omega_R]=|D| \quad \figbox{0.3}{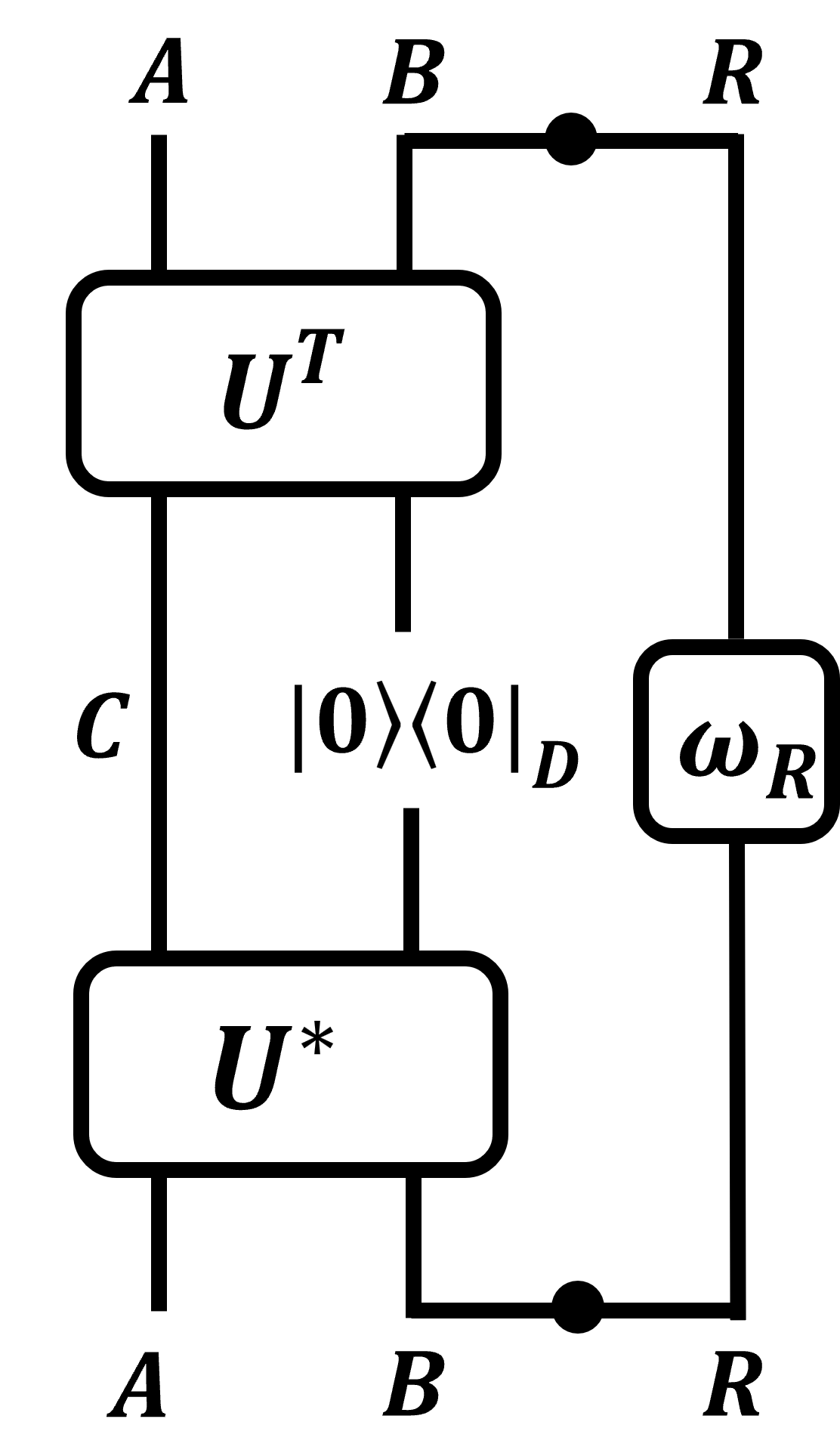}\;,
\end{eqnarray}
which can be rewritten as 
\begin{eqnarray}
    \mathcal{N}_{HP}^\dagger[\omega_R]=|D| \langle EPR|_{BR} U_{CD}^T\otimes I_{R}
\left( I_C\otimes |0\rangle_{D}\langle 0| \otimes\omega_{R} \right)U_{CD}^*\otimes I_{R} |EPR\rangle_{BR}\;.
\end{eqnarray}

Now let us consider the quantum channel corresponding to Yoshida-Kitaev protocol. From Figure \ref{YK_decoding_protocol}, the decoding protocol written as the form of quantum channel is given by 
\begin{eqnarray}
    \mathcal{R}_{YK}[\omega_R]= |D| \textrm{Tr}_{C'}\left[ \langle 0|_D  U_{RA'}^*\otimes I_{\overline{A}'}\left(\omega_R\otimes |EPR\rangle_{A'\overline{A}'}\langle EPR|\right) U_{RA'}^T\otimes I_{\overline{A}'} |0\rangle_{D} \right]\;. 
\end{eqnarray}
Again, using the graphical representation, it is also given as 
\begin{eqnarray}\label{YK_chg}
    \mathcal{R}_{YK}[\omega_R]=|D|\quad \figbox{0.3}{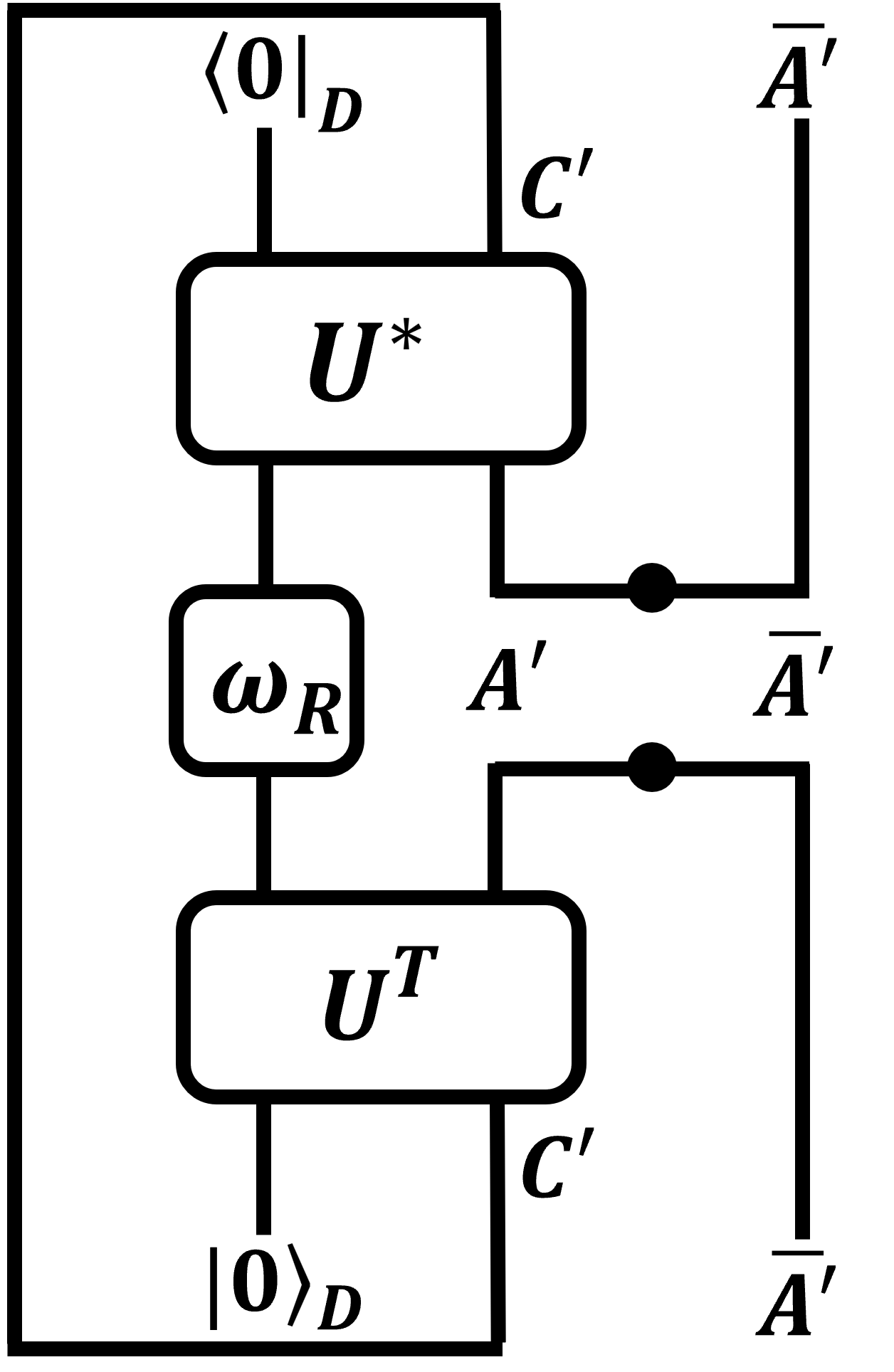}\;.
\end{eqnarray}
It is easy to see that the graph representation of the Yoshida-Kitaev channel is identical to the graph of the adjoint channel of Hayden-Preskill channel up to a factor $\frac{|B|}{|A|}$. Therefore, we have
\begin{eqnarray}
    \mathcal{R}_{YK}=\frac{|B|}{|A|}\mathcal{N}_{HP}^\dagger\;.
\end{eqnarray}
One may concern the output of the channel $\mathcal{N}_{HP}^\dagger$ is different from the output of the channel $\mathcal{R}_{YK}$ because one is a density matrix on $\mathcal{H}_{A}$ and another is a density matrix on $\mathcal{H}_{\overline{A}'}$. This can be easily fixed by noting that the two Hilbert spaces are isometric.

In summary, by using the graphical representations, we have shown that the Yoshida-Kitaev decoding protocol for the Hayden-Preskill experiment with the local projective measurement is equivalent to the Petz recovery channel.

\section{Information recovery with projective measurement in noise channels} 
\label{sec:IR_noise}

In this section, we consider the decoherence effects on the information recovery with projective measurement. For this aim, we model the decoherence with the most common quantum channels: depolarizing channel. A detailed discussion on the depolarizing channel is carried out, which shows the significance impact on the decoding probability and fidelity. We also briefly comment on another kind of quantum channel: dephasing channel, which shows that there is no effect on the decoding protocol. 

\subsection{Depolarizing channel}

The depolarizing channel is given by \cite{Wilde:2011npi}
\begin{eqnarray}
    \mathcal{Q}(\rho)=(1-p)U\rho U^\dagger + p\frac{I_{\tilde{d}}}{\tilde{d}}\;,
\end{eqnarray}
where $p$ is the probability of the decoherence, $\tilde{d}$ is the dimension of the density matrix $\rho$ and $I_{\tilde{d}}$ is the identity matrix of dimension $\tilde{d}$. The depolarizing channel is a ``worst-case scenario" channel. It is clear that the input density matrix is replaced by the maximally mixed state with the probability $p$. When $p = 0$, there is no decoherence. When $p = 1$, the channel is full depolarizing. 

In this case, the Hayden-Preskill state with the projective measurement is replaced with 
\begin{eqnarray}\label{HP_state_Q}
    \tilde{\rho}_{HP}&=&\mathcal{Q}\left(|EPR\rangle_{\overline{A}A}\langle EPR|\otimes |EPR\rangle_{BR}\langle EPR|\right)
    \nonumber\\
    &=&|D|\quad \figbox{0.3}{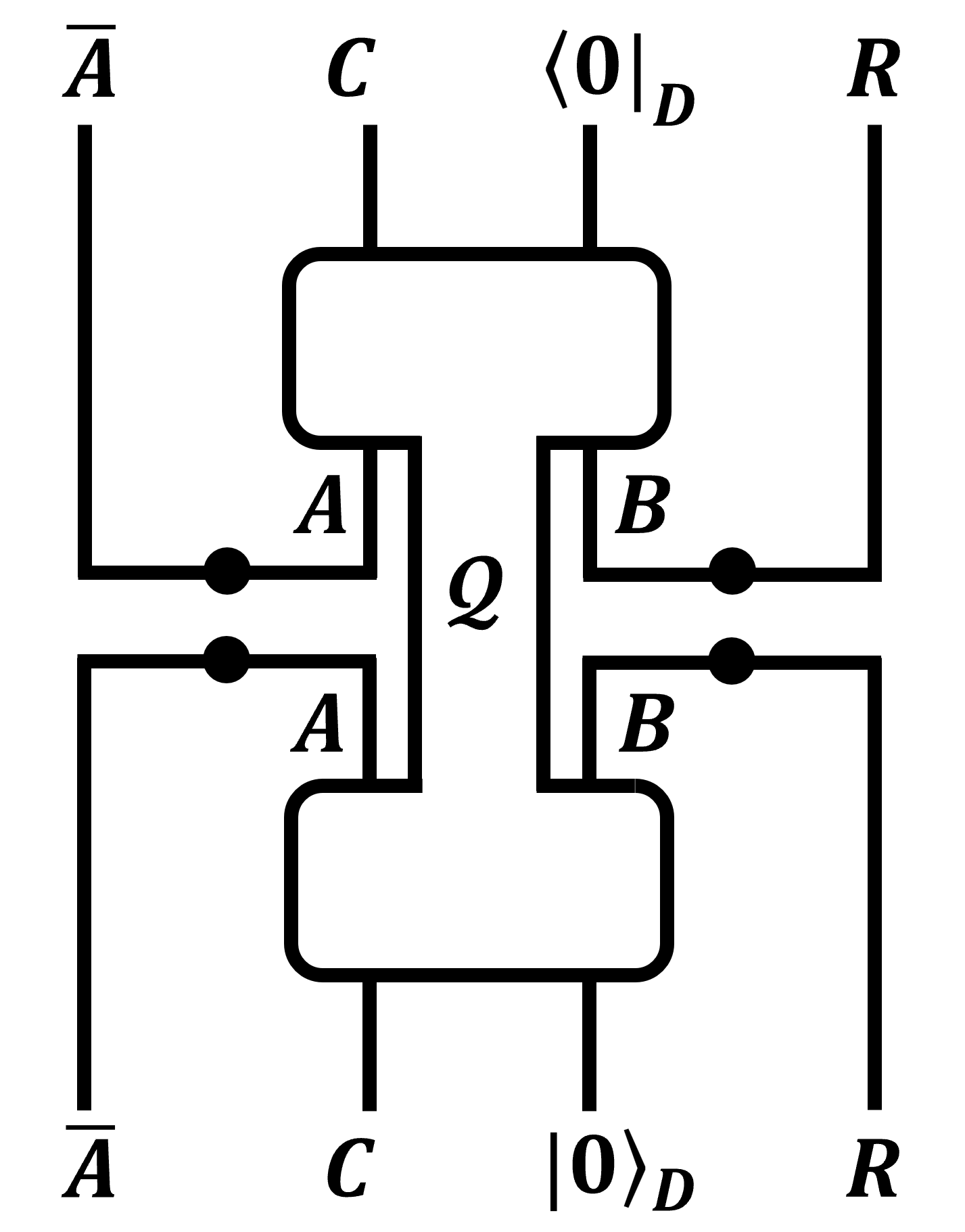}\;,
\end{eqnarray}
where the prerfactor is to preserve the normalization. As an intuitive understanding of how the depolarizing channel works, one can refer to Appendix \ref{sec:HPQ_norm} for the calculation of the normalization of this state.

It is expected that the decoupling condition may be changed. However, a detailed calculation shows that it is unchanged. Let us consider the following quantity as the one without decoherence noise 
\begin{eqnarray}
    \overline{\left\|\tilde{\rho}_{\overline{A}C}-\tilde{\rho}_{\overline{A}} \otimes \tilde{\rho}_C \right\|_1}\leq  \sqrt{|\overline{A}||C|} \left[\overline{\textrm{Tr}\left(\tilde{\rho}_{\overline{A}C}^2\right)}-\overline{\textrm{Tr}\left(\tilde{\rho}_{\overline{A}C}\cdot\left(\tilde{\rho}_{\overline{A}}\otimes\tilde{\rho}_{C}\right)\right)} \right]^{\frac{1}{2}}\;,
\end{eqnarray}
where the reduced density matrices are obtained from the Hayden-Preskill state in Eq.\eqref{HP_state_Q}. 

Using the results from Appendix \ref{sec:Haar_trace_tilde_rho_sq} and \ref{sec:Haar_trace_tilde_3rho}, the inequality can be estimated as 
\begin{eqnarray}
    \left(\overline{\left\|\tilde{\rho}_{\overline{A}C}-\tilde{\rho}_{\overline{A}} \otimes \tilde{\rho}_C \right\|_1}\right)^2&\leq& 
    \left[\frac{|A|^2}{d^2}-\frac{(1-p)^2}{(d^2-1)}\left(1-\frac{1}{|B|^2}\right)\right] \left(|C|^2-1\right)|D|   
    \nonumber\\
    &\approx&\frac{|A|^2}{|D|}\;,
\end{eqnarray}
where in the second step we have used the approximation condition that the subsystems are large enough. This result shows that in the depolarizing channel, the decoupling condition does not change, and is still given by Eq.\eqref{Decoupling_cond}.

We now proceed to apply the Yoshida-Kitaev protocol to recover the quantum information from the depolarizing channel. The protocol works the same manner as discussed in the previous section, except the scrambling operators $U$ and $U^\dagger$ are replaced with the completely positive and trace preserving quantum channel $\mathcal{Q}$.

The decoding probability $P$ can be graphically represented by 
\begin{eqnarray}
    P=\frac{1}{|A||C|}\quad \figbox{0.3}{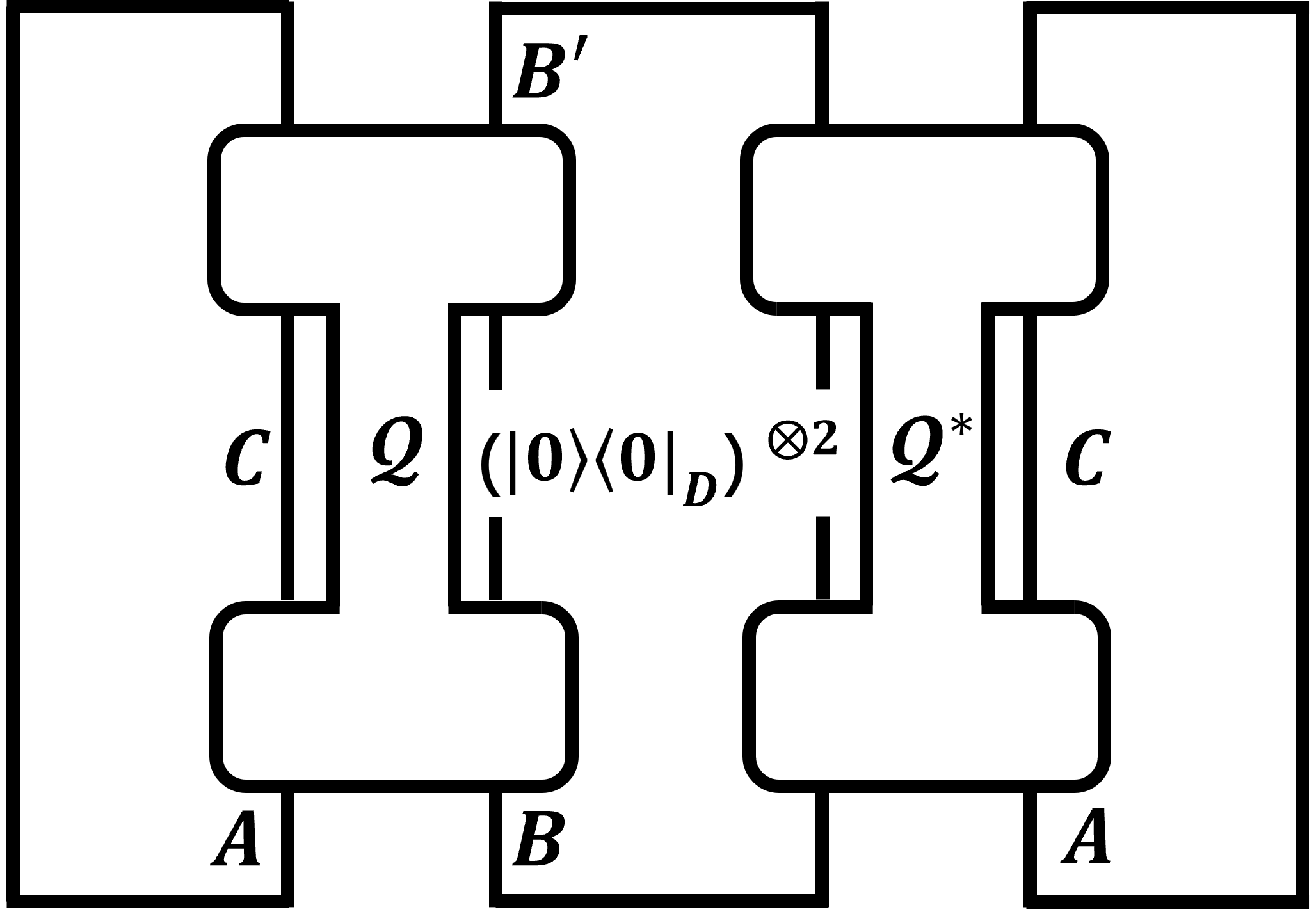}=\frac{|C|}{|A|}\textrm{Tr}\left(\tilde{\rho}_{\overline{A}C}^2\right)\;.
\end{eqnarray}
A straightforward graphical calculation then yields the Haar average of the probability as
\begin{eqnarray}\label{Decoding_Prob_channel}
\overline{P}&=&\frac{(1-p)^2}{(d^2-1)}\left(\frac{d^2}{|A|^2}+\frac{d^2}{|D|}-\frac{d^2}{|A|^2|D|}-1\right)+\frac{(2p-p^2)}{|D|}\nonumber\\&\approx&
\frac{(1-p)^2}{|A|^2}+\frac{1}{|D|}-\frac{(1-p)^2}{|A|^2|D|}-\mathcal{O}\left(\frac{1}{d^2}\right)\;,
\end{eqnarray}
where in the second step we have used the large $d$ approximation. When the decoupling condition is perfectly satisfied and the depolarizing probability is small, the decoding probability can be written as 
\begin{eqnarray}
    \overline{P}\approx \frac{(1-p)^2}{|A|^2}\;,
\end{eqnarray}
which is smaller than the decoding probability without the decoherence noise. As compared with the result for the decoding probability given in Eq.\eqref{Decoding_Prob}, this result indicates that the noise quantum channel can reduce the decoding probability.

In addition, in the limit case with the decoherence probability $p=1$, the average decoding probability is $\overline{P}=\frac{1}{|D|}$. However, this result is entirely unreasonable. Because, in this case, the initial entangled subsystems $\overline{A}$ and $A$ do not interact with the auxilliary subsystems $A'$ and $\overline{A}'$ any longer. The Yoshida-Kitaev protocol does not work.

Now let us turn to the calculation of the decoding fidelity. Using the graphical representation, the decoding fidelity can be straightforwardly calculated as follows  
\begin{eqnarray}
    F&=&\frac{|D|}{|A|^3|B|P} \quad \figbox{0.3}{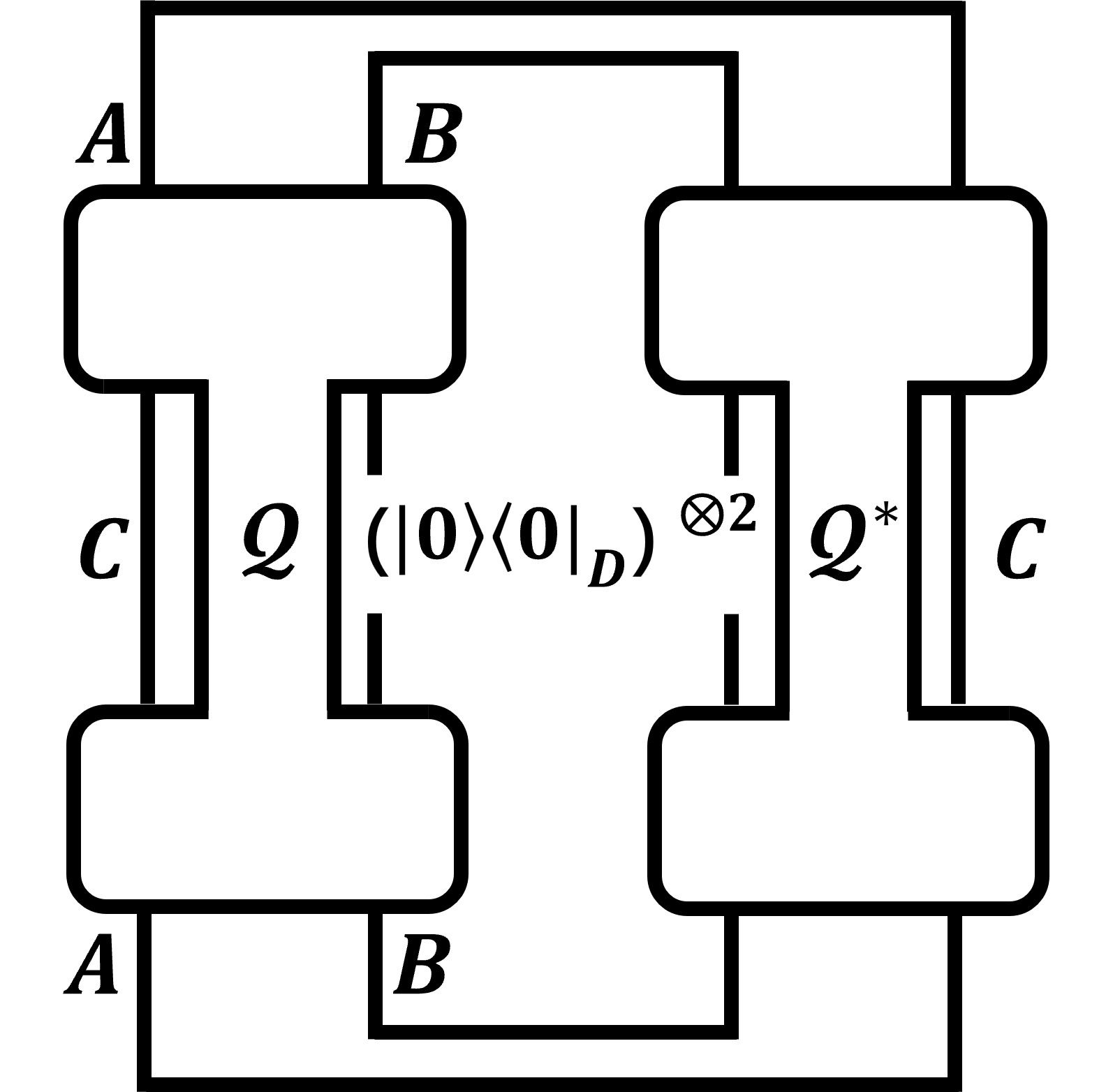}\nonumber\\
    &=&\frac{1}{|A|^2P}\left((1-p)^2+\frac{(2p-p^2)}{|D|}\right)\;.
\end{eqnarray}
This result suggests that 
\begin{eqnarray}\label{FP_eq}
    FP=\frac{1}{|A|^2} \left((1-p)^2+\frac{(2p-p^2)}{|D|}\right)\;.
\end{eqnarray}
By using the Haar average of the decoding probability given in Eq.\eqref{Decoding_Prob_channel}, one can get
\begin{eqnarray}
    F\approx 1-\frac{1}{(1-p)^2}\frac{\left(|A|^2-1\right)}{|D|}
    \approx 1-\frac{1}{(1-p)^2} \frac{|A|^2}{|D|}\;,
\end{eqnarray}
where the decoupling condition and the large subsystem approximation are used. The fidelity is always smaller than $1$ when the decoherence effects are considered. It is also obvious that the decoding fidelity decreases along with the increasing of the depolarizing probability $p$. This is to say that the decoherence also reduces the decoding fidelity.

Recall that the Yoshida-Kitaev protocol is analogous to the quantum teleportation \cite{PhysRevLett.70.1895}. It is known that for the random pure input state of an individual qubit \cite{PhysRevLett.74.1259}, the maximal average fidelity reachable by the perfect teleportation protocol is $\frac{2}{3}$. For a random mixed state \cite{Vidal:1998rk,Bussandri:2021wep}, the maximal mean fidelity is given by $F_{QT}=0.811$. As we have seen in the last section, for the perfect decoding protocol, the fidelity can reach nearly $1$. This means that the scrambling dynamics can improve the fidelity greatly compared with the classical quantum teleportation protocol. This is to say that a successful decoding always serves as a definite signature of quantum scrambling. However, when considering the decoherence effects, the fidelity is strictly less than unity. In this way, one can estimate the threshold value of the decoherence probability as
\begin{eqnarray}
    p=1-\frac{2}{\sqrt{|D|(1-F_{QT})}}\;,
\end{eqnarray}
where we have set $|A|=2$ for the case that only one qubit is thrown into the black hole. When the decoherence probability $p$ is greater than this threshold value, the advantage of scrambling dynamics in teleporting the quantum state by using the Hayden-Preskill protocol is not significant compared to that by using the classical teleportation protocol.

At last, let us comment on the decoding probability and the fidelity for the depolarizing channel without projective measurement. In \cite{Yoshida:2018vly}, the author considered the Yoshida-Kitaev protocol with the decoherence noise. It is argued that by jointly measuring both the decoding probability and the decoding fidelity, one can directly extract the precise “noise” parameter $p$ which quantitatively captures the nonscrambling-induced decay of out-of-time-order correlation functions, thereby characterizing the amount of noise in the quantum channel. The argument is based on the observation that the product of the probability and the fidelity is given by 
\begin{eqnarray}\label{FP_eq_nm}
    FP=\frac{1}{|A|^2} \left((1-p)^2+\frac{(2p-p^2)}{|D|^2}\right)\;.
\end{eqnarray}
This relation is for the decoding protocol without projective measurement, and is slightly different from the one given in Eq.\eqref{FP_eq}. However, the Haar average of the probability and the fidelity are not explicitly presented. Here, by using the graphical calculations, we can get
\begin{eqnarray}\label{Decoding_Prob_channel_nm}
\overline{P}&=&\frac{(1-p)^2}{(d^2-1)}\left(\frac{d^2}{|A|^2}+\frac{d^2}{|D|^2}-\frac{d^2}{|A|^2|D|^2}-1\right)+\frac{(2p-p^2)}{|D|^2}\nonumber\\&\approx&
\frac{(1-p)^2}{|A|^2}+\frac{1}{|D|^2}-\frac{(1-p)^2}{|A|^2|D|^2}-\mathcal{O}\left(\frac{1}{d^2}\right)\;,
\end{eqnarray}
and
\begin{eqnarray}
    F&=&\frac{1}{|A|^2P}\left((1-p)^2+\frac{(2p-p^2)}{|D|^2}\right)\approx 1-\frac{1}{(1-p)^2}\frac{|A|^2}{|D|^2}\;.
\end{eqnarray}
The last equation shows that, when the decoupling condition is perfectly satisfied, the measurement of the decoding fidelity is sufficient to extract the noise parameter.

\subsection{On dephasing channel} 

The dephasing channel is also one of the most common decoherence channels. It is a unital map that destroys the relative phases between the computational basis states $\{|k\rangle\}$. It is defined as \cite{Lautenbacher:2023dpn}
\begin{eqnarray}
    \Lambda(\rho)=(1-p)\rho+p \sum_{k=0}^{\tilde{d}-1}P_{k}\rho P_{k}\;,
\end{eqnarray}
with the projector $P_{k}=|k\rangle\langle k|$ and $\tilde{d}$ being the dimension of the Hilbert space. The parameter $p\in [0, 1]$ characterizes the noise strength. When $p=0$, there is no noise. When $p=1$, the noise is maximal and the coherence vanishes completely.

In general, the effect of the dephasing channel is to eliminate the off-diagonal terms of the density operator when represented with respect to the computational basis. For our case, we consider the dephasing channel acting on the following density matrix 
\begin{eqnarray}
    \rho=|EPR\rangle_{\overline{A}A}\langle EPR|\otimes |EPR\rangle_{BR}\langle EPR|\;.
\end{eqnarray}
It is obvious that the EPR state in computational basis is diagonal. Therefore, the resulting density operator in the dephasing channel is the same as the input one. We can conclude that the dephasing channel has no effect on the Yoshida-Kitaev decoding protocol.

\section{Quantum simulation of information recovery with projective measurement}
\label{sec:q_sim}

In this section, we will try to implement the Yoshida-Ketaev protocol on the quantum processors to verify the feasibility of the information recovery from the black hole with the projective measurement. The scrambling operator are modeled by the 3-qubit scrambling unitary \cite{Yoshida:2018vly}, which can disperse all single-qubit operators into three-qubit operators \cite{Li:2023rue}. We are examining two aspects: one for teleporting the quantum state and another for teleporting quantum entanglement.

\begin{figure}
    \centering
   \includegraphics[width=0.8\textwidth]{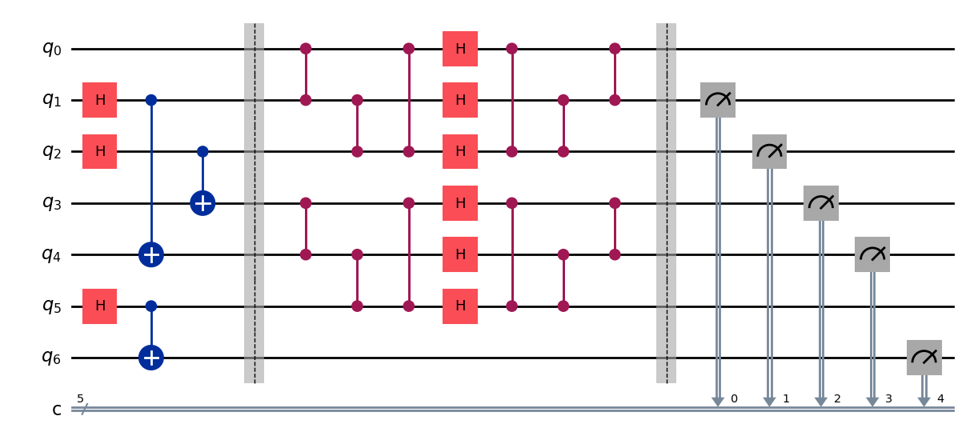} 
   \caption{Quantum circuit for teleporting quantum state.}
    \label{Circuit_state}
\end{figure}

\begin{figure}
    \centering
      \includegraphics[width=0.5\textwidth]{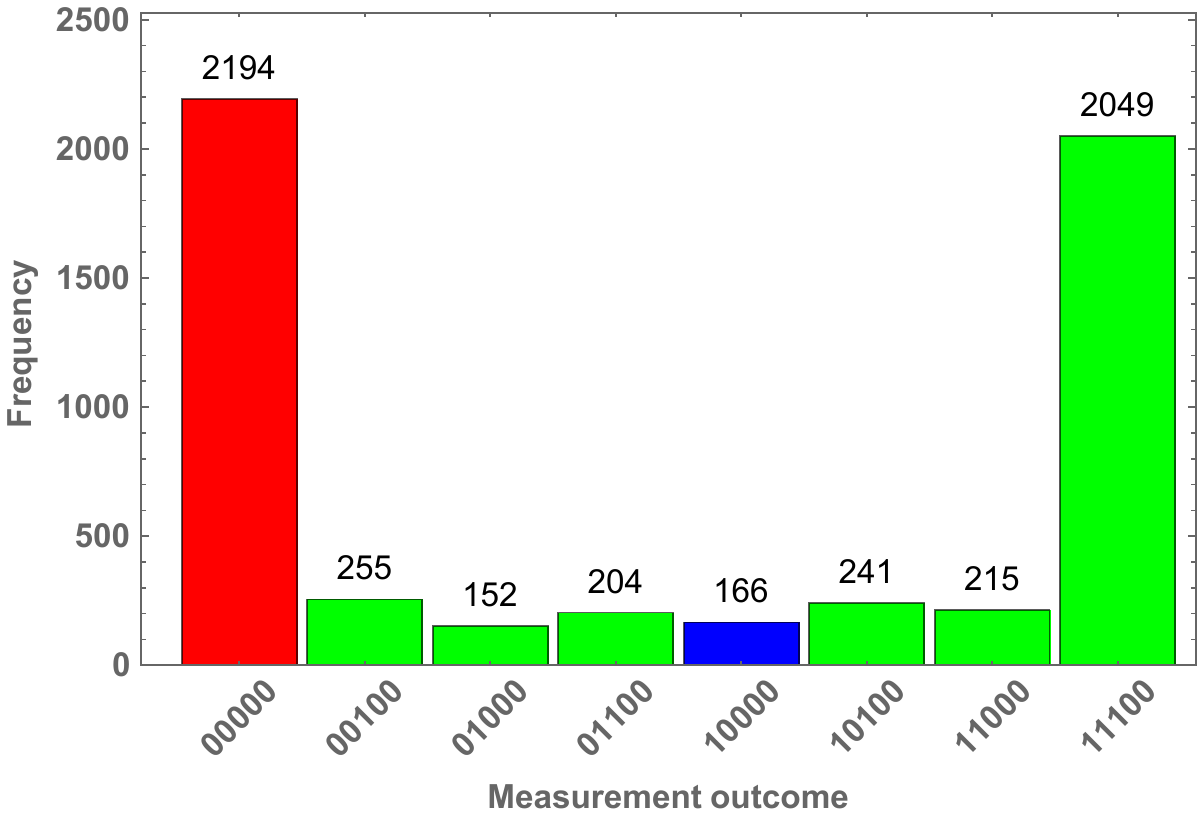}
       \caption{The measurement outcomes from executing the circuit in Figure \ref{Circuit_state} on $IBM-brisbane$ quantum processor. The horizontal axis represents the measurement outcome of $q_1q_2q_3q_4q_6$ and the vertical axis represents the corresponding frequency.}
    \label{Circuit_state_chart}
\end{figure}

The circuit for teleporting quantum state is shown in Figure \ref{Circuit_state}. It can be divided into three parts. In the first part, the entangled pairs are prepared by using the Hadamard gate and controlled-X gate. Then the the first six qubits are scrambled by a series of controlled-Z gates and Hadamard gates. At last, five qubits are measured. In this circuit, we want to teleport the quantum state of qubit $q_0$ to qubit $q_6$. The circuit was executed on the IBM-brisbane processor. The measurement outcomes are given in Figure \ref{Circuit_state_chart}.

Recall the decoding protocol presented in Eq.\eqref{psi_out}, the measurement on qubits $q_1$ and $q_2$ is to model the projective measurement of the subsystem $D$ and the measurement on $q_3$ and $q_4$ is to model the measurement on the subsystem $D'$. If the two measurement outcomes coincide, for instance, if both are $``00"$ as we have selected, it signifies that the state of $D'$ is projected onto the same state of $D$. The red and the blue bars in Figure \ref{Circuit_state_chart} represent this case. Thus we can compute the decoding probability as $P=43.1\%$. The successful decoding means recovering the quantum state of $q_0$ on qubit $q_6$. The measurement outcome of $q_6$ being $``0"$ means a successful decoding. Thus the red bar represents the frequency of successful decoding event. The decoding fidelity is computed as $F=92.9\%$, which means a high quality of teleporting quantum state towards this circuit.

\begin{figure}
    \centering
   \includegraphics[width=0.8\textwidth]{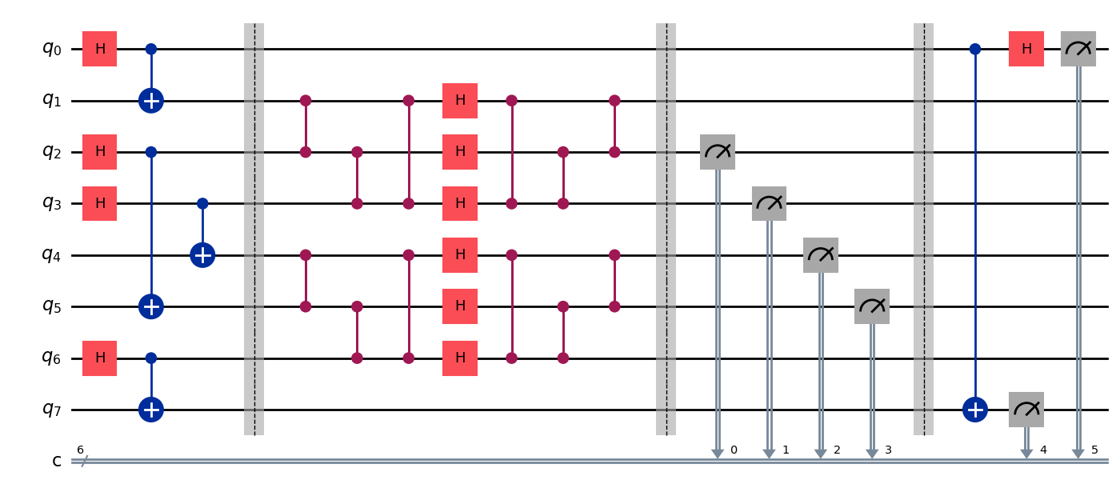}
    \caption{Quantum circuit for teleporting quantum entanglement.}
    \label{Circuit_entangle}
\end{figure}

\begin{figure}
    \centering
     \includegraphics[width=0.6\textwidth]{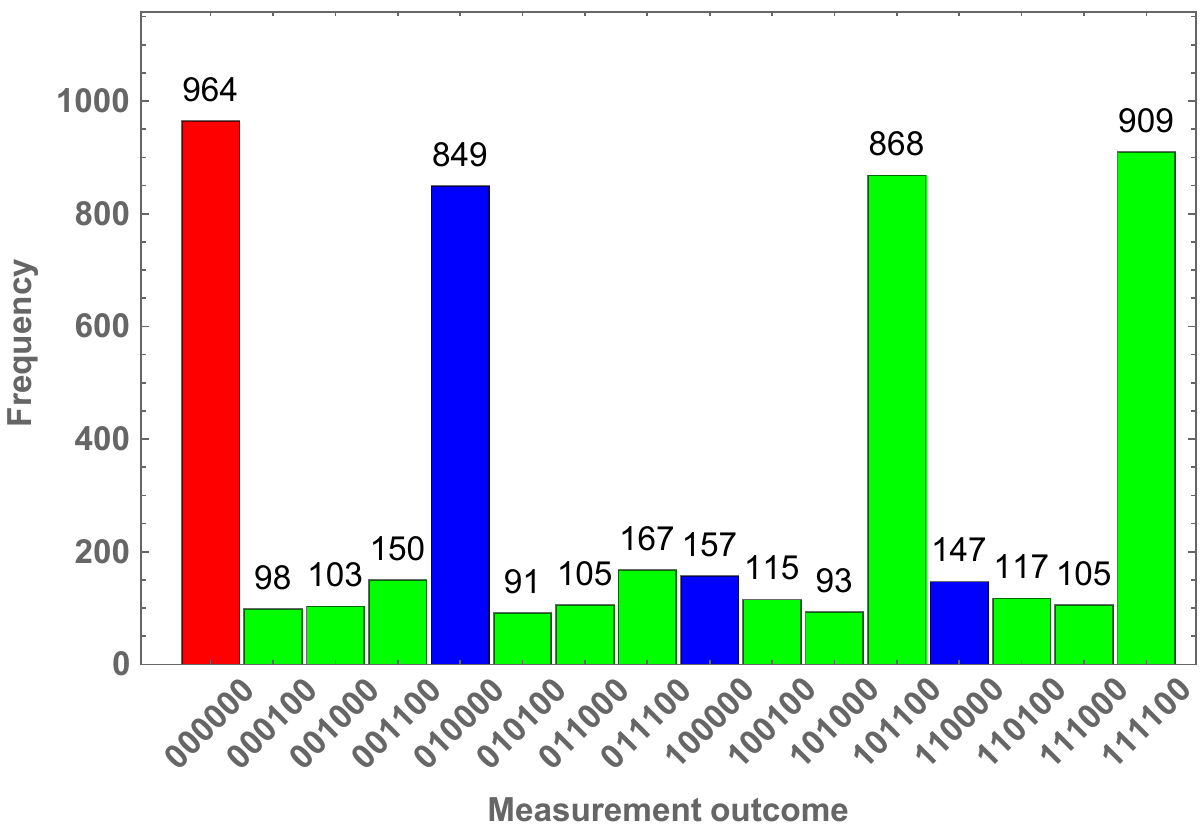}
    \caption{The measurement outcomes from executing the circuit in Figure \ref{Circuit_entangle} on $IBM-kyoto$ quantum processor. The horizontal axis represents the measurement outcome of $q_1q_2q_3q_4q_6q_0$ and the vertical axis represents the corresponding frequency.}
    \label{Circuit_entangle_chart}
\end{figure}

The circuit for teleporting quantum entanglement is shown in Figure \ref{Circuit_entangle}. This circuit is similar to that for teleporting the state. However, by using this circuit, we want to teleport the entanglement between $q_0$ and $q_1$ to $q_0$ and $q_7$. Therefore, in the final step, we perform an entanglement measurement on $q_0$ and $q_7$. The circuit was executed on the IBM-kyoto processor. The measurement outcomes are given in Figure \ref{Circuit_entangle_chart}.

In this case, the red and the blue bars in Figure \ref{Circuit_entangle_chart} also represent the successful projection of $D'$ onto the same state with $D$. The decoding probability is given by $P=42.0\%$. The successful teleportation of quantum entanglement is indicated by the measurement outcomes of $q_0$ and $q_7$ is $``00"$, which is represented by the red bar in the chart. Thus the decoding fidelity is given by $F=41.4\%$. It indicates that the quality of teleporting quantum entanglement is significantly lower than that of teleporting the quantum state.

The simulation results indicate that as long as we have an efficient quantum processor, the recovery of information from black hole is achievable. It is generally conjectured that black holes are the fast scramblers in nature \cite{Sekino:2008he,Lashkari:2011yi,Maldacena:2015waa}. In addition, from the perspective of quantum mechanics, the dynamics of black holes as an isolated objects must be unitary. Therefore, the interior dynamics of black hole is generally modeled by a random unitary operator \cite{Page:1993df,Page:1993wv}. Our simulations of the decoding protocols on the quantum processors employ the assumption that the interior unitary dynamics of black hole is known to the external observers. Recent study shows that the assumption is not necessary \cite{Oliviero:2022url,Leone:2022edn}. It is demonstrated that even without the prior knowledge of the internal dynamics, the information swallowed by a black hole can be recovered by the external observer using a strategy from quantum machine learning. By throwing test information into the black hole and analyzing the outgoing Hawking radiation, the decoder can in principle learn how to construct a Clifford circuit to decode the information. Therefore, this strategy may provide an essential tool for decoding Hawking radiation. On the other hand, it challenges the assumption of black hole scrambling.

\section{Conclusion and discussion} 
\label{Sec:con_dis}

In summary, we have explored various aspects related to the Hayden-Preskill thought experiment with the local projective measurement. The measurement is assumed to be applied on the Hawking radiation that was emitted after throwing the quantum diary into the black hole. First, in section \ref{sec:IR_PM}, we have shown that the Yoshida-Kitaev probabilistic decoding strategy can be properly applied to recover information from the Hayden-Preskill protocol with the projective measurement. The decoding probability and fidelity are explicitly calculated by using the graphical representation technique, which shows a small decoding probability resulting in the high decoding quality.

Then we discuss the relation between the model with projective measurement and the black hole final state model. For the generic final state model, we discuss two types of decoding protocols, distinguished by how much degrees of freedom are projected. For the first protocol, all the output degrees of freedom from the scrambling dynamics are projected, which shows the similar decoding probability and fidelity as obtained in section \ref{sec:IR_PM}. For the second protocol, only a portion of degrees of freedom is projected. The results for the decoding probability and fidelity show that the probability can be enhanced with the cost of decoding quality.

We also illustrated that the Yishida-Kitaev decoding protocol treated as a quantum channel is equivalent to the Petz recovery channel for the Hayden-Preskill protocal with the local projective measurement. The derivation is mainly completed by using the graphical representations, which indicates that this technique is a powerful tool in studying the problems related to the black hole information.

Then by taking the decoherence or the noise into account, we studied their effects on the projective measurement model. We have investigated two types of decoherence channels: the depolarizing channel and the dephasing channel. Towards explicit calculations, we show that the depolarizing channel can reduce the decoding probability as well as the fidelity. Conversely, the dephasing channel has no impact on the decoding procedure.

Finally, we have conducted an experimental simulation of the decoding protocol using the quantum processors. We execute two circuits on the quantum processor, one for teleporting the quantum state and another for teleporting the quantum entanglement. It is shown that the Yoshida-Kitaev protocol for teleporting quantum state is superior to the protocol for teleporting quantum entanglement.

\appendix
\section{Calculations of the Haar averages using the graphical representations}
\label{sec:Haar_graphs}

In this appendix, we give the graphical representations for the calculations of the Haar average. The Haar integral formulas used in our computation are given by \cite{collins2002moments,Collins_2006}
\begin{eqnarray}
    \int dU U_{ij} U^\dagger_{j'i'}&=&\frac{1}{d}\delta_{ii'}\delta_{jj'} \;,\label{2U_integral}\\
    \int dU\  U_{i_1j_1}U_{i_2j_2}U^\dagger_{j_1'i_1'}U^\dagger_{j_2'i_2'}
&=&\frac{1}{(d^2-1)}\left(\delta_{i_1i_1'}\delta_{i_2i_2'}\delta_{j_1j_1'}\delta_{j_2j_2'} + \delta_{i_1i_2'}\delta_{i_2i_1'}\delta_{j_1j_2'}\delta_{j_2j_1'}\right)
 \nonumber\\
 &-&\frac{1}{d(d^2-1)}\left(\delta_{i_1i_1'}\delta_{i_2i_2'}\delta_{j_1j_2'}\delta_{j_2j_1'} + \delta_{i_1i_2'}\delta_{i_2i_1'}\delta_{j_1j_1'}\delta_{j_2j_2'}\right)\,\label{4U_integral}.
\end{eqnarray}
It is convenient to represent the integral formulas in the graphical form 
\begin{eqnarray}
    \int dU \left(\figbox{0.2}{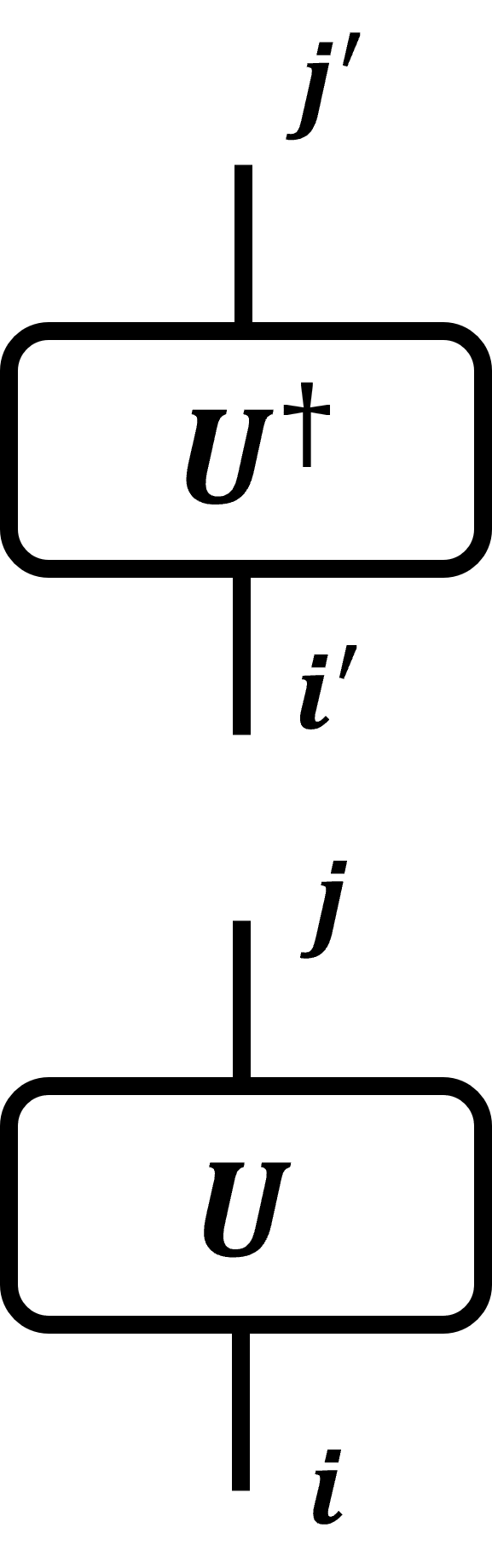}\right)&=&\frac{1}{d}\left(\figbox{0.2}{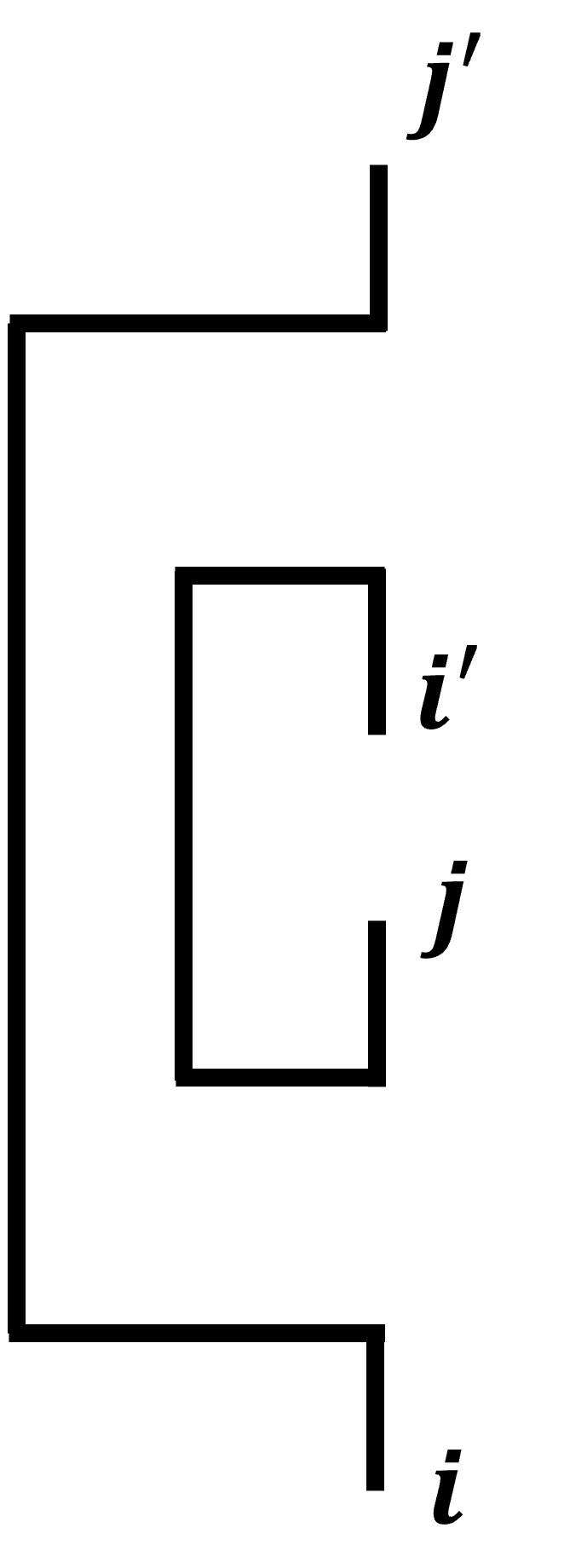}\right)\;,\label{2U_graph}\\
    \int dU \left(\figbox{0.2}{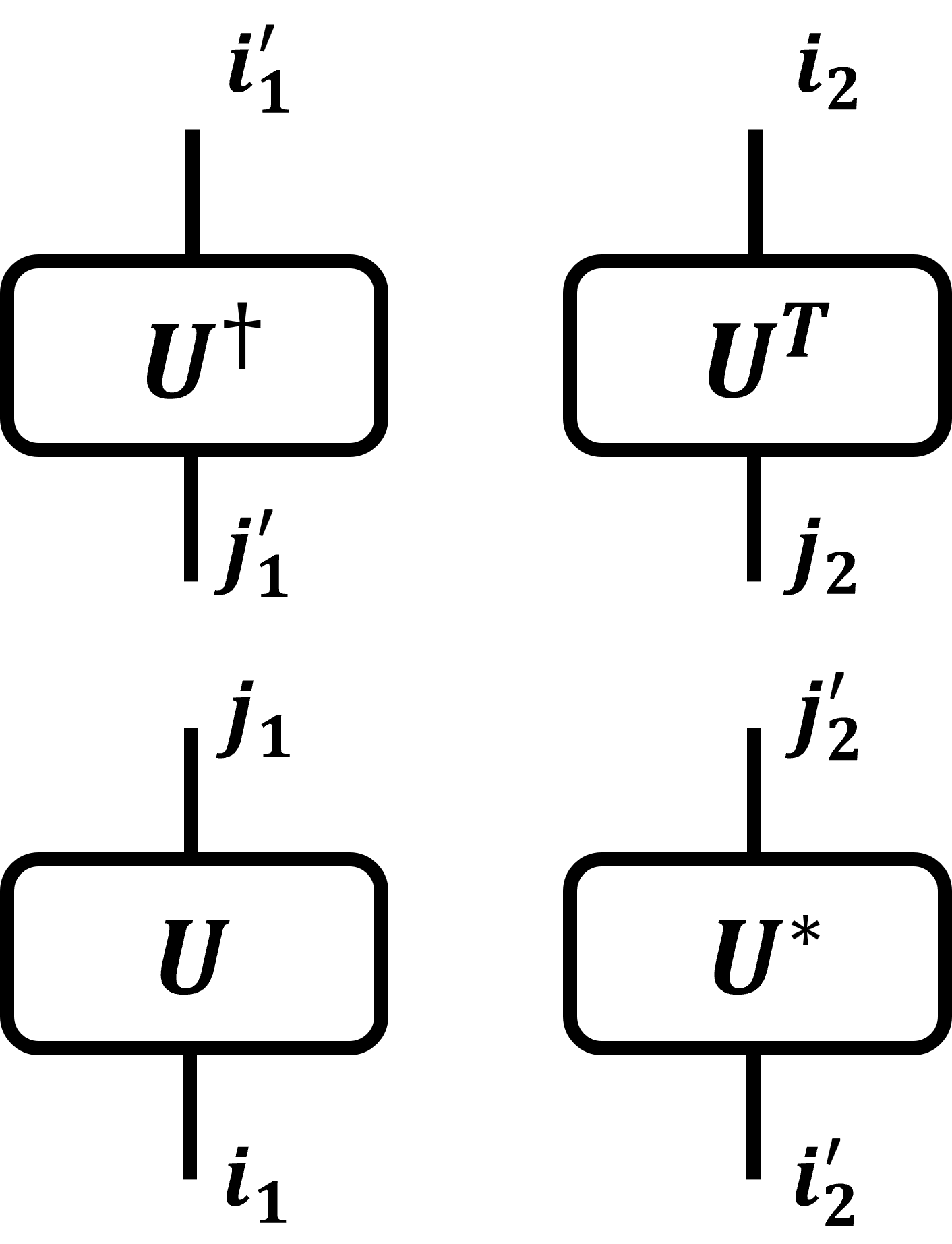}\right) &=&\frac{1}{(d^2-1)}\left(\figbox{0.2}{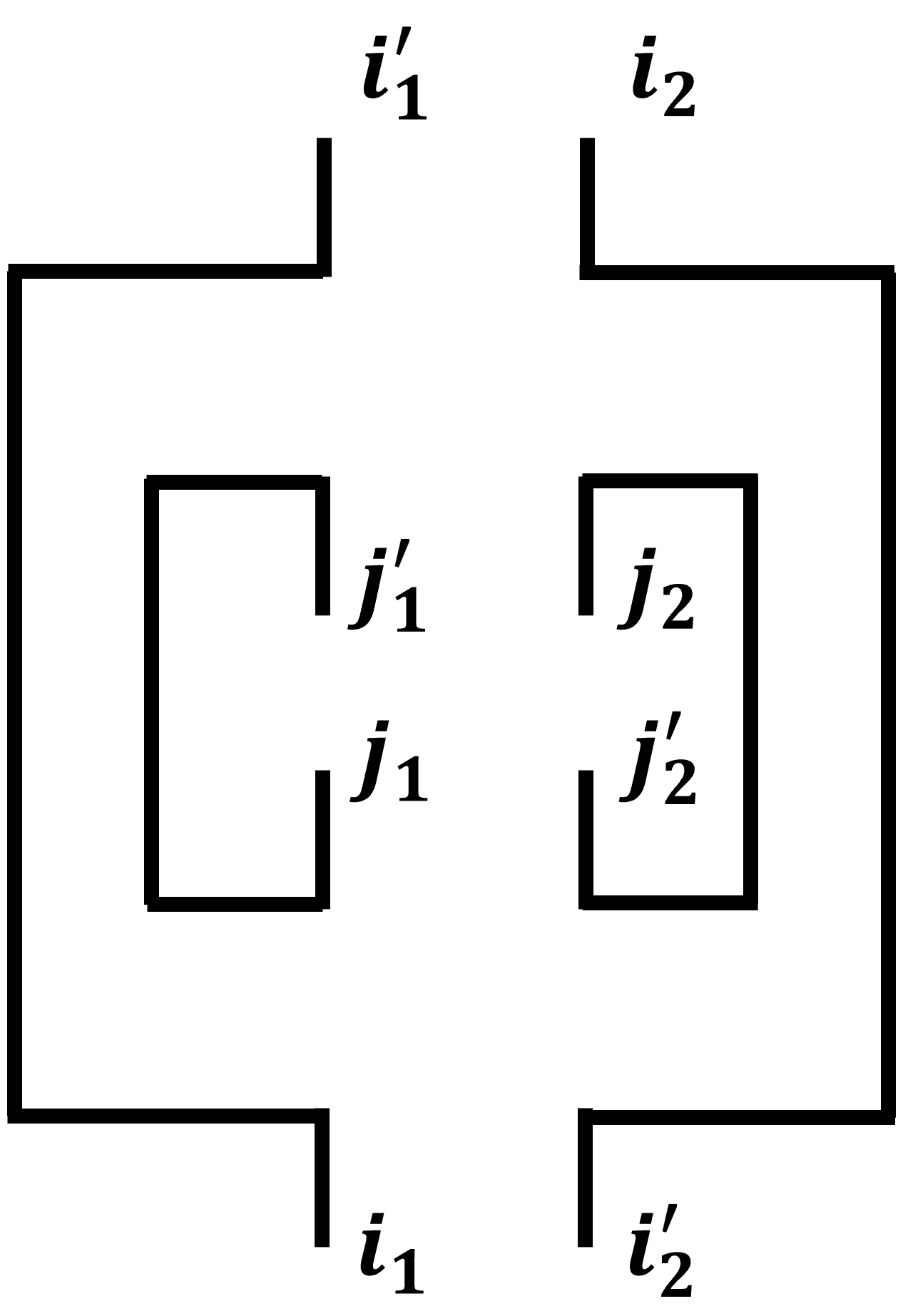}+\figbox{0.2}{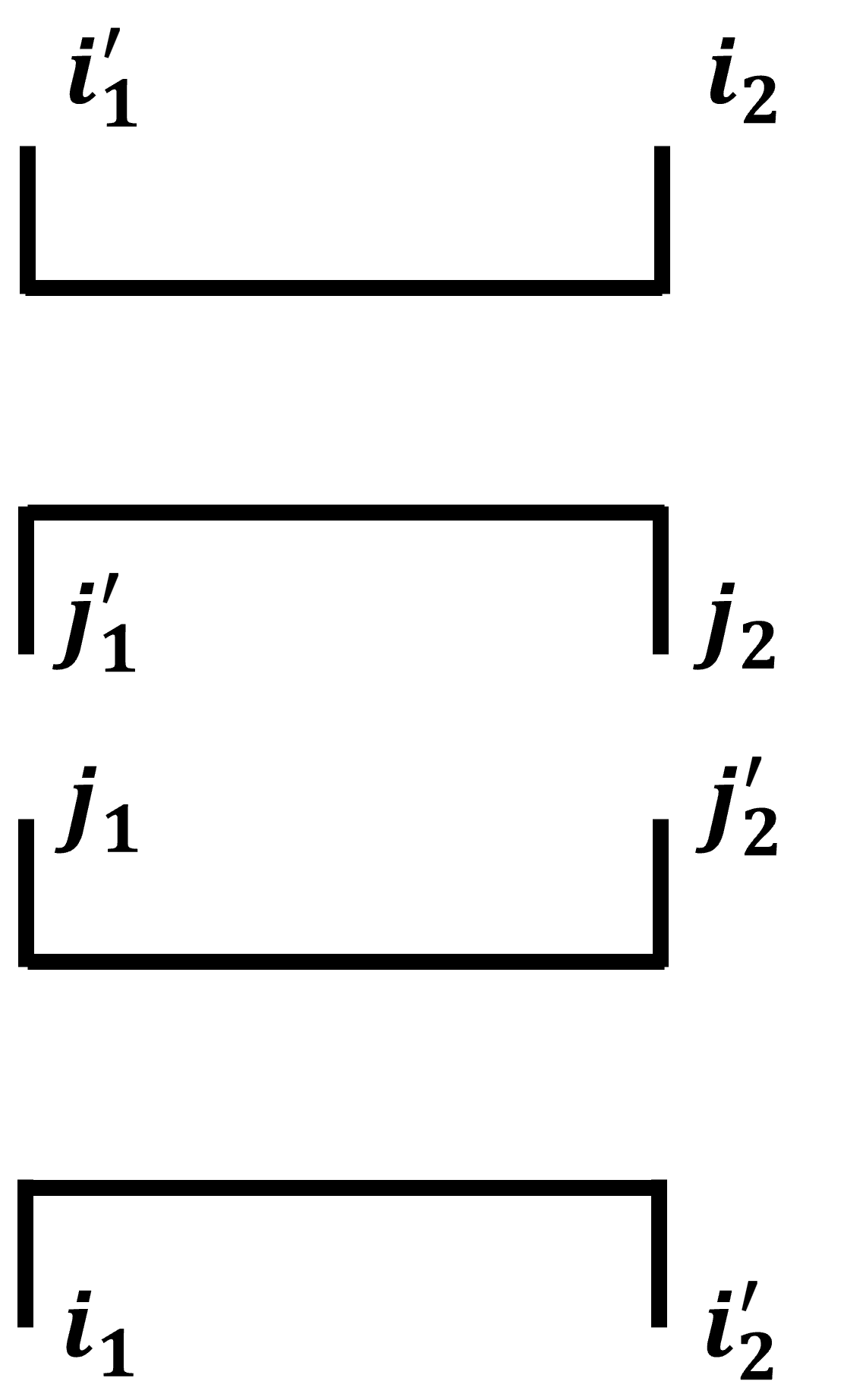}\right)\nonumber\\
    &+&\frac{1}{d(d^2-1)} \left(\figbox{0.2}{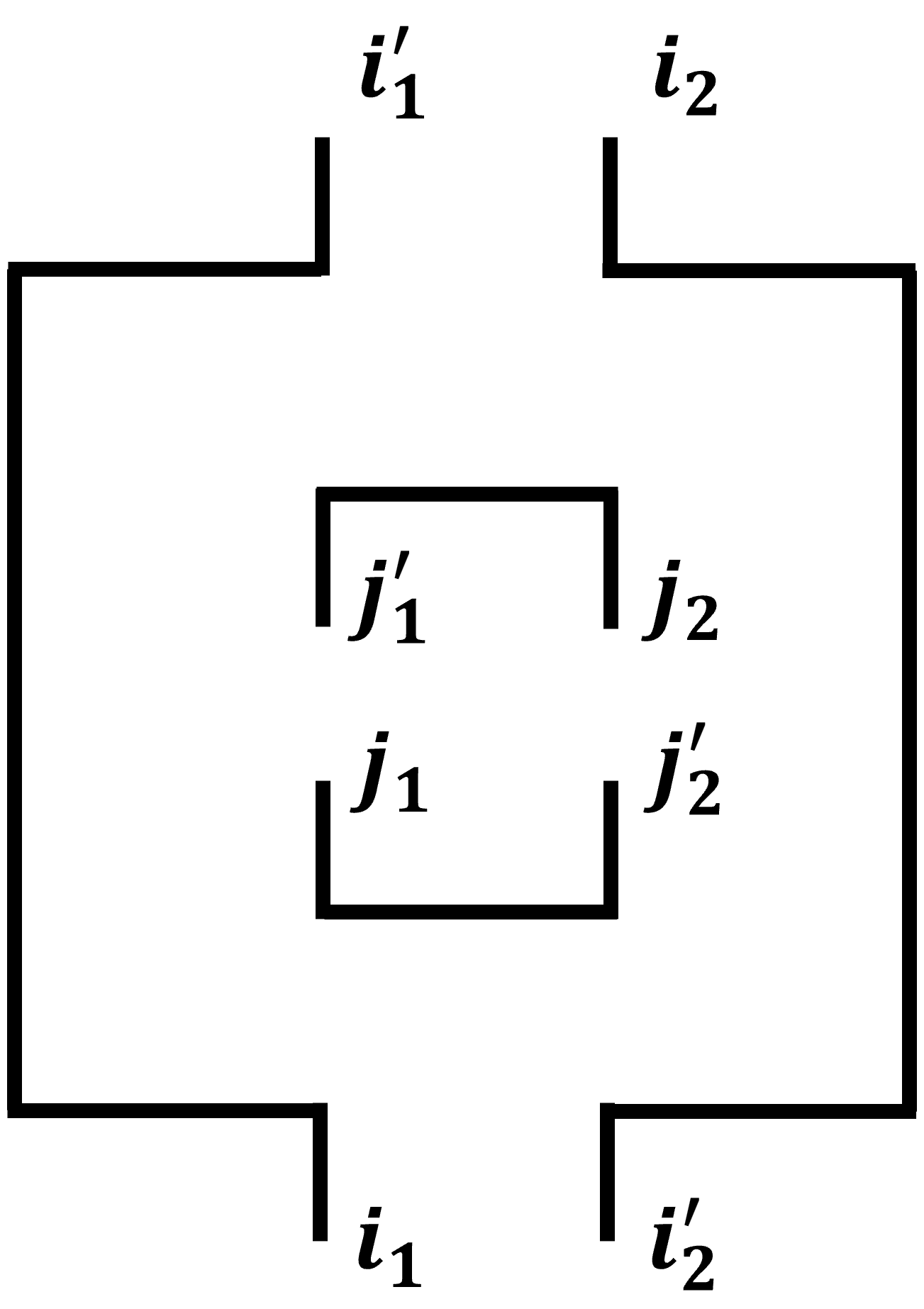}+\figbox{0.2}{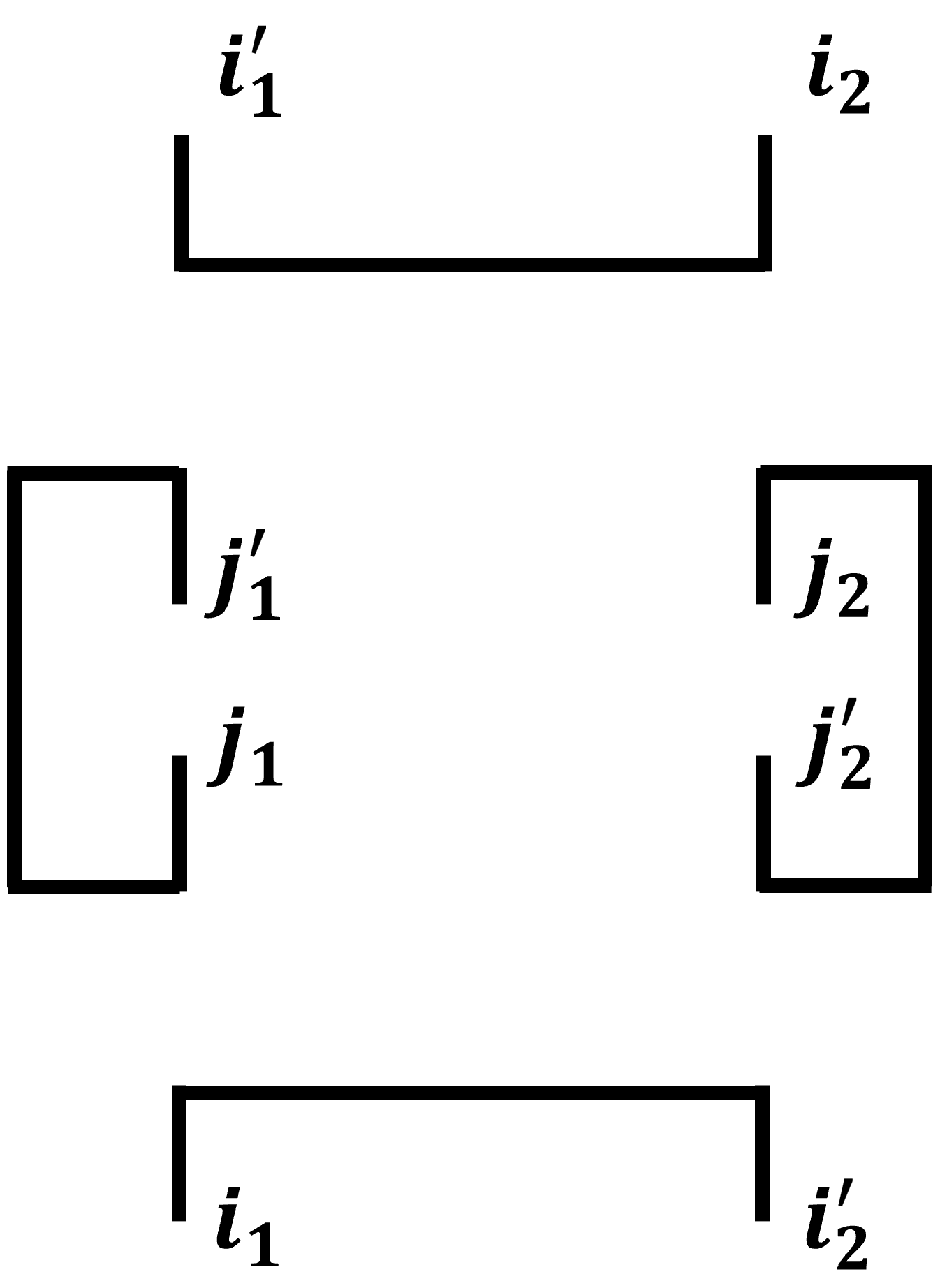}\right)\;.\label{4U_graph}
\end{eqnarray}
These graphical representations provide a convenient way to calculate the Haar averages. In the following, we will present the detailed calculations of some of the results presented in the main text. 

\subsection{The normalization of the Hayden-Preskill state with projective measurement} 
\label{sec:HP_norm}

For the Hayden-Preskill state $|\Psi_{HP}\rangle$ in Eq.\eqref{HP_state_eq}, its conjugate state is given by reversing the graphical representation of $|\Psi_{HP}\rangle$ 
\begin{eqnarray}
    \langle \Psi_{HP}|=\sqrt{|D|}\quad \figbox{0.3}{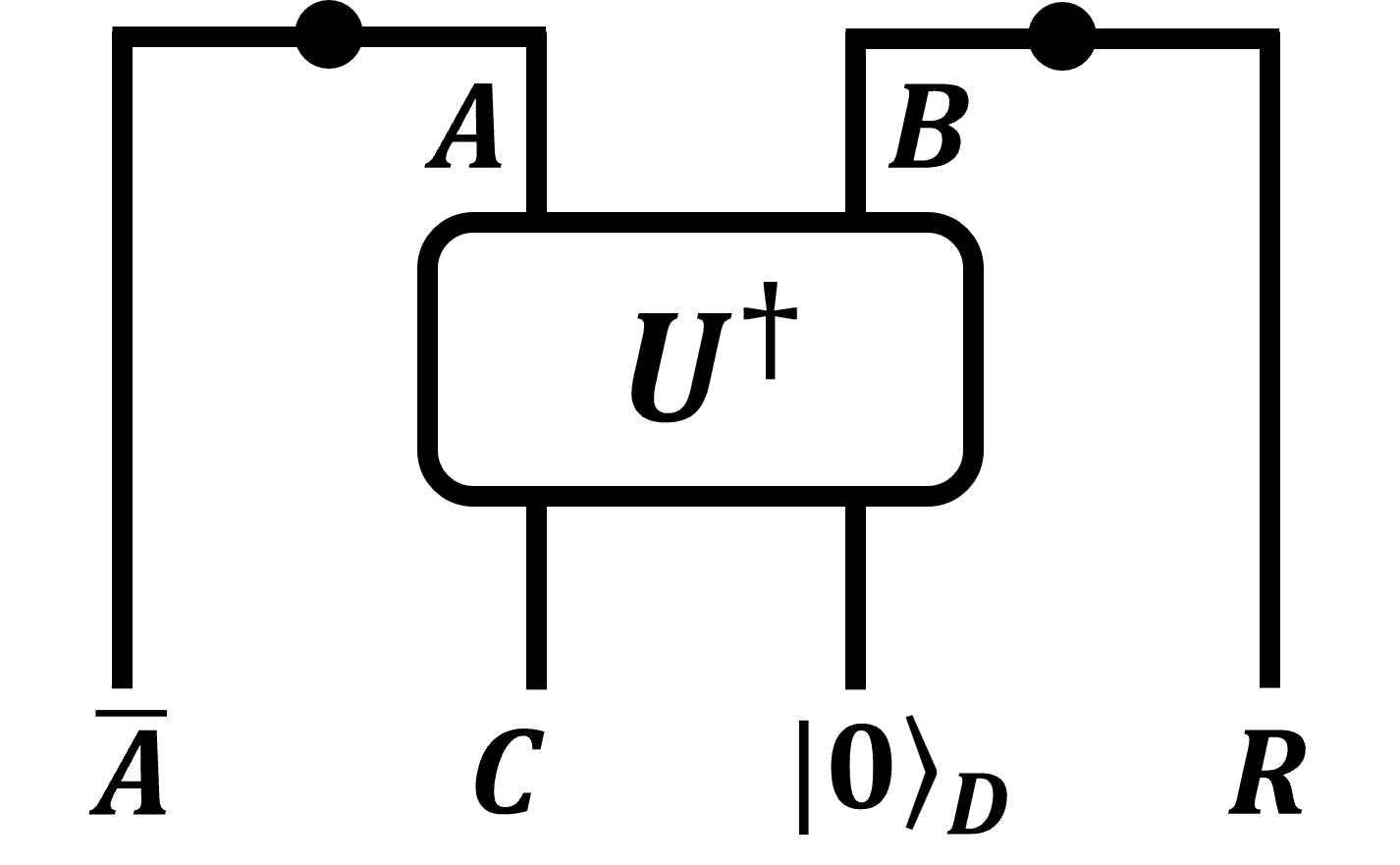}  \;.
\end{eqnarray}
The norm of the Hayden-Preskill state is then given by
\begin{eqnarray}
    \langle \Psi_{HP}|\Psi_{HP}\rangle&=&|D|\quad \figbox{0.3}{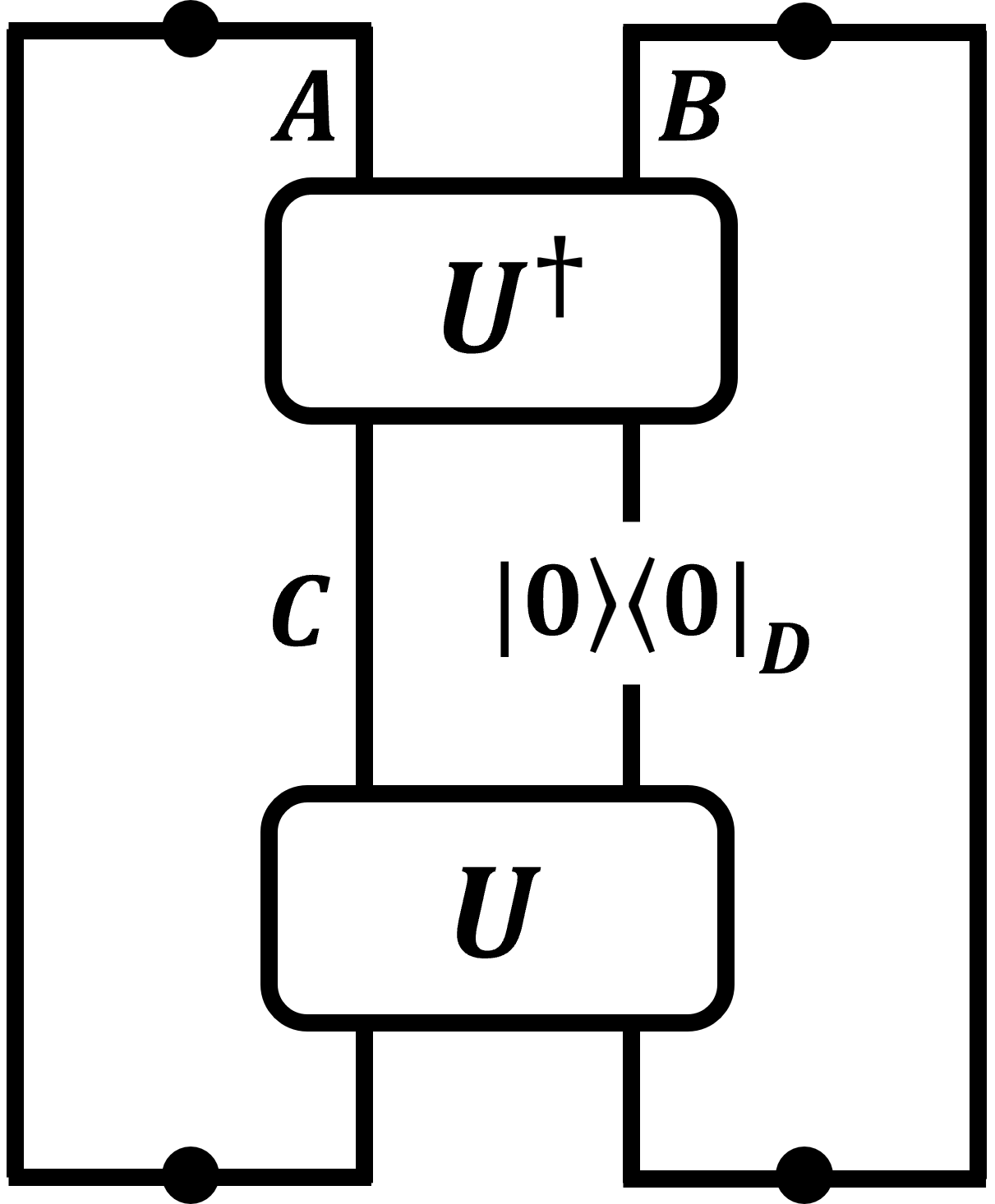} =\frac{|D|}{|A||B|}\quad\figbox{0.3}{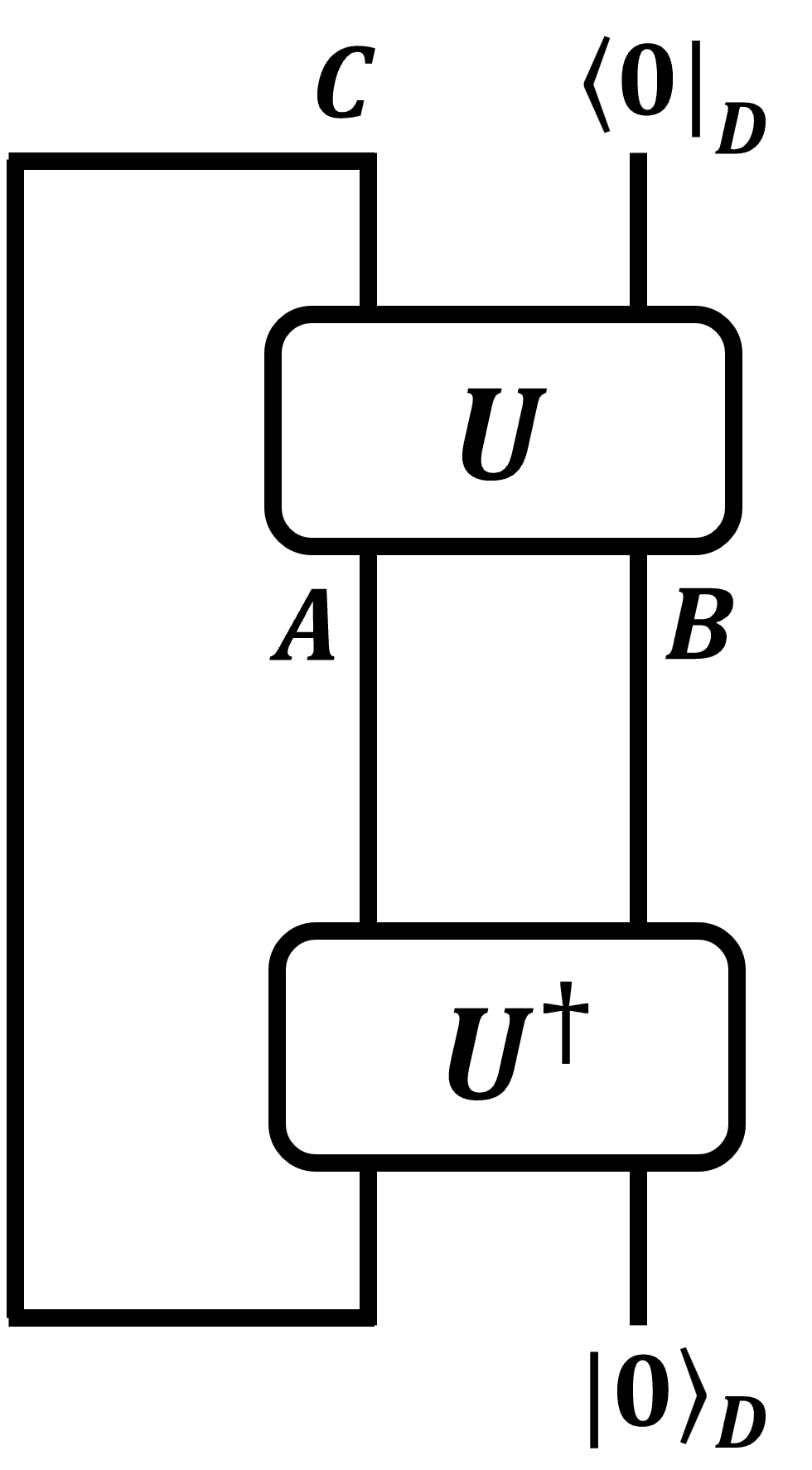}
    =\frac{|D|}{|A||B|}\left(\figbox{0.3}{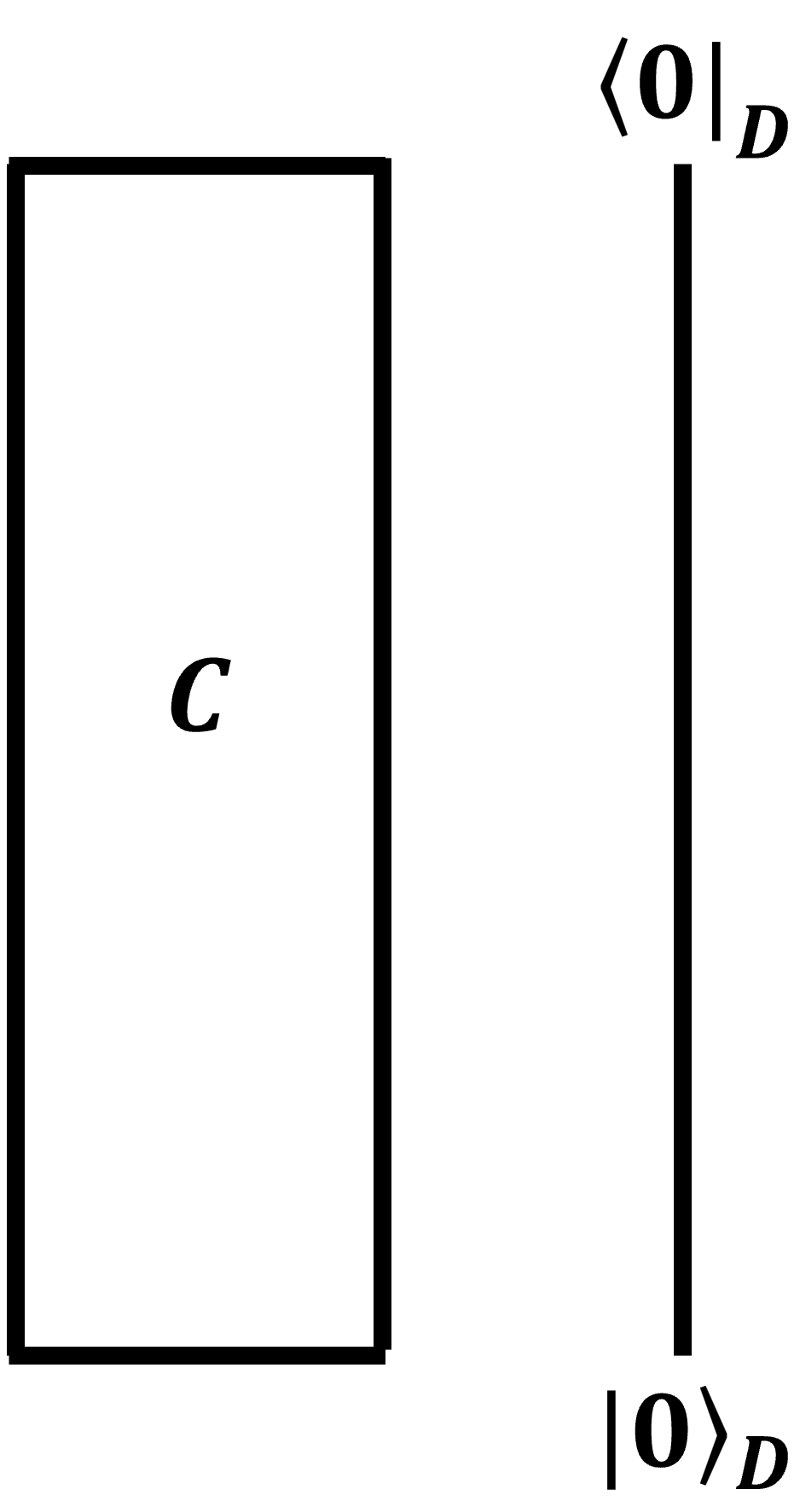}\right)
    \nonumber\\
    &=&\frac{|C||D|}{|A||B|}=1\;.
\end{eqnarray}
Note that the topology and connectivity of the identical graphs remain unchanged throughout the transformation between them. Additionally, we also use the fact that a loop effectively contributes the result with a factor corresponding to the dimension of the associated Hilbert space.

\subsection{Calculation of Haar average of \texorpdfstring{$\textrm{Tr}\left(\rho_{\overline{A}C}^2\right)$}{Lg}} 
\label{sec:Trac_rho2}

The density matrix for the Hayden-Preskill state with the projective measurement is given by
\begin{eqnarray}
    \rho_{HP}=|\Psi_{HP}\rangle \langle \Psi_{HP}|= |D| \quad \figbox{0.3}{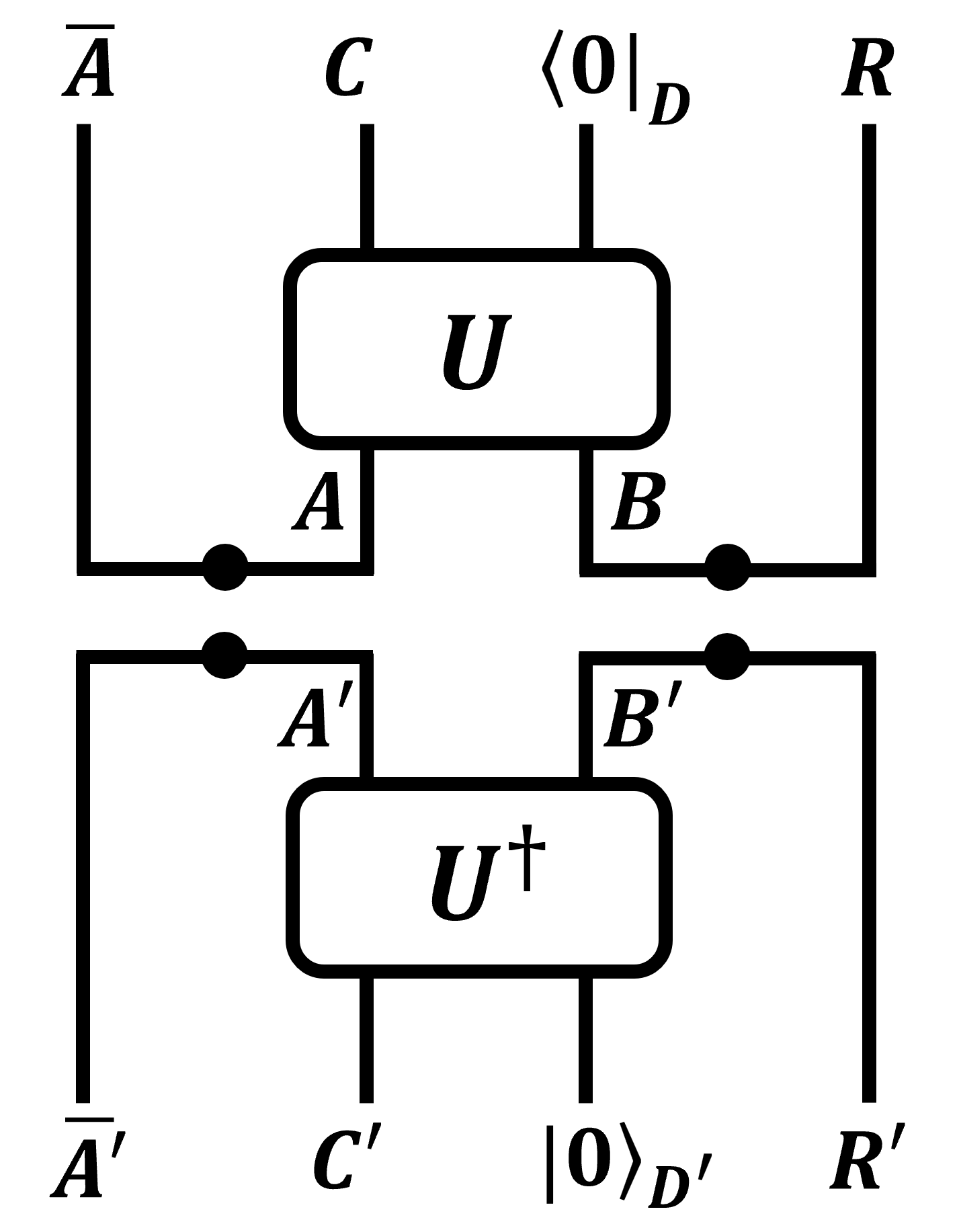}\;.
\end{eqnarray}
The reduced density matrix of for the combined subsystem $\overline{A}C$ is given by tracing out the subsystem $R$ as  
\begin{eqnarray}
    \rho_{\overline{A}C}&=& \textrm{Tr}_{RR'}  |\Psi_{HP}\rangle \langle \Psi_{HP}| = \frac{|D|}{|R|} \quad \figbox{0.3}{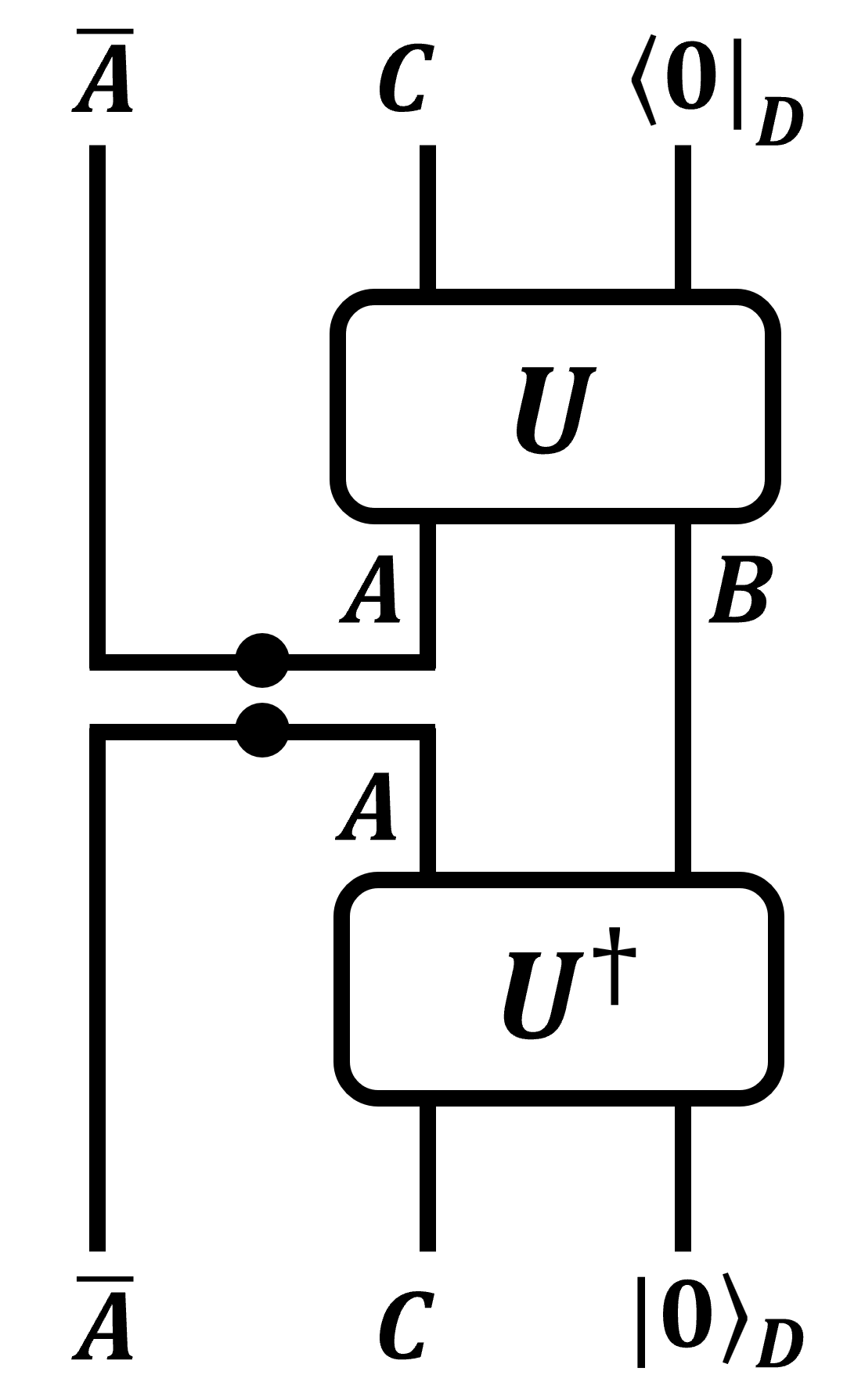}\;.
\end{eqnarray}
For $\rho_{\overline{A}C}^2$, we have  
\begin{eqnarray}
   \rho_{\overline{A}C}^2=\frac{|D|^2}{|R|^2}\quad 
   \figbox{0.3}{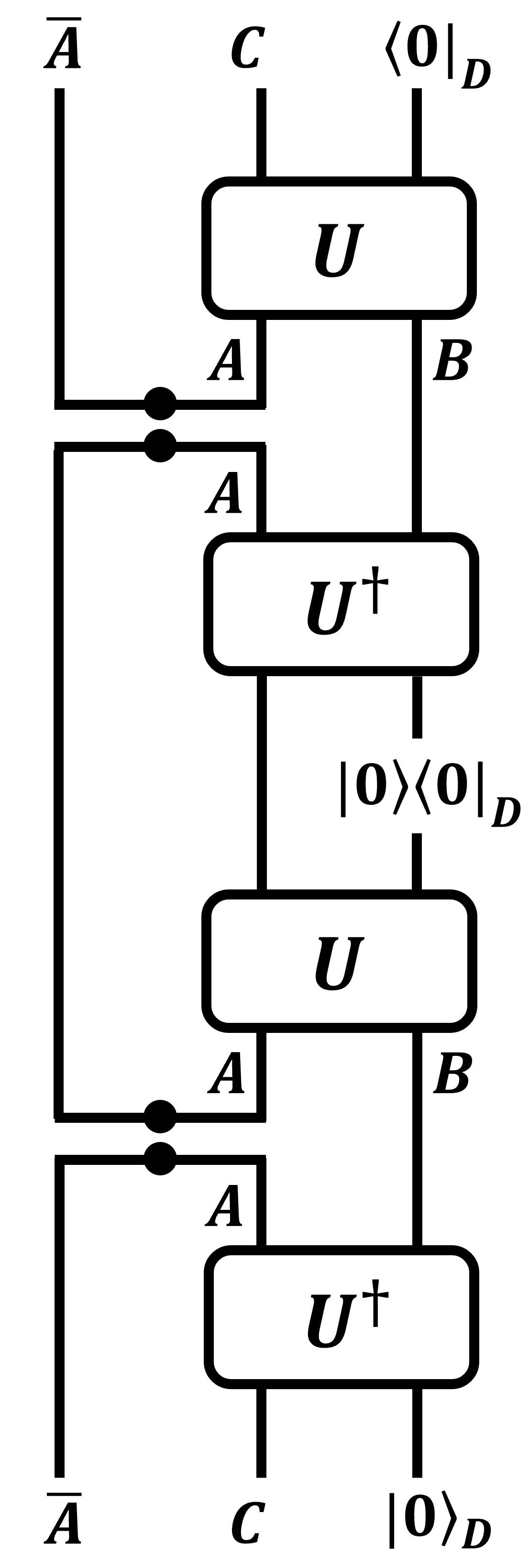}\;.
\end{eqnarray}
With some rearrangement, the trace of $\rho_{\overline{A}C}^2$ is then given by 
\begin{eqnarray}
    \textrm{Tr}\left(\rho_{\overline{A}C}^2\right)&=&\frac{|D|^2}{|A|^2|R|^2}\quad 
   \figbox{0.3}{Trace_rho_sqare.png}\;.
\end{eqnarray}
In arriving at the above representation, we have used the fact that 
\begin{eqnarray}
    \figbox{0.3}{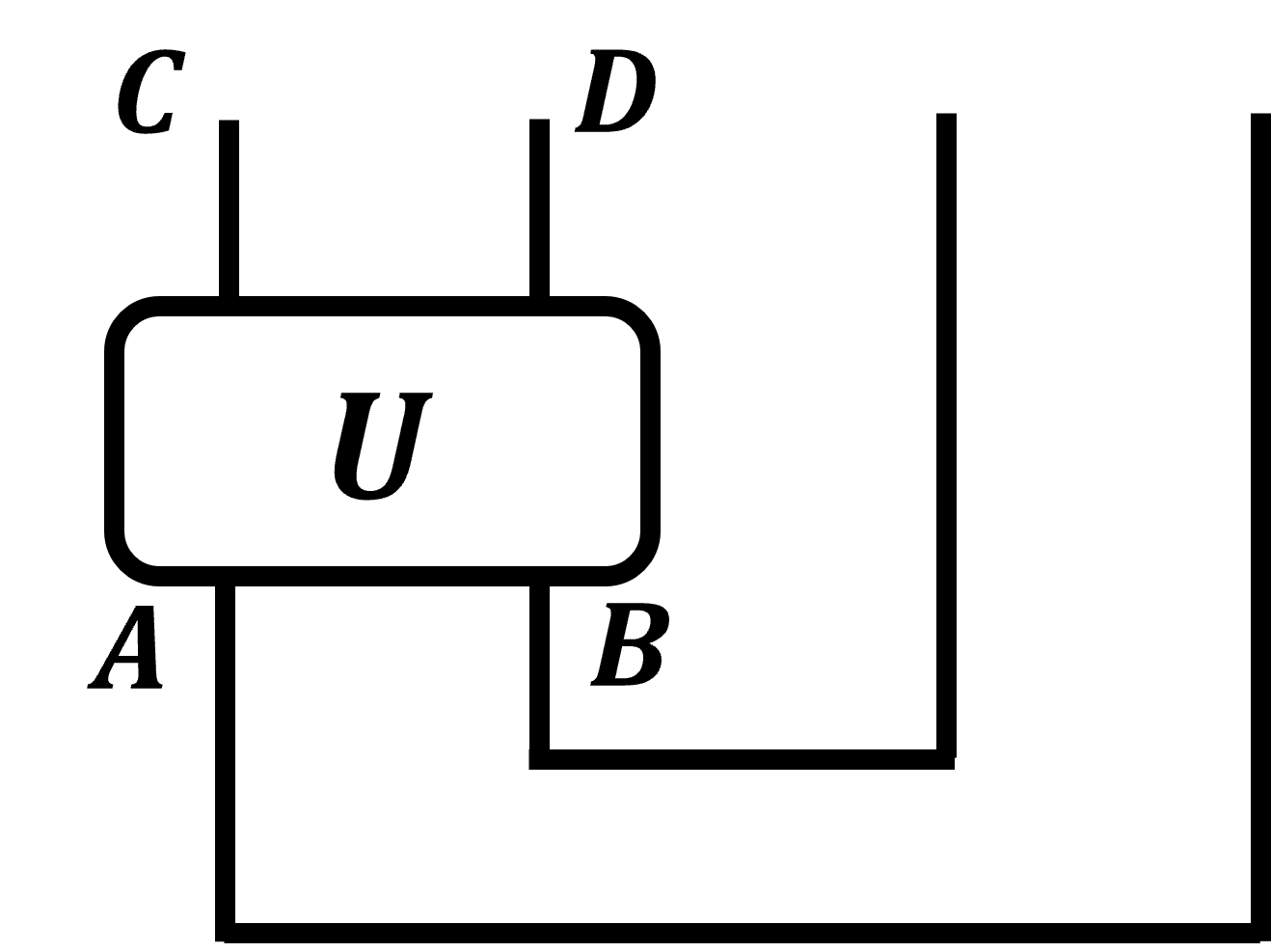}=\figbox{0.3}{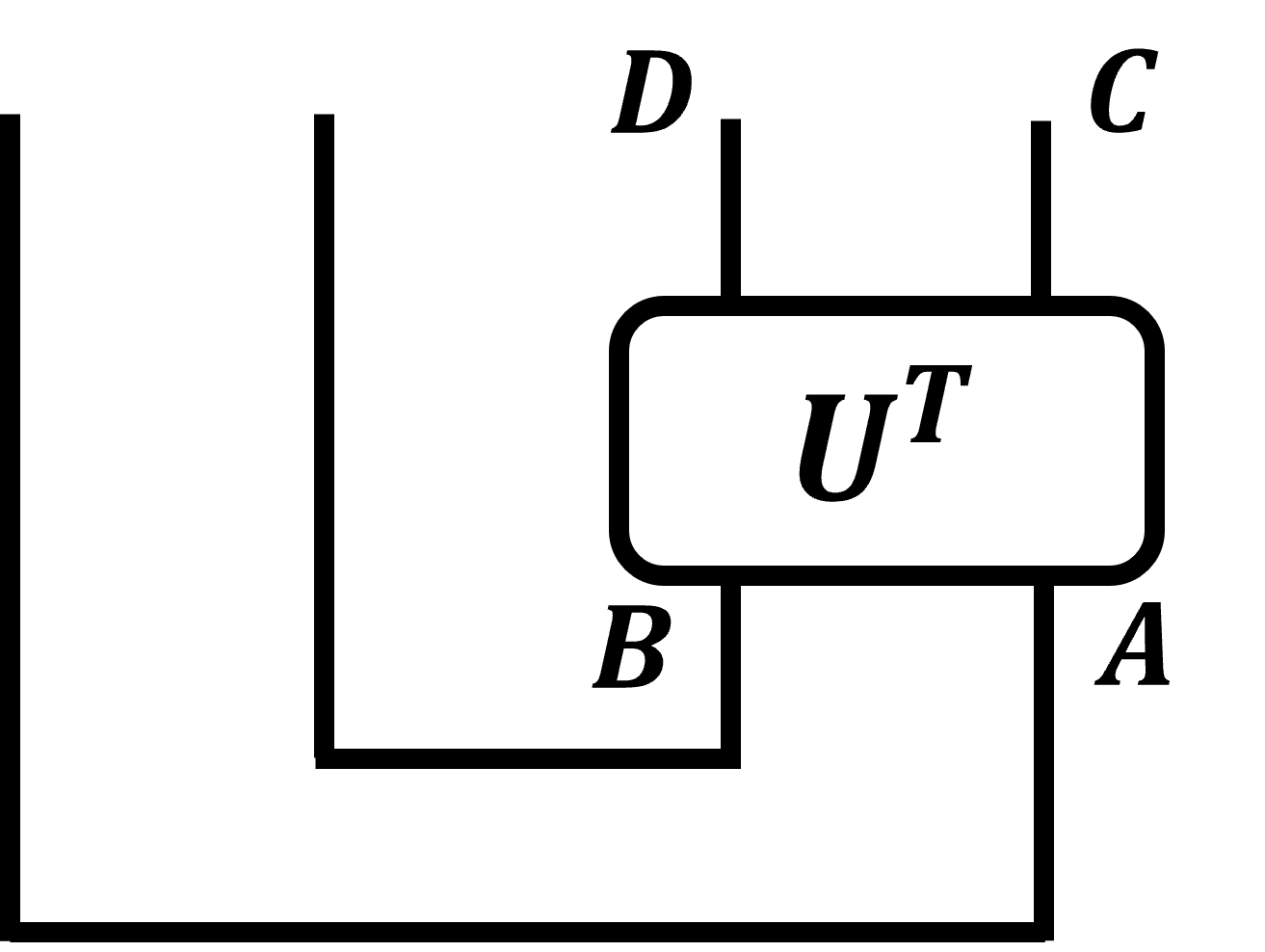}\;.
\end{eqnarray}

The average can be early carried out by using the graphical representation of the Haar integral as  
\begin{eqnarray}\label{Haar_aver_rho_ACsq}
    \int dU \textrm{Tr}\left(\rho_{\overline{A}C}^2\right)&=&
    \frac{|D|^2}{|A|^2|B|^2}\left[
    \frac{1}{(d^2-1)}\left( \figbox{0.2}{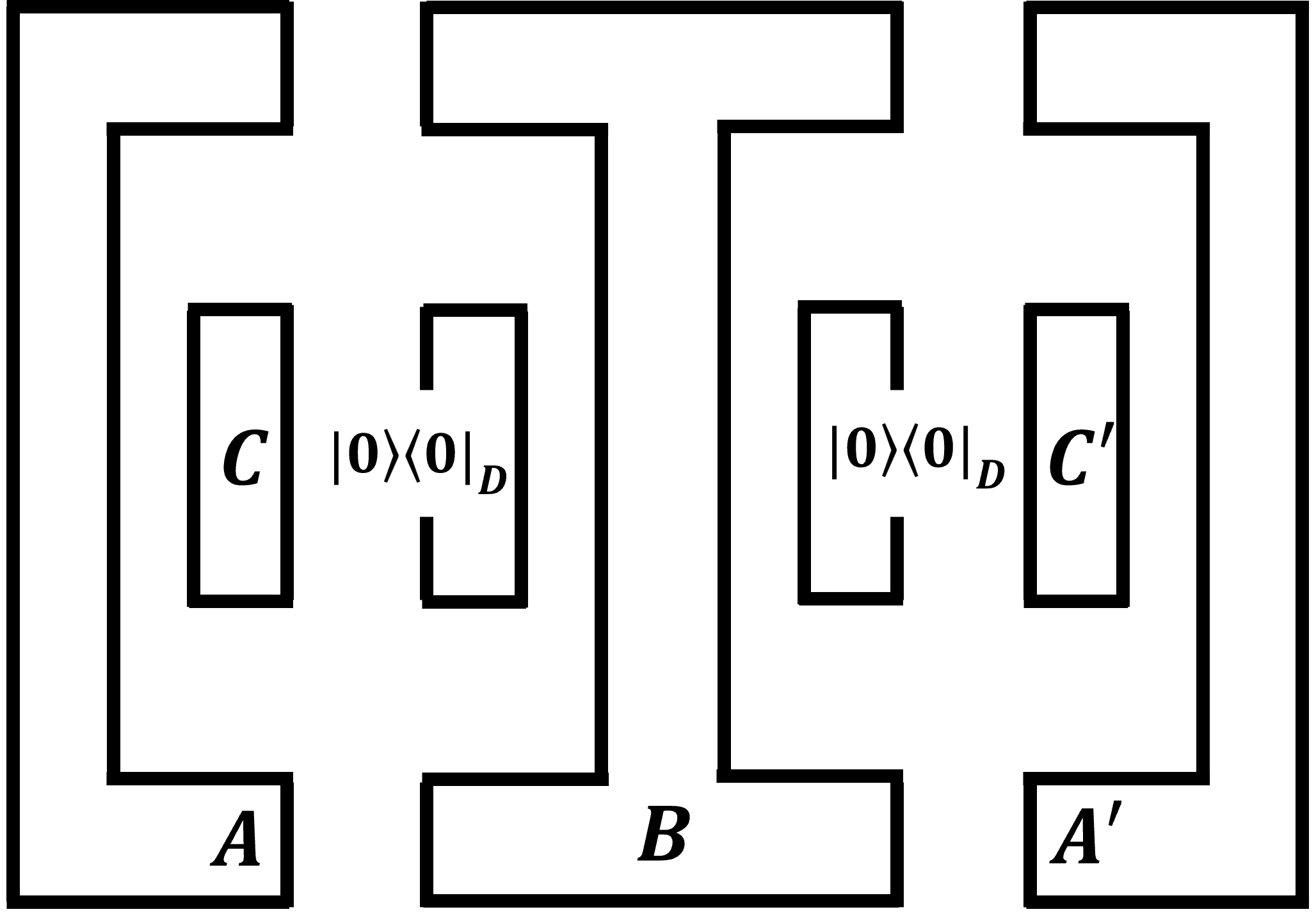}+\figbox{0.2}{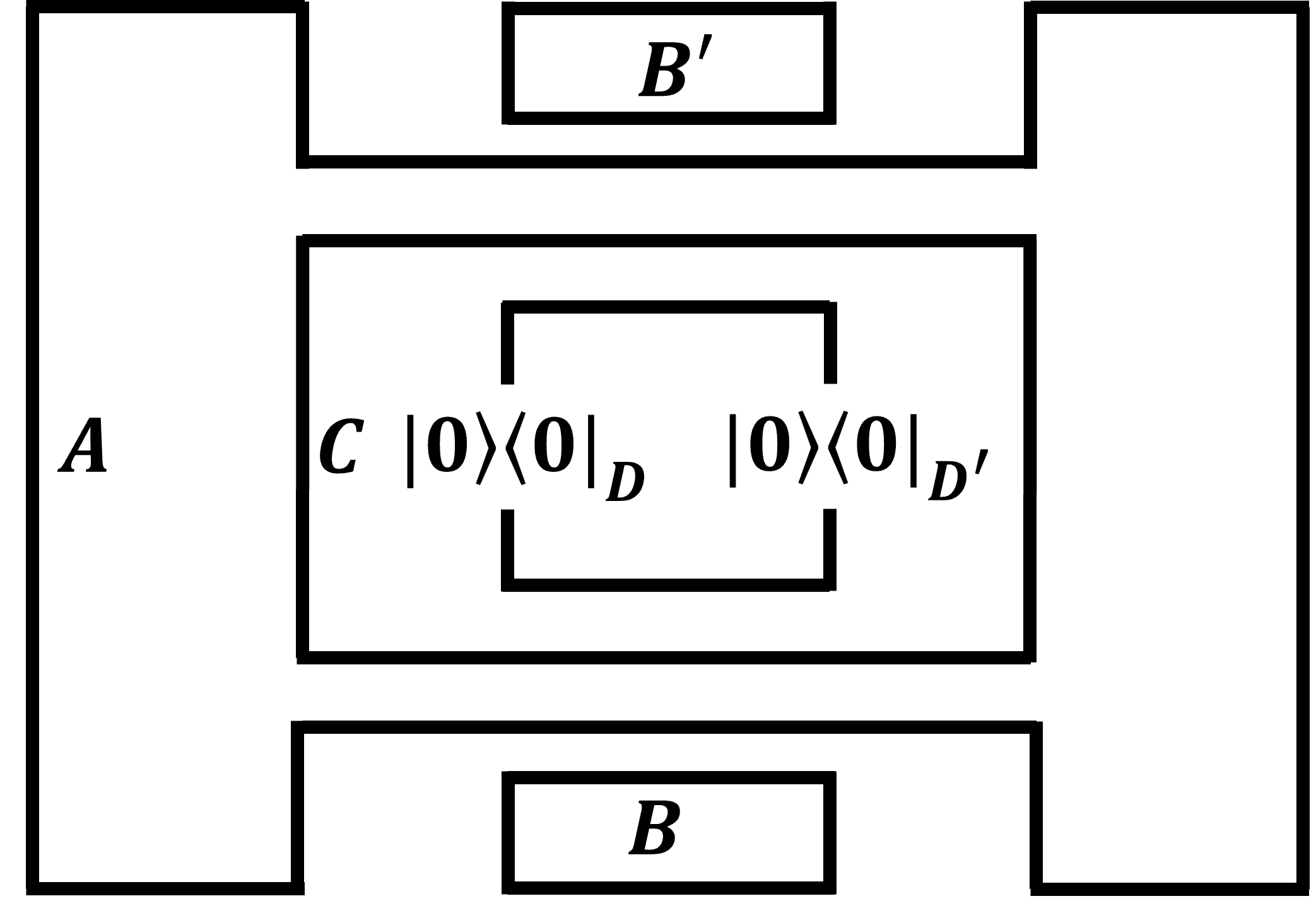}\right)\right.\nonumber\\
    &&-\left.\frac{1}{d(d^2-1)}
    \left( \figbox{0.2}{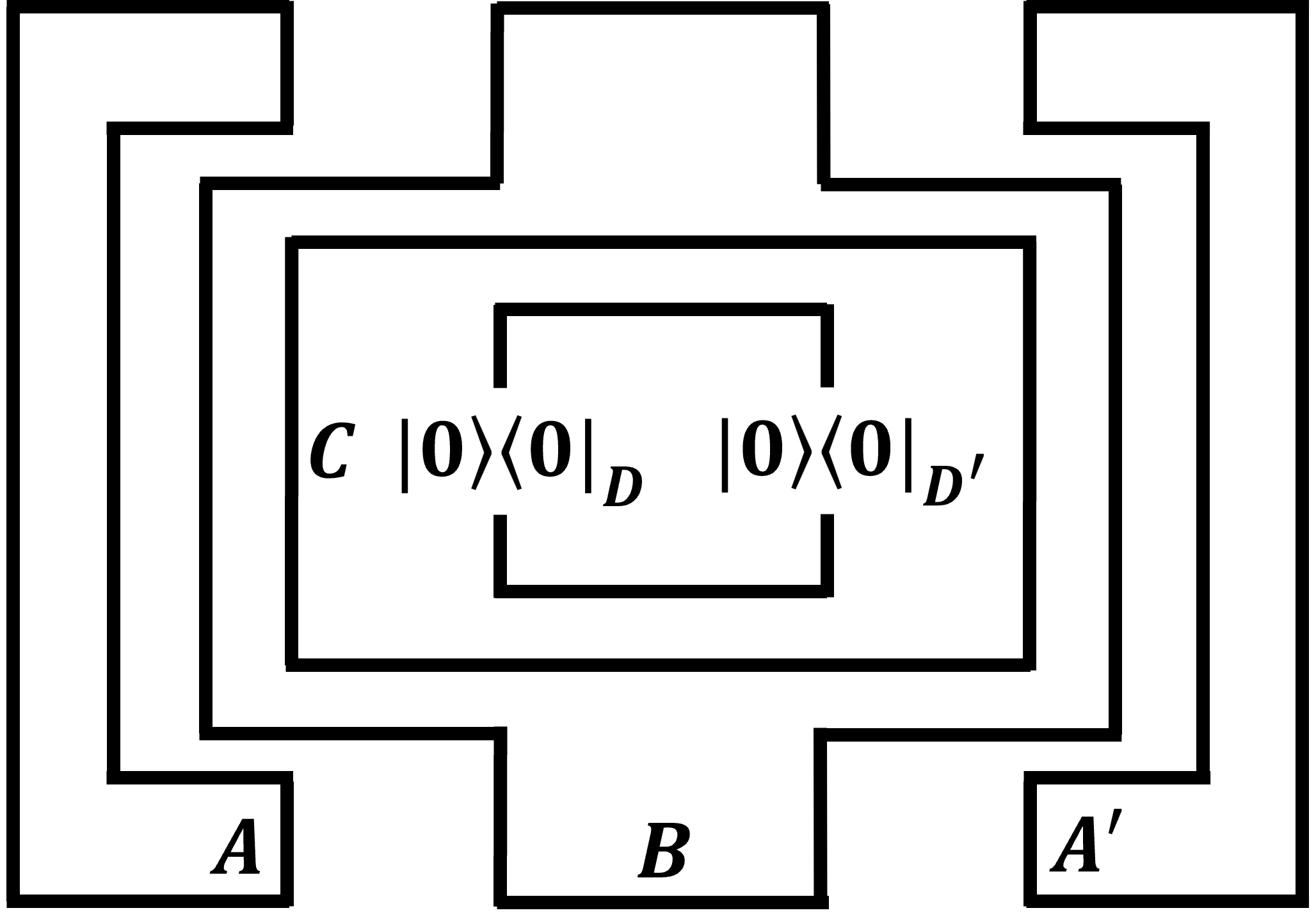}+\figbox{0.2}{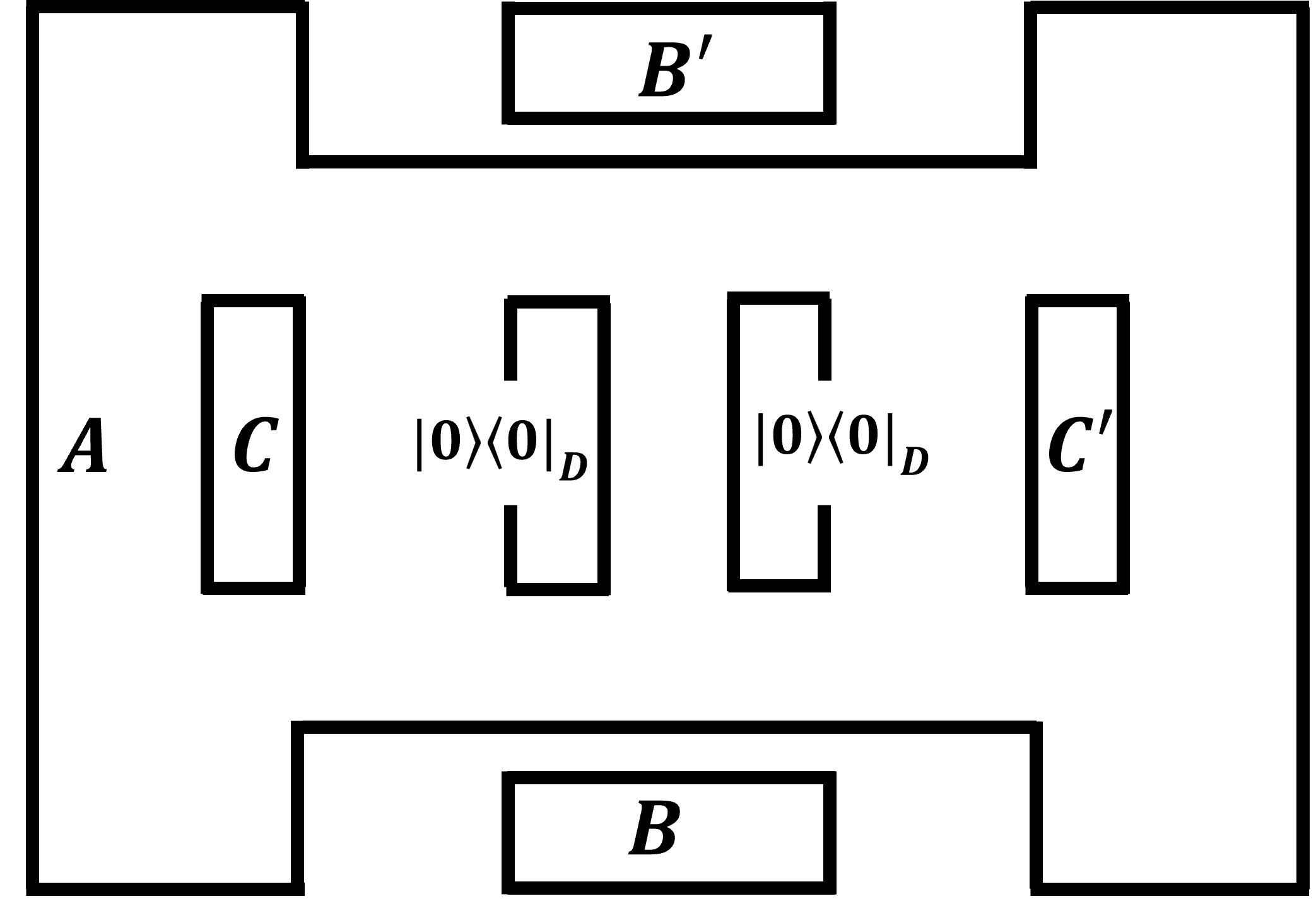}\right)
    \right] \nonumber\\
    &=&\frac{|D|^2}{|A|^2|B|^2}\left[\frac{1}{(d^2-1)}\left(|A|^2|B||C|^2+|A||B|^2|C|\right)\right.\nonumber\\
    &&\left.
    -\frac{1}{d(d^2-1)}\left(|A|^2|B||C|+|A||B|^2|C|^2\right)\right]\nonumber\\
    &=& \frac{1}{(d^2-1)}\left(|A|d+ \frac{|D|}{|A|}d -\frac{|A||D|}{d} -\frac{d}{|A|} \right)\;.
   \end{eqnarray}
where we have used the identity $d=|A||B|=|C||D|$ to eliminate the unrelated factor $|B|$ and $|C|$. This is the result presented in Eq.\eqref{trace_result1}.

\subsection{Haar average of \texorpdfstring{$\textrm{Tr}\left[\rho_{\overline{A}C}\cdot\left(\rho_{\overline{A}}\otimes\rho_{C}\right)\right]$}{Lg}}
\label{sec:Haar_3rho}

As noted in the main text, the reduced density matrix of the subsystem $C$ is maximally mixed, i.e. $\rho_{C}=\frac{I_C}{|C|}$. However, $\rho_{\overline{A}}$ is not, which can be explicitly given by 
\begin{eqnarray}
    \rho_{\overline{A}}=\frac{|D|}{|B|}\figbox{0.3}{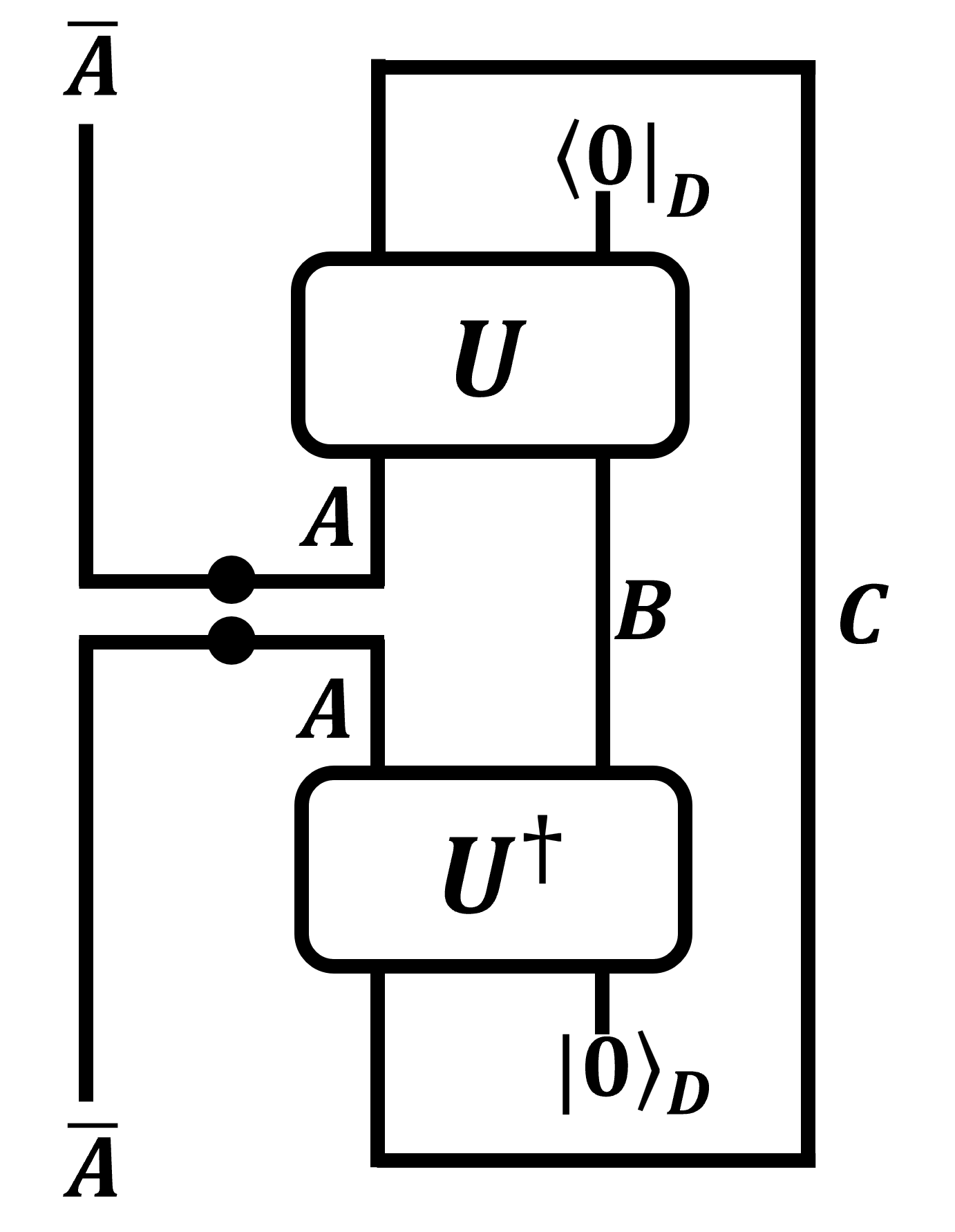}\;.
\end{eqnarray}
Then, $\textrm{Tr}\left[\rho_{\overline{A}C}\cdot\left(\rho_{\overline{A}}\otimes\rho_{C}\right)\right]$ can be graphically given by 
\begin{eqnarray}
    \textrm{Tr}\left[\rho_{\overline{A}C}\cdot\left(\rho_{\overline{A}}\otimes\rho_{C}\right)\right]=\frac{1}{|C|^3} \figbox{0.3}{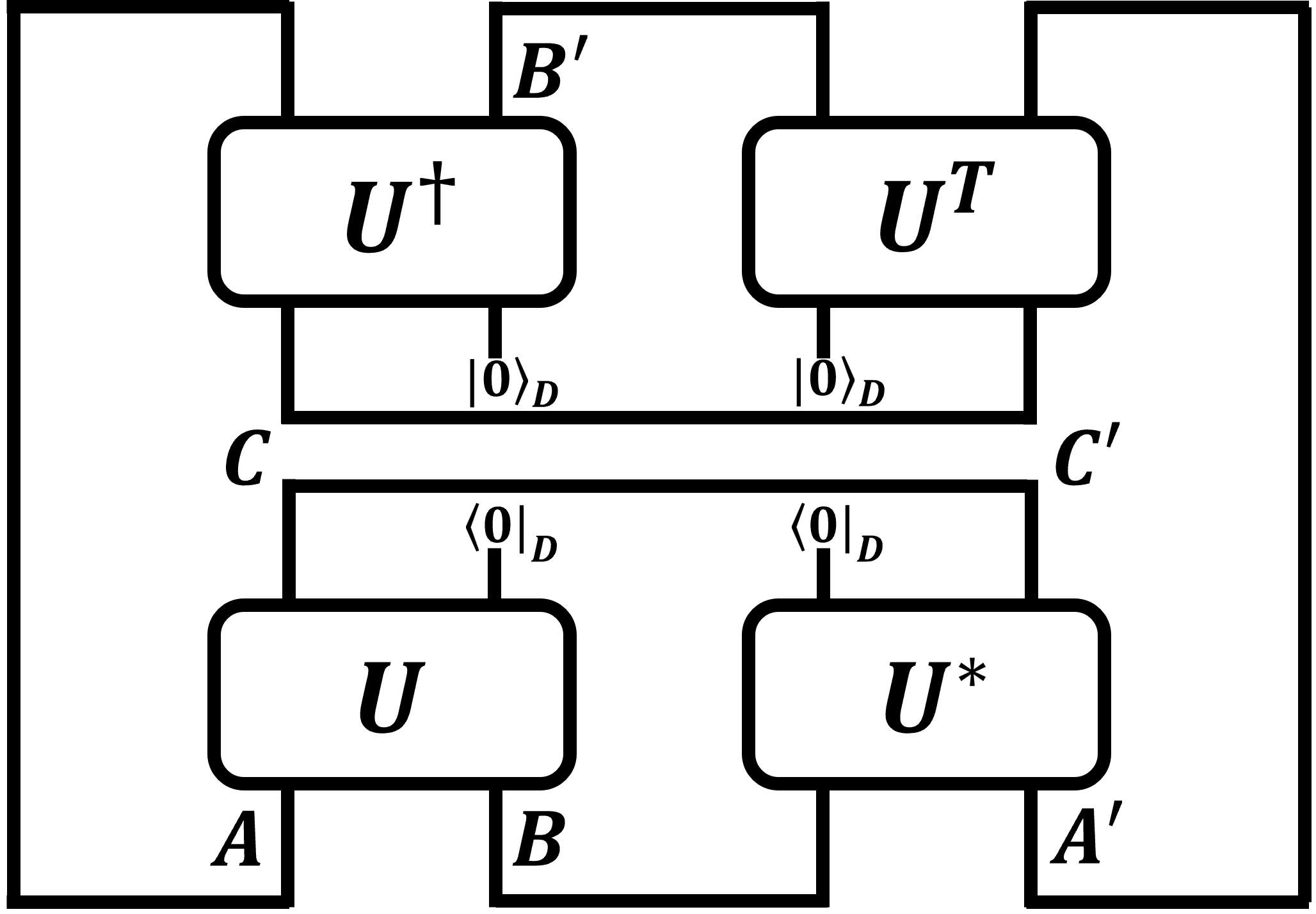}\;.
\end{eqnarray}
Using the graphical representation of Haar integral, one can get
\begin{eqnarray}\label{Haar_avr_3rho}
    \overline{\textrm{Tr}\left[\rho_{\overline{A}C}\cdot\left(\rho_{\overline{A}}\otimes\rho_{C}\right)\right]}&=&\frac{1}{|C|^3} \left[\frac{1}{(d^2-1)}\left(|A|^2|B||C|+|A||B|^2|C|^2\right)
    \right.\nonumber\\
    &&\left.-\frac{1}{d(d^2-1)}\left(|A|^2|B||C|^2+|A||B|^2|C|\right)\right]\;.
\end{eqnarray}
With some simplification, one can get the result presented in Eq.\eqref{trace_result2}.

\subsection{The normalization of \texorpdfstring{$|\Psi_{in}\rangle$}{Lg}}
\label{sec:psi_in_norm}

The normalization of the state $|\Psi_{in}\rangle$ is easy to verify, which is given by as follows
\begin{eqnarray}
    \langle\Psi_{in}|\Psi_{in}\rangle&=&\frac{|D|}{|A|^2|B|}\quad \figbox{0.3}{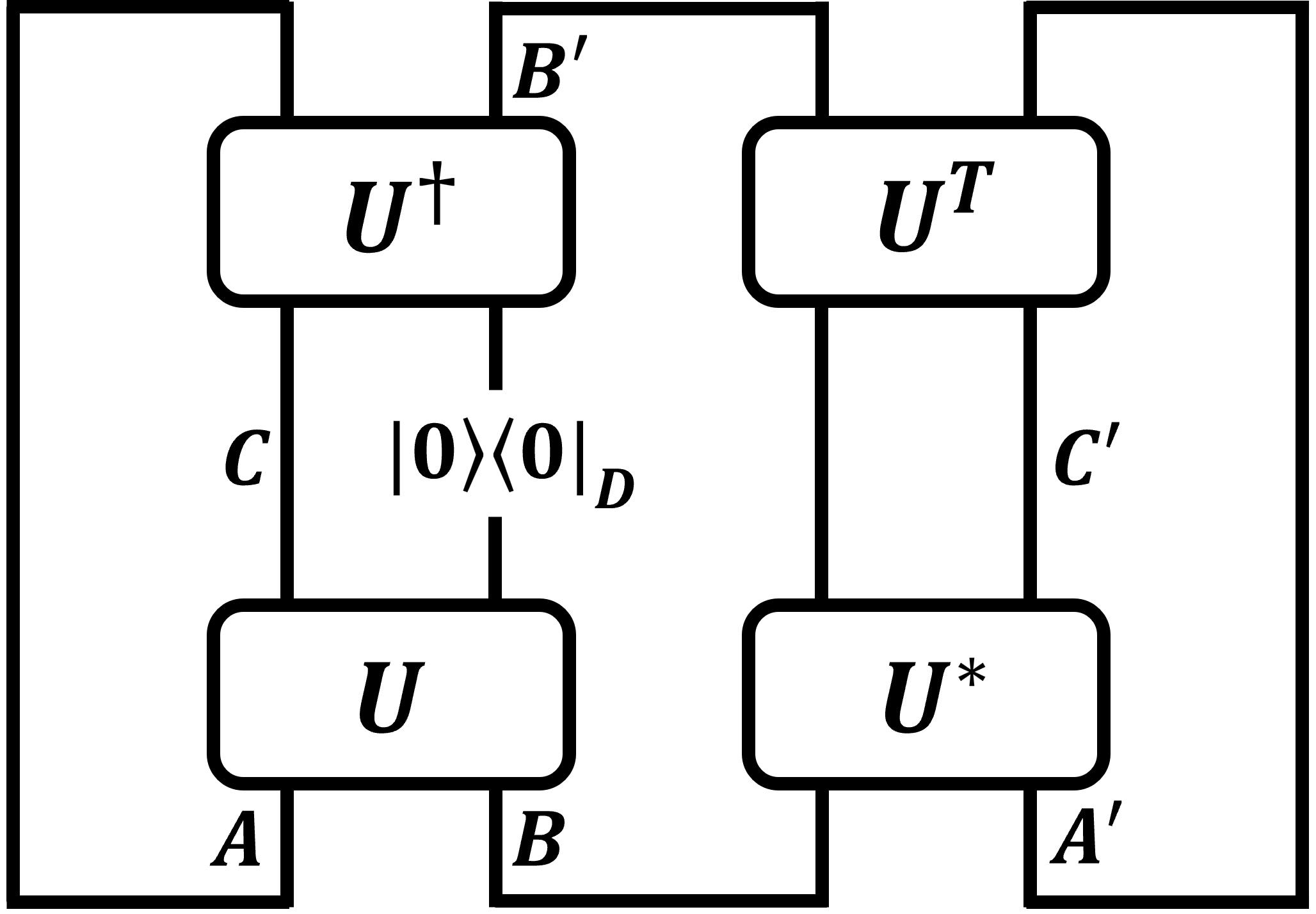}=
    \frac{|D|}{|A|^2|B|}\quad \figbox{0.3}{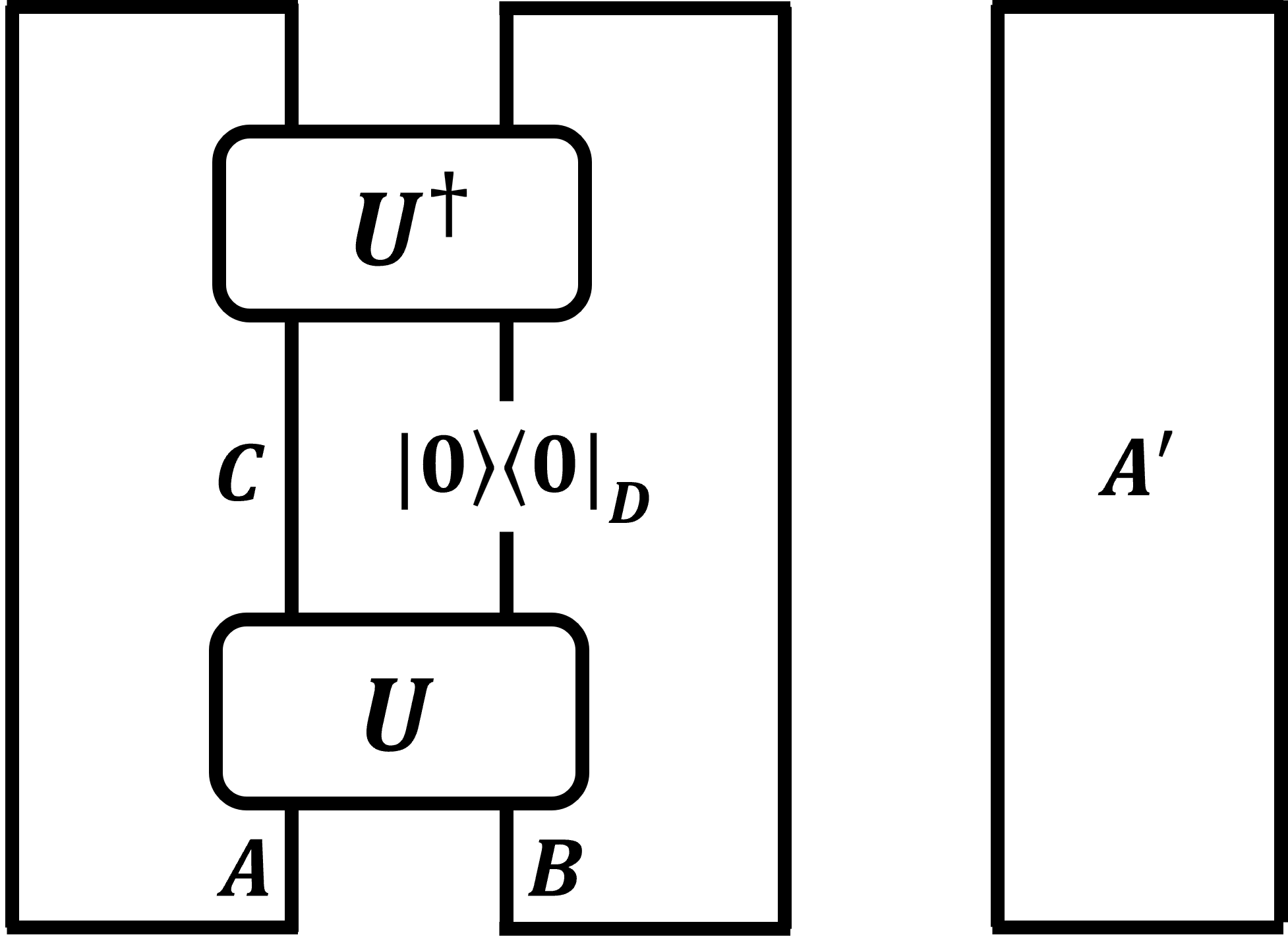}\nonumber\\
    &=&\frac{|D|}{|A||B|}\quad \figbox{0.3}{HP_state_norm_1.png}
    =\frac{|C||D|}{|A||B|}=1\;.
\end{eqnarray}

\subsection{The normalization of \texorpdfstring{$\tilde{\rho}_{HP}$}{Lg}}
\label{sec:HPQ_norm}

The acting of the depolarizing channel on the Hayden-Preskill state can be graphically represented as 
\begin{eqnarray}\label{Q_decomposition}
    \Tilde{\rho}_{HP}&=&|D|\quad \figbox{0.3}{HP_density_matrix_Q.png}\nonumber\\
    &=&|D|\quad\left[(1-p)\left(\figbox{0.3}{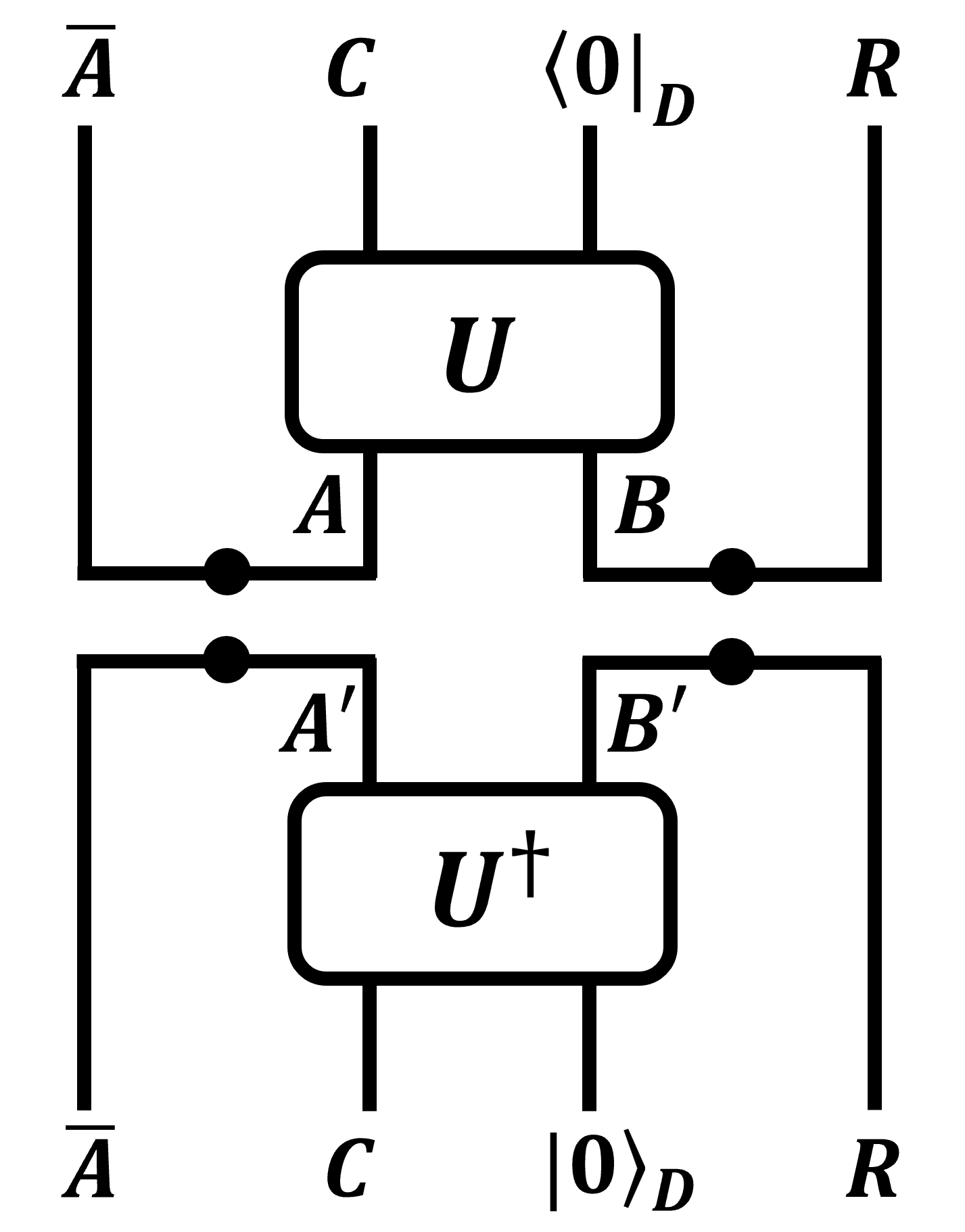}\right) 
   +p \left(\figbox{0.3}{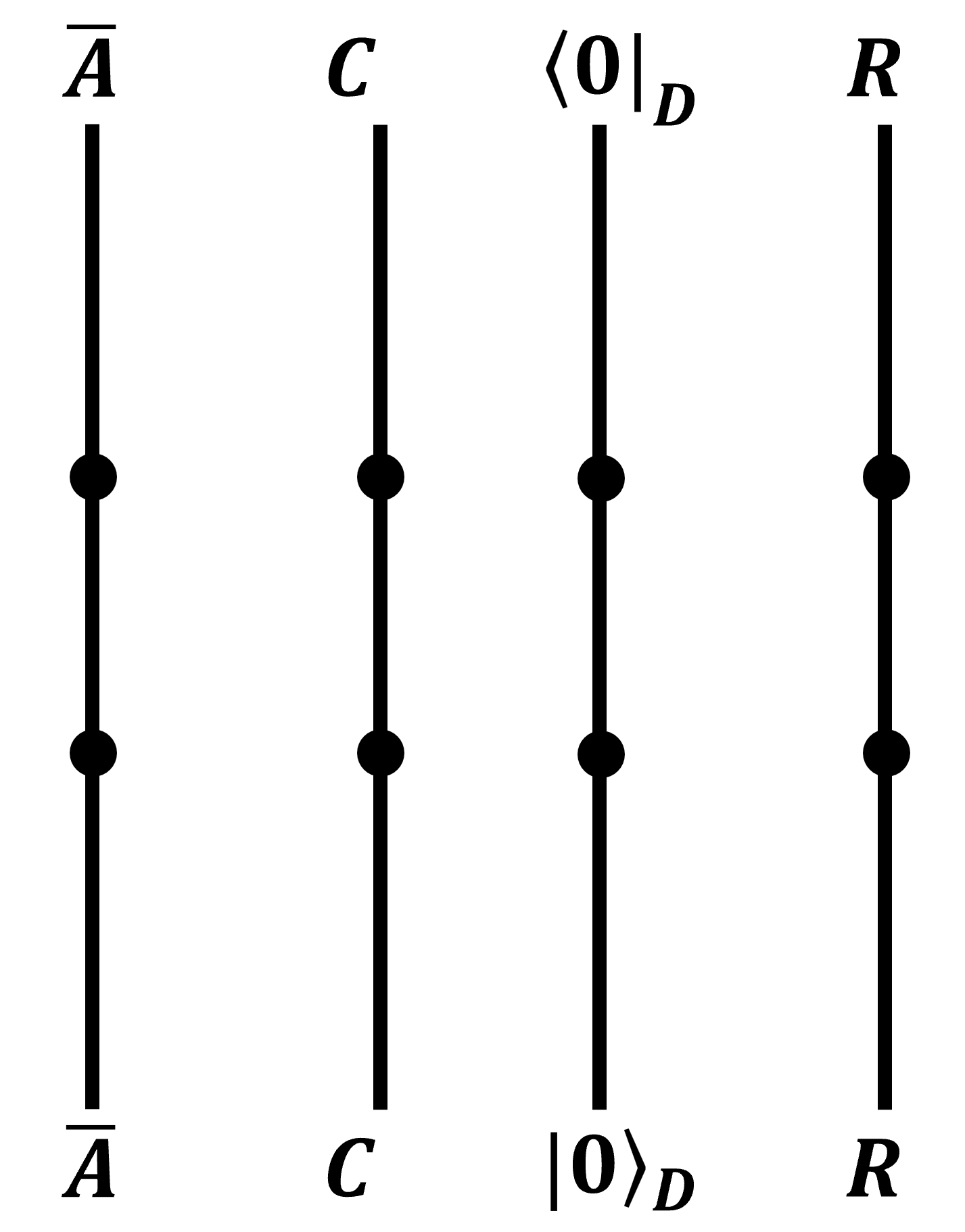}\right)    
    \right] \;,
\end{eqnarray}
The trace of $\tilde{\rho_{HP}}$ can be obtained by connecting the corresponding endpoints. It is easy to see that the two graphs in the above equation all give the same factor $\frac{1}{|D|}$. Then the trace of $\tilde{\rho}_{HP}$ is given by 
\begin{eqnarray}
    \textrm{Tr} \tilde{\rho}_{HP} = |D|\left[(1-p)\frac{1}{|D|}+\frac{p}{|D|}\right]=1\;. 
\end{eqnarray}
From the above illustration, we can see that the depolarizing channel $\mathcal{Q}$ can be decomposed into two parts: the first is original density matrix without coherence and the second is the completely depolarizing part which corresponds a maximally mixed state.

\subsection{Haar average of \texorpdfstring{$\textrm{Tr}\left(\tilde{\rho}_{\overline{A}C}^2\right)$}{Lg}} 
\label{sec:Haar_trace_tilde_rho_sq}

From the Hayden-Preskill state in Eq.\eqref{HP_state_Q}, one can get the reduced density matrix of the subsystem $\overline{A}C$ as
\begin{eqnarray}
    \tilde{\rho}_{\overline{A}C}=\frac{1}{|C|}\quad \figbox{0.3}{tilde_rho_Ac.png}\;.
\end{eqnarray}
Then $\tilde{\rho}_{\overline{A}C}^2$ is given by 
\begin{eqnarray}
    \tilde{\rho}_{\overline{A}C}^2=\frac{1}{|C|^2}\quad \figbox{0.3}{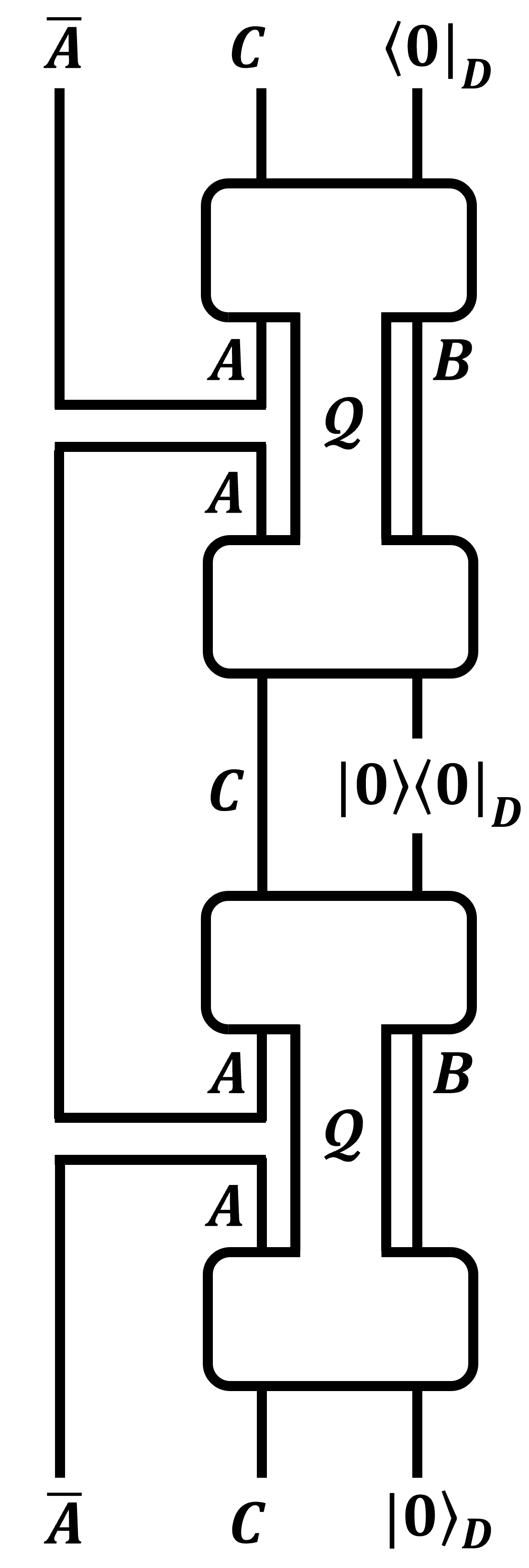}\;. 
\end{eqnarray}
By noting that 
\begin{eqnarray}
    \figbox{0.3}{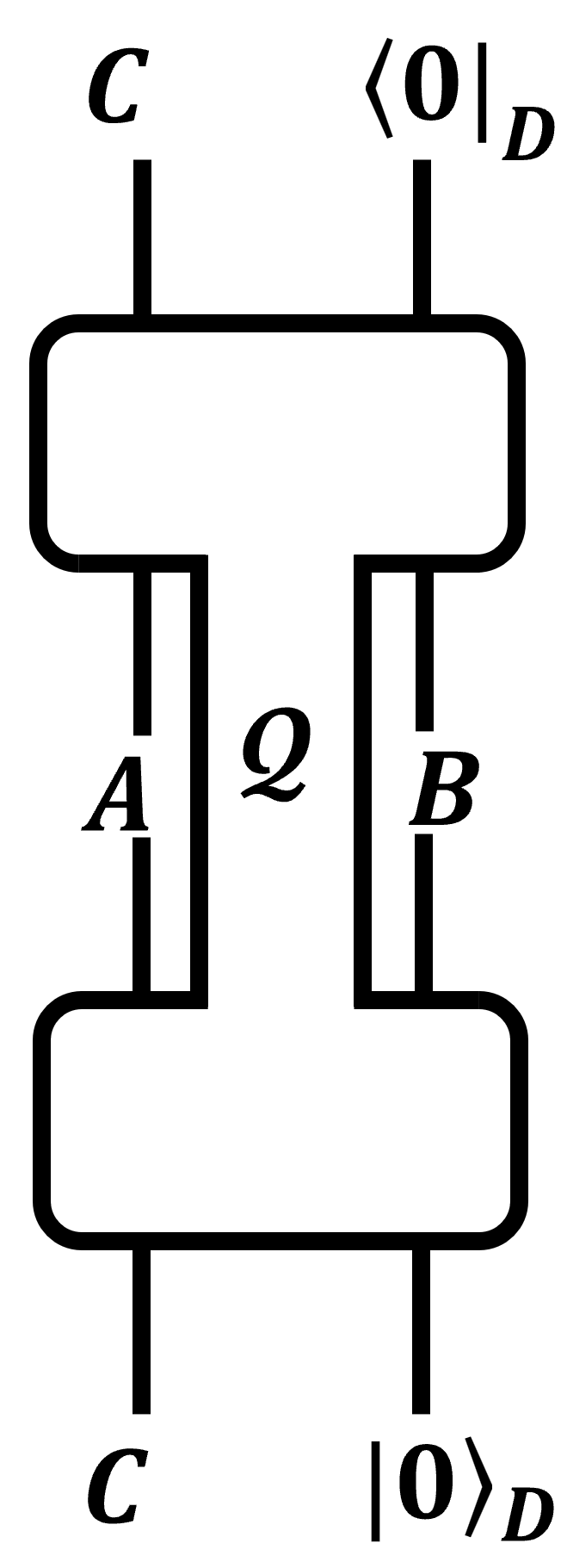}=\figbox{0.3}{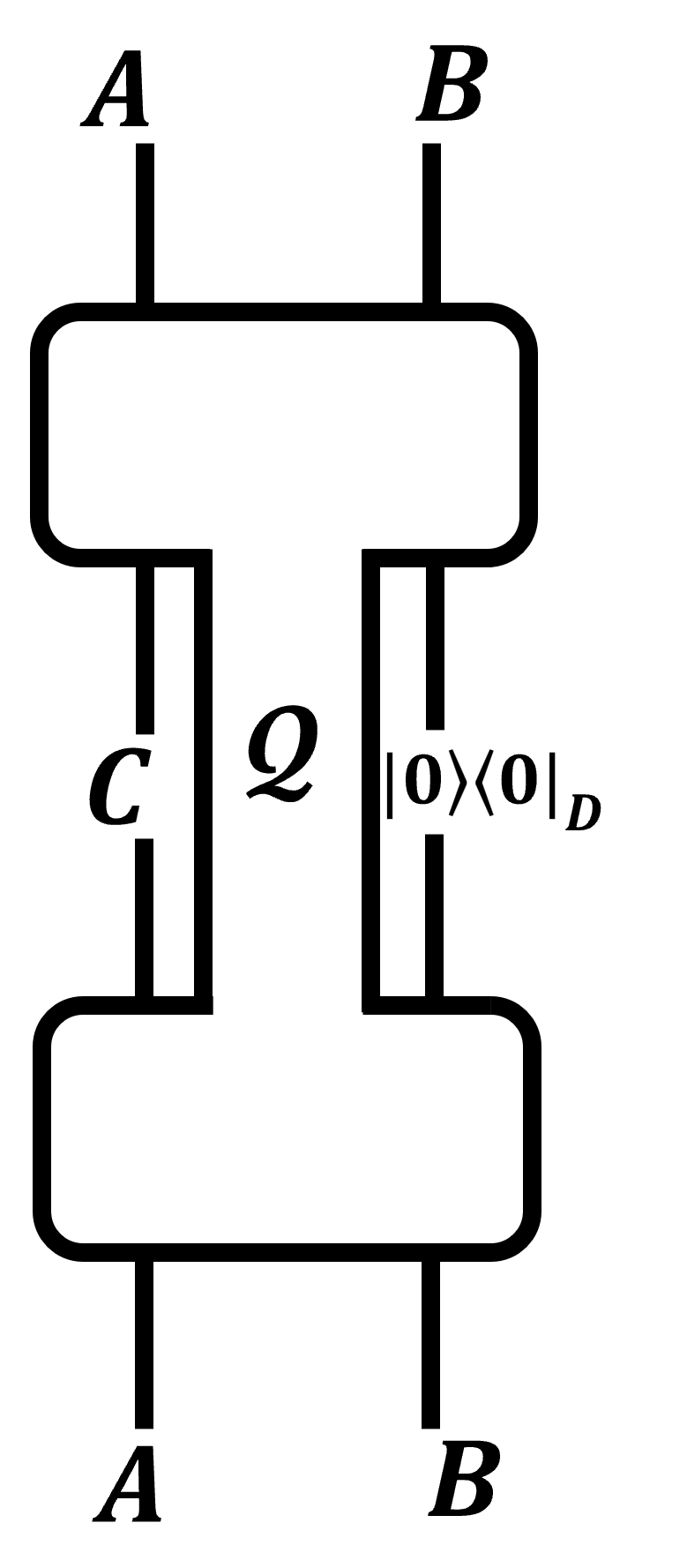}\;,
\end{eqnarray}
the trace of $\tilde{\rho}_{\overline{A}C}^2$ is given by 
\begin{eqnarray}
    \textrm{Tr}\left(\tilde{\rho}_{\overline{A}C}^2\right)=\frac{1}{|C|^2}\quad \figbox{0.3}{Trace_tilde_rho_AC_sq.png}\;.
\end{eqnarray}

By using the decomposition similar to Eq.\eqref{Q_decomposition}, one can calculate the Haar average of $\tilde{\rho}_{\overline{A}C}^2$. The procedure is rather troublesome but the result can be written in a compact form. The final result is given by 
\begin{eqnarray}
    \overline{ \textrm{Tr}\left(\tilde{\rho}_{\overline{A}C}^2\right)}=(1-p)^2\;\overline{ \textrm{Tr}\left(\rho_{\overline{A}C}^2\right)}+\frac{(2p-p^2)}{|B|}\;,
\end{eqnarray}
where $\overline{ \textrm{Tr}\left(\rho_{\overline{A}C}^2\right)}$ has been calculated in Eq.\eqref{Haar_aver_rho_ACsq}.

\subsection{Haar average of \texorpdfstring{$\textrm{Tr}\left[\tilde{\rho}_{\overline{A}C}\cdot\left(\tilde{\rho}_{\overline{A}}\otimes\tilde{\rho}_C\right)\right]$}{Lg}} 
\label{sec:Haar_trace_tilde_3rho}

The calculation is similar to that is performed in Appendix \ref{sec:Haar_3rho}. We briefly show the procedure. It is easy to see that the reduced density matrix of the subsystem $C$ is a maximally mixed, i.e. $\tilde{\rho}_C=\frac{I_C}{|C|}$. Therefore the following equation is satisfied 
\begin{eqnarray}
    \textrm{Tr}\left[\tilde{\rho}_{\overline{A}C}\cdot\left(\tilde{\rho}_{\overline{A}}\otimes\tilde{\rho}_{C}\right)\right]=\textrm{Tr}\left[\left(\tilde{\rho}_{\overline{A}}\otimes\tilde{\rho}_{C}\right)^2\right]\;.
\end{eqnarray}
The Haar average of $ \textrm{Tr}\left[\tilde{\rho}_{\overline{A}C}\cdot\left(\tilde{\rho}_{\overline{A}}\otimes\tilde{\rho}_{C}\right)\right]$ can be calculated by using the following graphical representation 
\begin{eqnarray}
    \textrm{Tr}\left[\tilde{\rho}_{\overline{A}C}\cdot\left(\tilde{\rho}_{\overline{A}}\otimes\tilde{\rho}_{C}\right)\right]=\frac{1}{|C|^3}\quad \figbox{0.3}{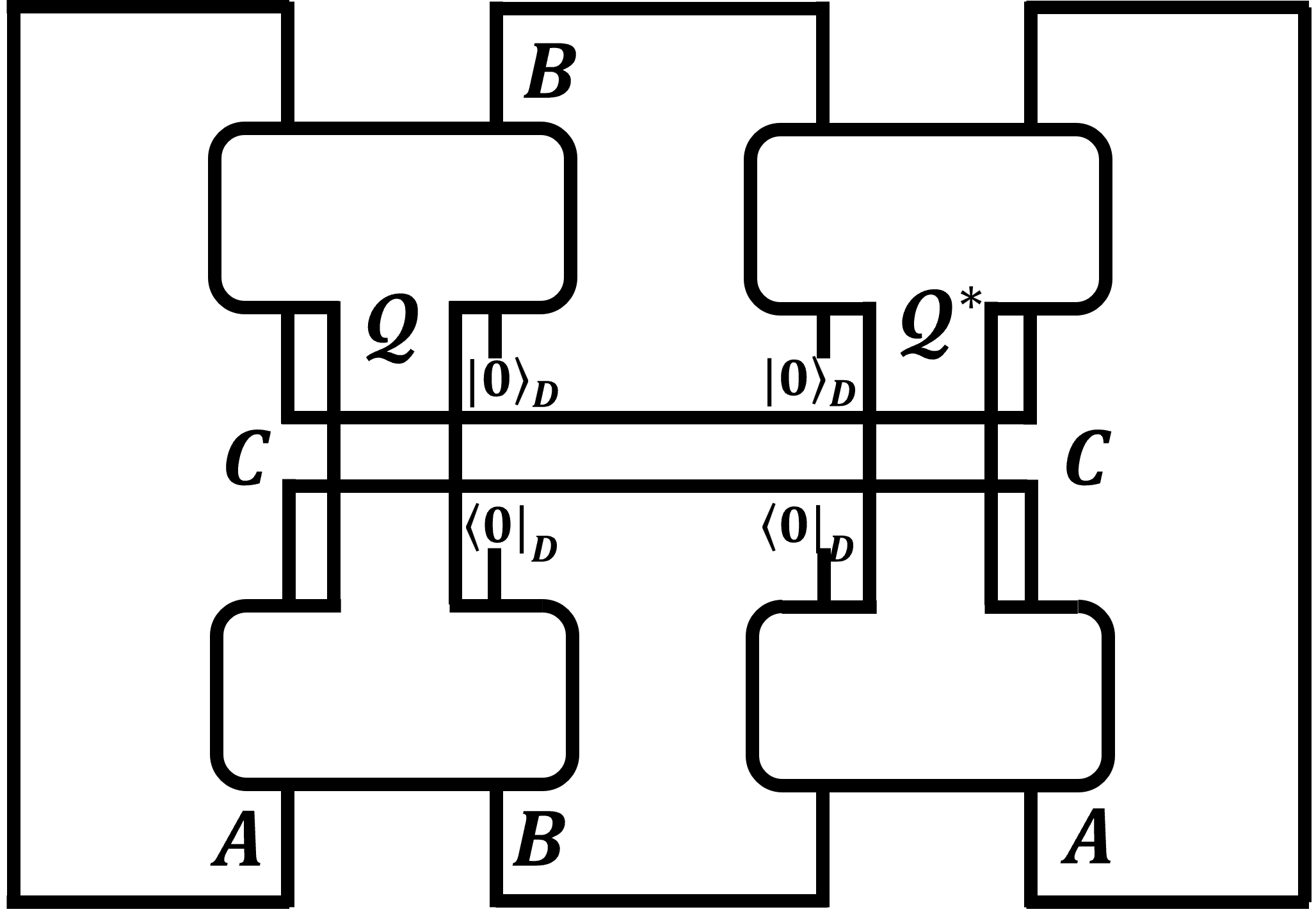}\;.
\end{eqnarray}
The result is given by
\begin{eqnarray}
    \overline{\textrm{Tr}\left[\tilde{\rho}_{\overline{A}C}\cdot\left(\tilde{\rho}_{\overline{A}}\otimes\tilde{\rho}_{C}\right)\right]} = (1-p)^2\;  \overline{\textrm{Tr}\left[\rho_{\overline{A}C}\cdot\left(\rho_{\overline{A}}\otimes\rho_{C}\right)\right]}+\frac{(2p-p^2)}{|B||C|^2}\;,
\end{eqnarray}
where the first term on the right hand side is given by Eq.\eqref{Haar_avr_3rho}.

\acknowledgments

We would like to thank Kun Zhang and Xuanhua Wang for useful discussions. We also acknowledge the service of IBM Quantum for this work.



\bibliographystyle{JHEP}
\bibliography{biblio.bib}

\end{document}